\documentclass[12pt,preprint]{aastex}
\usepackage{mathrsfs}
\begin{document}

\title{
    Dynamics of the Sharp Edges of Broad Planetary Rings\vspace*{0.25in}
}

\author{
    Joseph M. Hahn
}
\affil{
    Space Science Institute\\
    10500 Loring Drive\\
    Austin, TX, 78750\\
    email: jhahn@spacescience.org\\
    phone: 512-291-2255\vspace*{0.55in}
}

\author{
    Joseph N.\ Spitale
}
\affil{
    CICLOPS\\
    Space Science Institute\\
    4750 Walnut Street\\
    Suite 205\\
    Boulder, CO, 80301\\
    email: spitale@pirl.lpl.arizona.edu\\
    phone: 520-207-8782\vspace*{0.55in}
}

\author{
    Carolyn C.\ Porco
}
\affil{
    CICLOPS\\
    Space Science Institute\\
    4750 Walnut Street\\
    Suite 205\\
    Boulder, CO, 80301\\
    email: carolyn@ciclops.org\\
    phone: 720-974-5849\vspace*{0.25in}
}

\author{
    Submitted for publication\\
    in the {\it Astrophysical Journal}\\
    December 29, 2008\\
    Accepted April 3, 2009
}

\begin{abstract}

The following describes a model of a broad planetary ring whose sharp edge is
confined by a satellite's $m^{\mbox{\scriptsize th}}$ Lindblad resonance (LR).
This model uses the streamline formalism of \cite{BGT82, BGT85} to
calculate the ring's internal forces, namely, ring gravity, pressure,
and viscosity. The model also allows for the possibility of a drag force
that can affect small ring particles directly, and large ring particles
indirectly via collisions with the small. The model calculates
the streamlines' forced eccentricities $e$, their longitudes of peripase
$\tilde{\omega}$, and the surface density $\sigma$ throughout the perturbed ring.
This model is then applied to the outer edge of Saturn's B ring, which is maintained by
an $m=2$ inner LR with the satellite Mimas. A suite of ring models are
used to illustrate how a ring's perturbed state depends on
the ring's physical properties: its surface density, its viscosity,
the ring particles' dispersion velocity, and the strength of the hypothetical
drag force. A comparison of model results to the outer B ring's observed
properties suggests
that the ring's surface density there is $10\lesssim\sigma\lesssim280$ gm/cm$^2$
in the ring's outermost $\sim40$ km.

The ring's sharp edge identifies the site where the ring's viscous
torque precisely counterbalances the 
perturbing satellite's gravitational torque on the ring.
However, an examination of several 
seemingly conventional viscous B ring models  shows that they all
fail, by wide margins, to balance these torques at the ring's outer edge.
This is partly due to the ring's self-gravity,
which tends to reduce forced eccentricities near the resonance. But this is also
due to the fact that a viscous ring tend to be nearly peri-aligned with the satellite.
Both effects conspire to reduce the satellite's torque on the ring,
which in turn makes the ring's edge more difficult to maintain.
Nonetheless, the following shows
that a torque balance can still be achieved in a viscous B ring, but only
in an extreme case where the ratio of the ring's
bulk/shear viscosities satisfy $\nu_b/\nu_s\sim10^4$.
However, if the dissipation of the ring's forced motions
is instead dominated by 
a weak drag force,  then the satellite can exert a much stronger 
torque across a wider annulus in the ring, which can
successfully counterbalance the ring's viscous torque there.
We also show how this streamline model can be adapted to study other 
interesting ring phenomena,
such as narrow eccentric ringlets and nonlinear spiral density waves.\vspace*{2ex}

\end{abstract}


\section{Introduction}
\label{intro}

The outer edge of Saturn's main B ring is confined
by an $m=2$ inner Lindblad resonance (ILR) with the satellite Mimas,
while the outer edge of the main A ring is confined by $m=7$ ILRs
with the coorbital satellites Janus and Epimetheus (Porco et al 1984).
Ring particles orbiting near a resonance  
execute $m$ radial excursions, or epicycles,
which gives a ring's edge a scalloped, $m$-lobed appearance.
The ring particles' epicyclic amplitude $R_m$, which is the half-amplitude of the 
particles' radial excursions, is obviously governed in part by the mass of the
perturbing satellite. However the ring's internal
forces---self gravity, pressure, and viscosity---also play a role here by
tempering the ring's response to the satellite's resonant gravitational perturbations.
Consequently, modeling these rings in a way that accounts for all of these perturbations,
and then comparing those models to observations of the ring's edge, should 
allow one to assess the relative importance of a ring's various internal forces.
This activity will also allow one to make estimates of, or place limits on, the ring's
physical parameters, such as its surface density $\sigma$, viscosity $\nu$, 
and the ring particles' dispersion velocity $c$. Also note that small ring particles 
are susceptible to drag forces---plasma drag, Poynting-Robertson drag,
and/or the Yarkovsky effect---whose significance can also
be assessed by this kind of modeling, too.

The epicyclic amplitudes of the outer A and B rings are quite small,
$R_m\sim10$'s of km, while the ring's radii are $r\sim10^5$ km, so the
ring particles' noncircular motions are only $\sim0.01\%$ in a fractional sense.
The smallness of those motions also makes any time-dependent ring models, 
such as Nbody, SPH, or hydrodynamic treatments, very difficult, 
due to the very many particles or hydrodynamic cells
needed to simulate the ring-edge's entire circumference.
Also keep in mind that one does not know the ring's equilibrium state in advance,
so  simulations of a ring's time-evolution would initially be dominated by
transient motions that can easily mask the ring's small forced motions.
Consequently, time-dependent models must also evolve the system
until the ring's dissipation has damped out those transients, which can also cost a
lot of CPU time.

Semi-analytic methods instead provide a much more efficient and illuminating
method for studying perturbed planetary rings. These methods are enabled by
the {\em streamline} concept, which is a very powerful tool for studying planetary
rings. A streamline is the epicyclic path that is traced by numerous ring particles that 
all have a common semimajor axis. A planetary ring
can then be thought of as a sum of many such streamlines. This streamline approach
also allows one to calculate the forces that any one streamline exerts on any
one ring particle. Summing over all the forces exerted by all of the ring's
streamlines, and inserting those forces 
into Newton's equations of motion, then provides
a model  that can be used to calculate
the perturbed ring's equilibrium state.

Streamlines were first used to show how a narrow ringlet's self-gravity can counter
the differential precession that occurs when orbiting an oblate planet
\citep{GT79}. A streamline approach was also used to show how viscosity
helps a satellite maintain a planetary ring's sharp edge \citep{BGT82},
and streamlines were used to examine how the gap-embedded satellite Pan
maintains the Encke gap in Saturn's A ring \citep{BGT89}. Streamlines are
also used to study nonlinear spiral density waves \citep{BGT86, LB86}. Evidently,
the streamline concept is a very general tool that can be used to
study a variety of interesting ring phenomena. And in the following, we use a
streamline formalism to examine in detail how a satellite's
$m^{\mbox{\scriptsize th}}$ ILR can disturb as well as
maintain the sharp outer edges of Saturn's main rings.

Section \ref{EOM} reviews the subject in detail, and derives all
of the equations that will be needed to assemble a comprehensive model
of a broad planetary ring whose edge is confined by a satellite's
$m^{\mbox{\scriptsize th}}$ Lindblad resonance. However, a reader who is uninterested
in the many details and derivations can skip ahead to Section \ref{sims},
which examines a suite of B ring models that illustrate how
the perturbed ring's state---its epicyclic amplitude, its orientation, and it surface density
variations---all depend on the ring's physical properties: $\sigma$, $\nu$, and $c$.
This section also shows how observations can be used to infer a ring's
physical properties, which for the B ring are quite unknown.
Section \ref{discussion} then discusses some important side issues,
while Section \ref{results} summarizes our main findings.
 
\section{Equation of motion}
\label{EOM}

This Section derives the equation of motions that will be used to
calculate the motion of an orbiting ring particle while it is
perturbed by an orbiting satellite and the planetary ring. The particle's position vector
$\mathbf{r}(t)$ evolves over time $t$ according to
Newton's second law of motion,
\begin{equation}
    \label{N2}
    \ddot{\mathbf{r}} = \frac{d^2\mathbf{r}}{dt^2} = -\nabla\Phi + \mathbf{a}
\end{equation}
where $\Phi(\mathbf{r})=\Phi_p+\Phi_s$ is the
total gravitational potential that is
due to the central planet $\Phi_p$ and satellite $\Phi_s$, and $\mathbf{a}$
is the acceleration of the particle that is due to the forces exerted by
the planetary ring. For simplicity the following assumes that the satellite's orbit is
circular and coplanar with the ring plane, 
but these results are unchanged if the satellite's
orbit is instead slightly noncircular or inclined. Polar coordinates will also be
used, where $\mathbf{r}=(r, \theta)$
with $r(t)$ being the particle's distance from the planet's center,
and $\theta(t)$ its longitude measured relative to some $\mathbf{\hat{x}}$
axis, with the $\mathbf{\hat{z}}$ axis perpendicular to the orbital
plane. Solutions to the equation of motion (\ref{N2}) are then obtained
after Fourier expanding the perturbations that are
acting on the orbiting ring particle.

\subsection{Fourier expansions of the perturbations}

Planet-centered coordinates will be used, so
the satellite's gravitational potential has direct and indirect parts
that are
\begin{equation}
    \label{Phi_s indirect}
    \Phi_s(r, \theta) = -\frac{Gm_s}{|\mathbf{r-r_s}|} + 
        \frac{Gm_s}{r_s^3}\mathbf{r\cdot r}_s,
\end{equation}
where $G$ is the gravitational constant, $m_s$ is the satellite's mass,
and $\mathbf{r}_s=(a_s, \theta_s)$ is its position vector in
polar coordinates, with $a_s$ being the satellite's semimajor axis
and $\theta_s$ its longitude. A Fourier expansion of that potential is
\begin{equation}
    \label{Phi_s exact}
    \Phi_s(r, \theta) = \frac{1}{2}\phi_s^0(r) + 
        \Re e\left(  \sum_{m=1}^\infty\phi_s^m(r)e^{im(\theta-\theta_s)}  \right)
\end{equation}
where the potential's Fourier amplitudes $\phi_s^m$ are
\begin{equation}
    \label{phi_s^m}
    \phi_s^m(r) = \frac{1}{\pi}\int_{-\pi}^\pi \Phi_s(r, \varphi)\cos(m\varphi)d\varphi =
        -\frac{Gm_s}{a_s}\left[ b^{(m)}_{1/2}(\beta) -\beta\delta_{m1}\right]
\end{equation}
where $\varphi=\theta-\theta_s$ is the particle's longitude relative to the satellite's,
and the Laplace coefficient
\begin{equation}
    \label{lap_coeff}
    b^{(m)}_{s}(\beta)=
        \frac{2}{\pi}\int_0^\pi\frac{\cos(m\varphi)d\varphi}
       {(1+\beta^2-2\beta\cos\varphi)^s}
\end{equation}
is a function of the ratio $\beta=r/a_s$. The Kronecker delta $\delta_{m1}$
in Eqn.\ (\ref{phi_s^m}) is due to the indirect part of the potential in
Eqn.\ (\ref{Phi_s indirect}), 
which only contributes to the $m=1$ part of the satellite's gravity.

The acceleration that the ring exerts on the particle is
$\mathbf{a}(r, \theta)=a_r\mathbf{\hat{r}} + a_\theta\mathbf{\hat{\theta}}$,
where $a_r$ and $a_\theta$ are the radial and tangential 
components. A Fourier expansion of those accelerations will also have the form
\begin{mathletters}
    \begin{eqnarray}
        a_r(r, \theta) &=& A_r^0(r) + 
            \Re e\left(  \sum_{m=1}^\infty A_r^m(r)e^{im(\theta-\theta_s)}  \right)\\
        a_\theta(r, \theta) &=&  A_\theta^0(r) + 
            \Re e\left(  \sum_{m=1}^\infty A_\theta^m(r)e^{im(\theta-\theta_s)}  \right).
    \end{eqnarray}
\end{mathletters}
These Fourier expansions are
convenient since each of the $m\ge1$ terms Eqn.\ (\ref{Phi_s exact})
correspond to Lindblad resonances that are all spatially segregated.
Consequently, when solving the equation of motion for the particle's motion,
we only need to retain a single $m^{\mbox{\scriptsize th}}$ 
term in the expression for $\Phi_s$. This is also true for the ring's internal
accelerations $a_r$ and $a_\theta$, since they are excited by the satellite's
$m^{\mbox{\scriptsize th}}$ resonant perturbation of the ring.
In light of this, write
\begin{mathletters}
    \label{perturbations}
    \begin{eqnarray}
        \label{Phi_s}
        \Phi_s(r, \theta) &\simeq& \phi_s^m(r)e^{im(\theta-\theta_s)}\\
        \label{a_r}
        a_r(r, \theta) &\simeq&  A_r^0(r) + A_r^m(r)e^{im(\theta-\theta_s)}\\
        \label{a_theta}
        a_\theta(r, \theta) &\simeq&  A_\theta^0(r) + 
       A_\theta^m(r)e^{im(\theta-\theta_s)}
    \end{eqnarray}
\end{mathletters}
with the $\Re e()$ notation dropped henceforth,
so it is to be understood that one is to preserve only the real parts of the following
equations. Also note that the axisymmetric part of the satellite's potential, $\phi_s^0$,
was omitted from Eqn.\ (\ref{Phi_s}), since it is convenient to combine it 
with the planet's
potential, $\Phi_p\rightarrow \Phi_p + \phi_s^0$, which is also axisymmetric.
Lastly, note that $\phi_s^m$, $A_r^0$, and $A_\theta^0$ are all real, while the 
$A_r^m$ and $A_\theta^m$ can be complex.

\subsection{motion near a Lindblad resonance}
\label{LR-section}

The radial and tangential parts of the particle's equation of motion are
\begin{mathletters}
    \label{eom}
    \begin{eqnarray}
        \label{r_eom}
        \ddot{r}-r\dot{\theta}^2 &=& -\frac{\partial\Phi}{\partial r} + a_r\\
        \label{theta_eom}
        \frac{1}{r}\frac{d}{dt}(r^2\dot{\theta}) &=& 
            -\frac{1}{r}\frac{\partial\Phi}{\partial \theta} + a_\theta
    \end{eqnarray}
\end{mathletters}
when Eqn.\ (\ref{perturbations}) is inserted into Eqn.\ (\ref{N2}).
Similar equations are solved in \cite{GT82}\ for an isolated
particle that does not experience a perturbation from the ring
({\it i.e.}, $\mathbf{a}=0$), so the following solution to the more
general $\mathbf{a}\ne0$ problem will use a strategy and notation
similar to that given in \cite{GT82}. A ring particle's
orbit will be nearly circular, so its trajectory has the form
\begin{equation}
    \label{displacements}
    r(t) = r_0 + r_1(t) \qquad\mbox{and}\qquad
        \theta(t) = \theta_0 + \Omega_0t + \theta_1(t)
\end{equation}
where the constant $r_0$ is the particle's mean distance from the planet,
$ \Omega_0$ is its mean angular velocity about the planet,
and $\theta_0$ is an arbitrary phase.
Since the perturbing accelerations $\mathbf{a}$ and $-\nabla\Phi_s$
are all small compared to the central planet's gravity $-\nabla\Phi_p$, 
the particle's displacements from a purely
circular orbit will be small such that
$|r_1|\ll r_0$ and $|\theta_1|\ll 1$,
which then allows the equation of motion to be linearized.

The satellite's orbital angular velocity is $\Omega_s$, so its longitude is
$\theta_s=\Omega_s t$ when time $t=0$ is chosen to be
the time when it traverses the $\mathbf{\hat{x}}$ axis.
This also means that the particle's relative longitude that
appears in Eqn.\ (\ref{perturbations}) is
$m(\theta-\theta_s) = m\theta_0 +\omega_m t + m\theta_1$
where $\omega_m(r) = m(\Omega_0 - \Omega_s)$ is the particle's
Doppler-shifted forcing frequency. But this quantity is usually needed
only to lowest order in the small angles, so
\begin{equation}
    \label{relative-long}
    m(\theta-\theta_s) \simeq m\theta_0 +\omega_m t.
\end{equation}

The particle's
specific angular momentum is $h=r^2\dot{\theta}$, and the time-evolution
of that quantity is obtained from Eqn.\ (\ref{theta_eom}) with $\dot{\theta}=h/r^2$:
\begin{equation}
    \label{dh/dt}
    \frac{dh}{dt} = -\frac{\partial\Phi}{\partial \theta} + ra_\theta
        = -im\phi_s^me^{im(\theta-\theta_s)} + rA_\theta^0
        + rA_\theta^me^{im(\theta-\theta_s)}.
\end{equation}
The first term is the specific torque that the satellite exerts on the ring particle,
while the other terms are the specific torques that are due to the ring's internal
forces. Evidently, the total torque $dh/dt$ 
is the sum of secular ({\it i.e.}, non-oscillatory) 
terms like $r_0A_\theta^0$ plus other oscillatory terms. 
In light of this, write $h(t)=h_0(t) + h_1(t)$ where $h_0=r_0^2\Omega_0$
is the secular part of the particle's specific angular momentum
$h(t)$, while $h_1(t)$ is the oscillatory part. The secular and oscillatory torques
on the particle, $dh_0/dt$ and $dh_1/dt$, are then
\begin{mathletters}
    \label{dh0h1/dt}
    \begin{eqnarray}
        \label{dh0/dt}
        \frac{dh_0}{dt} &\simeq& r_0A_\theta^0 + T_s\\
        \label{dh1/dt}
        \frac{dh_1}{dt}  &\simeq&  
            (-im\phi_s^m +r_0A_\theta^m)e^{i(\theta_0+\omega_m t)}
    \end{eqnarray}
\end{mathletters}
when Eq.\ (\ref{dh/dt}) is
written to lowest order in the particle's coordinates, {\it i.e.}, with
$r\simeq r_0$ and $m(\theta-\theta_s) \simeq m\theta_0 +\omega_m t$.
Note that we have also added to Eqn.\ (\ref{dh0/dt})
an additional term $T_s$ to represent the secular part of the specific torque that
the satellite exerts on the particle; Section \ref{ang-mom-lum} will show 
that this second--order term is important only near the ring's sharp edge.
And if the ring particle's orbit is to be static such that its mean orbit radius $r_0$
is constant, then the secular torque on the particle, $dh_0/dt$, must
be zero, for otherwise that particle (as well as its neighboring ring particles) 
would drift radially. Consequently, equilibrium thus requires
all particles to satisfy the torque-balance equation, $T_s=-r_0A_\theta^0$. 

Integrating Eqn.\ (\ref{dh1/dt}) with respect to time $t$
provides the oscillatory part of the particle's specific angular momentum,
\begin{equation}
    \label{h1}
    h_1 = -\left( \frac{m\phi_s^m}{\omega_m} +\frac{ir_0A_\theta^m}{\omega_m}
            \right) e^{i(m\theta_0+\omega_m t)}.
\end{equation}
Also note that $|h_1|$ is small compared to $h_0=r_0^2\Omega_0$.
The quantity $h_1$ is then used to solve the radial part of the equation of motion,
which is
\begin{equation}
    \label{r_eom2}
    \ddot{r} - \frac{h^2}{r^3} = -\frac{\partial\Phi_p}{\partial r} + A_r^0 +
        \left(-\frac{\partial\phi_s^m}{\partial r} + A_r^m\right)
            e^{i(m\theta_0+\omega_m t)}
\end{equation}
when $\dot{\theta}=h/r^2$ and Eqns.\ (\ref{perturbations}) are substituted into
Eqn.\ (\ref{r_eom}). Inserting $r=r_0+r_1$ and $h=h_0+h_1$ into the above,
Taylor-expanding to first order in the small quantities $r_1$ and $h_1$,
and then inserting $h_0=r_0^2\Omega_0$ and Eqn.\ (\ref{h1}) into that result
then yields
\begin{equation}
    \label{r_eom3}
    \ddot{r}_0 + 
        \left( \frac{\partial\Phi_p}{\partial r}  - r_0\Omega_0^2 - A_r^0 \right)
       + \ddot{r}_1 
       + \left( 3\Omega_0^2 + \frac{\partial^2\Phi_p}{\partial r^2} \right)r_1
       \simeq  \left( -\frac{\partial\phi_s^m}{\partial r} 
       -  \frac{2m\Omega_0}{r_0\omega_m}\phi_s^m
       -\frac{2i\Omega_0}{\omega_m}A_\theta^m + A_r^m \right)
       e^{i(m\theta_0+\omega_m t)},
\end{equation}
where it is understood that all quantities in the above are to be evaluated at $r=r_0$.

The ring particle is assumed to be in torque balance, so $r_0$ is constant and
$\ddot{r}_0=0$. And since the terms in the first set of parentheses in 
Eqn.\ (\ref{r_eom3}) are secular while the remaining terms are oscillatory,
that parentheses, which is the condition for centrifugal equilibrium,
must separately sum to zero, which provides the particle's mean angular 
velocity $\Omega_0=\Omega(r_0)$ where
\begin{equation}
    \label{Omega^2}
    \Omega^2 = \frac{1}{r}\frac{\partial\Phi_p}{\partial r}  - \frac{A_r^0}{r}.
\end{equation}
The constant in the second parentheses in Eqn.\ (\ref{r_eom3}) is the particle's
epicyclic frequency, $\kappa_0=\kappa(r_0)$, where
\begin{equation}
    \label{kappa^2}
    \kappa^2 = 3\Omega^2 + \frac{\partial^2\Phi_p}{\partial r^2}.
\end{equation}
Since the ring's radial acceleration $|A_r^0|$ is small compared to the
central planet's gravity, these angular frequencies are
\begin{mathletters}
    \label{Omega_kappa_approx}
    \begin{eqnarray}
        \label{Omega_approx}
        \Omega &\simeq& \Omega_p\left(1-\frac{A_r^0}{2r\Omega_p^2}\right) \\
        \label{kappa_approx}
        \mbox{and} \qquad \kappa &\simeq&
            \kappa_p\left(1-\frac{3A_r^0}{2r\kappa_p^2}\right)
    \end{eqnarray}
\end{mathletters}
where $\Omega_p=\sqrt{r^{-1}\partial\Phi_p/\partial r}$ would be the particle's
angular velocity if ring forces were absent, and
$\kappa_p=\sqrt{4\Omega_p^2+r\partial\Omega_p^2/\partial r}$
would be its epicyclic frequency when $A_r^0=0$.

The terms on the right of Eqn.\ (\ref{r_eom3})
that involve $\phi_s^m$ are the satellite's forcing function,
\begin{equation}
    \label{ff}
    \Psi_s^m(r) = -\frac{\partial\phi_s^m}{\partial r} 
       -  \frac{2m\Omega}{r\omega_m}\phi_s^m,
\end{equation}
which accounts for the satellite's radial and tangential forcings. All the coefficients
on the right hand side of Eqn.\ (\ref{r_eom3}) will be known as the system's
complex forcing function,
\begin{equation}
    \label{cff}
    \Psi_c^m(r) = \Psi_s^m - \frac{2i\Omega_0}{\omega_m}A_\theta^m + A_r^m.
\end{equation}
Inserting this into the above then casts Eqn.\ (\ref{r_eom3}) in its simplest form,
\begin{equation}
    \label{r_eom_final}
    \ddot{r}_1 + \kappa_0^2 r_1 \simeq  \Psi_c^m(r_0) e^{i(m\theta_0+\omega_m t)}.
\end{equation}

\subsection{single particle motion}
\label{spm}

When the ring's internal forces are absent, {\it i.e.}, $\mathbf{a}=0$,
then $\Psi_c^m=\Psi_s^m$ is a constant, and Eqn.\ (\ref{r_eom_final})
describes a driven simple harmonic oscillator whose solution is
\begin{equation}
    \label{r1_single}
    r_1(t) = -R_me^{i(m\theta_0+\omega_m t)}.
\end{equation}
This solution is examined in \cite{GT82}, which is summarized here
since those results are used throughout this study.
Inserting Eqn.\ (\ref{r1_single}) into Eqn.\ (\ref{r_eom_final}) then yields
the particle's epicyclic amplitude $R_m$, which is
\begin{mathletters}
    \begin{eqnarray}
        \label{R_epi_single}
        R_m &=& -\frac{\Psi_s^m}{D(r_0)} \\
        \mbox{where} \quad D(r) &=& \kappa^2 - \omega_m^2
    \end{eqnarray}
\end{mathletters}
is the particle's distance from resonance in frequency-squared units.
When a particle is far
from a resonance, $|D|$ is of order $\Omega^2$, and the particle's epicyclic
amplitude $R_m$ is negligibly small. However, when the particle is near a resonance,
$|D|\ll\Omega^2$, and the particle's response to the satellite's resonant 
forcing is much larger. Exact resonance is the site where $D(r_r)=0$,
or where $\kappa=\epsilon\omega_m$ with $\epsilon\pm1$. If the central planet's
potential $\Phi_p$ were Keplerian, then $\Phi_p=-GM_p/r$
where $M_p$ is the planet's mass and
$\kappa=\Omega=\sqrt{GM_p/r^3}$. Inserting
this into the resonance condition
$\kappa=\epsilon\omega_m=\epsilon m(\Omega-\Omega_s)$
then yields the radius of the Lindblad resonance,
\begin{equation}
    \label{approximate resonance}
    r_r = (1-\epsilon/m)^{2/3}a_s
\end{equation}
where $a_s$ is the satellite's orbit radius. Resonances having $\epsilon=+1$ are
inner Lindblad resonances (ILRs) since they reside interior to the satellite's
orbit, while those with $\epsilon=-1$ are outer Lindblad resonances (OLRs).
The focus of this work will be on the outer edge of Saturn's main B ring,
which is confined by an ILR with a satellite that orbits exterior to the ring, so
$\epsilon=+1$ here. Lastly, note that if the central planet is oblate, then the
particle's epicyclic frequency $\kappa$ differs slightly from its angular
velocity $\Omega$, so Eqn.\ (\ref{approximate resonance}) is
only approximately true. However, a precise calculation of the resonance
location is also given in Section \ref{resonance-location}.

For a particle orbiting near a LR it is sufficient to linearize $D(r)$ via
$D\simeq x {\cal D}$ where $x=(r-r_r)/r_r$ is the particle's fractional
distance from resonance, and
\begin{equation}
    \label{cal-D-repeat}
   {\cal D} = \left. r\frac{dD}{dr}\right|_{r_r}=3\epsilon(m-\epsilon)\Omega_0^2.
\end{equation}
Inserting  Eqn.\ (\ref{phi_s^m}) into (\ref{ff}) and noting that
$\omega_m\simeq\epsilon\Omega$ near a LR also provides the 
satellite's forcing function, which is
\begin{equation}
    \label{ff2-repeat}
    \Psi_s^m(r_r) = \epsilon f_\epsilon^m \mu_s r_0\Omega_0^2
\end{equation}
where $\mu_s$ is the satellite's mass in units of the central planet's,
and $f_\epsilon^m$ is
\begin{equation}
    \label{f_eps^m}
    f_\epsilon^m = \epsilon\beta^2\frac{\partial b^{(m)}_{1/2}}{\partial\beta}
        + 2m\beta b^{(m)}_{1/2} - (2m+\epsilon)\beta^2\delta_{m1},
\end{equation}
which is a positive numerical coefficient that
depends on the resonance in question. For instance, this study is interested
in Mimas' $m=2$ ILR in the B ring, which has $\epsilon=+1$
and $f_\epsilon^m=1.500$. Inserting these quantities into
Eqn.\ (\ref{R_epi_single}) then provides the particle's epicyclic amplitude,
\begin{equation}
    \label{R_epi_single2}
     R_m = -\frac{f_\epsilon^m\mu_sr_0}{3(m - \epsilon) x},
\end{equation}
as well as its forced eccentricity 
\begin{mathletters}
    \begin{eqnarray}
        \label{e_forced_single}
        e &=& \frac{|r_1|}{r_0} =
           \frac{f_\epsilon^m\mu_s}{3(m - \epsilon)|x|} =\left| \frac{\psi_s}{x}\right| \\
        \label{psi}
        \mbox{where} \quad \psi_s &\equiv& \frac{\Psi_s^m}{|r_0{\cal D}| }
            =\frac{\epsilon f_\epsilon^m\mu_s}{3(m - \epsilon)}
    \end{eqnarray}
\end{mathletters}
is a dimensionless version of the satellite's forcing function.

Note that $R_m$ is real and has the opposite sign as $x$. This then allows a
parametrization of the particle's radial displacement  $r_1$
in terms of its longitude $\theta$,
\begin{equation}
    \label{r1_single2}
    r_1(\theta) \simeq \mbox{sgn}(x)|R_m|\cos m(\theta-\theta_s),
\end{equation}
when Eqn.\ (\ref{relative-long}) is utilized. Also
note that the magnitude of the particle's radial excursion $|r_1|$ is maximal
when it arrives at that satellite's longitude, $\theta=\theta_s$. Consequently,
if the particle's orbit lies interior to the resonance with $x<0$, then $r_1<0$,
so the particle must also be at periapse since $r(\theta=\theta_s)=r_0+r_1=r_0-|R_m|$.
Similarly, a particle orbiting exterior to resonance would be at apoapse,
$r=r_0+|R_m|$, when in conjunction with the satellite.
So in summary, orbits interior to a LR 
are expected to be peri-aligned with the satellite's
longitude, while orbits exterior should be apo-aligned.

Interestingly, the outer edge of Saturn's B ring does not behave according to these
expectations. Voyager observations revealed that the B ring's outer edge actually
lies about 24km exterior to the resonance \citep{PDG84}.
This is quite a curiosity, since, if Mimas'
$m=2$ ILR is indeed responsible for maintaining this ring's outer edge, then one 
might expect the ring's outer edge to lie at or just interior to the resonance.
Another curiosity is that the ring is peri-aligned with Mimas, whereas
one would expect the ring to be apo-aligned if it truly extended beyond the resonance.
\cite{PDG84} suggest that this interesting behavior might be a consequence
of the ring's internal forces, which are considered in Section \ref{internal forces}.

\subsubsection{resonance location}
\label{resonance-location}

A Lindblad resonance is the site $r=r_r$ that satisfies $D(r_r)=0$,
which is also where 
\begin{equation}
    \label{resonance-criterion}
    \kappa(r_r)=\epsilon\omega_m(r_r)=\epsilon m[\Omega(r_r)-\Omega_s],
\end{equation}
with the frequencies $\Omega$ and $\kappa$ defined by
Eqns.\ (\ref{Omega^2}) and (\ref{kappa^2}).
Those quantities also depend on gradients in the
central planet's gravitational potential $\Phi_p$, which is
\begin{equation}
    \label{Phi_p}
    \Phi_p(r)=-\frac{GM_p}{r}\left[ 1 
        - \sum_{k=1}^\infty P_{2k}(0)J_{2k}\left(\frac{r}{R_p}\right)^{-2k}\right]
\end{equation}
for bodies orbiting in an oblate planet's equatorial plane, 
where the $J_{2k}$ are the planet's zonal harmonics and the $P_{2k}$
are Legendre polynomials \citep{MD99}. 
All of the Saturnian system's physical constants
that are used in the model, $GM_p$, $J_{2k}$, etc.,
are listed in Table \ref{Saturn-table}. For instance,
solving Eqn.\ (\ref{resonance-criterion}) for 
the location of Mimas' $m=2$ ILR yields
$a_r=117,553.71$ km when the standard $R_p=60,330$ km is adopted
as Saturn's radius. The position $a_r$ will be referred to as the
{\em nominal} resonance position, since exact resonance can  
be shifted further by the ring's internal forces, which are assessed below.

\subsection{the ring's internal forces}
\label{internal forces}

A ring particle is also subject to additional forces, such as the gravity exerted
by the entire ring, and collisions with other ring particles. The cumulative
effect of those collisions will be modeled here as if they were due to
pressure (c.f., \citealt{GT78}). A ring particle is also subject to viscous effects, which
could be due to collisions among ring particles \citep{GT82}, or perhaps due to
transient self-gravitating wakes in the ring  \citep{DTI01}
and/or the viscous overstability \citep{SSS01}. 
Small ring particles are also susceptible to a number of 
other drag forces \citep{GT82} that can be
be accounted for here, too.
The following subsections will assess the ring's internal forces, the calculation
of which is greatly facilitated by the streamline concept.
\cite{BGT82} note that Eqn.\ (\ref{r1_single2})
also represents a streamline in the ring,
which is a closed trajectory that is traced
by numerous ring particles having a common semimajor axis $r_0$.
A broader planetary ring can then be regarded as being comprised
of numerous streamlines having distinct semimajor axes $r_0$.
As \cite{BGT85} show, the calculation of the acceleration
$\mathbf{a}$ that the entire ring exerts on a single particle
can be simplified substantially by considering the particle's interaction
with individual streamlines. But doing that will first require understanding
the relationship between the ring's surface density $\sigma(r, \theta)$
and a streamline's semimajor axis $a=r_0$, eccentricity $e$,
and its longitude of periapse $\tilde{\omega}$.

\subsubsection{streamlines}
\label{streamlines}

Equation (\ref{cff}) shows that if the ring's internal
forces have a tangential component, then the
equation of motion (\ref{r_eom_final}) for a ring
particle's radial displacement $r_1$ is complex, so write
\begin{equation}
    \label{r1}
    r_1(t) = -{\cal R}_me^{i(m\theta_0+\omega_m t)}
        \simeq -{\cal R}_me^{im(\theta-\theta_s)}
        = -R_me^{im(\theta-\theta_s-\tilde{\omega})}
\end{equation}
where ${\cal R}_m(a)\equiv R_me^{-im\tilde{\omega}}$ is the particle's
(or streamline's) complex epicyclic amplitude, and $\tilde{\omega}(a)$
its longitude of periapse, both of which are to be regarded as functions of
the streamline's semimajor axis $r_0=a$. The particle's total planetocentric
distance is then 
\begin{equation}
    \label{r}
    r(a, \theta) = a + r_1 \simeq a -\Re e[{\cal R}_m(a)e^{im(\theta-\theta_s)}]
        = a -R_m(a)\cos m(\theta-\theta_s-\tilde{\omega}).
\end{equation}
Also identify $e(a)=R_m/a$ as the particle's forced eccentricity,
with $e$ and $R_m$ to be 
regarded as positive real quantities.

The following will also need the ring particle's longitudinal displacement, $\theta_1$,
which is obtained by linearizing its
total specific angular momentum $h=r^2\dot{\theta}\simeq
r_0^2\Omega_0+r_0^2\dot{\theta}_1+2r_0\Omega_0r_1=h_0+h_1$,
so $h_0=r_0^2\Omega_0$ as expected and
\begin{equation}
    \label{dtheta1}
    \dot{\theta}_1 = -\frac{2\Omega_0r_1}{r_0}+\frac{h_1}{r_0^2}
        =\left( \frac{2\Omega_0{\cal R}_m}{r_0} -\frac{m\Phi_s^m}{r_0^2\omega_m}
         + \frac{iA_\theta^m}{r_0\omega_m} \right)e^{i(m\theta_0+\omega_mt)},
\end{equation}
which is integrated to obtain
\begin{equation}
    \label{theta1}
    \theta_1 \simeq 
        \left( -\frac{2i\epsilon{\cal R}_m}{r}
          + \frac{im\Phi_s^m}{r^2\Omega^2}
         + \frac{A_\theta^m}{r\Omega^2} \right)e^{im(\theta-\theta_s)}
\end{equation}
since $\omega_m\simeq\epsilon\kappa\simeq\epsilon\Omega$ near a LR.
However the particle's epicyclic amplitude ${\cal R}_m$
will be large near a resonance, dwarfing the other terms, so
\begin{equation}
    \label{theta1_approx}
    \theta_1 \simeq  \frac{2i\epsilon\ r_1}{a}
\end{equation}
and
\begin{equation}
    \label{dtheta2}
    \dot{\theta}_1 \simeq 2e\Omega_0\exp{[im(\theta-\theta_s - \tilde{\omega})]}.
\end{equation}

\subsubsection{nonlinear surface density}
\label{surface density}

To calculate a perturbed ring's surface density $\sigma$,
it is convenient to make the `local' approximation,
which assumes that the interparticle forces are exerted primarily by
nearby ring particles that lie a small distance $\ell$ away from
the perturbed particle, where $\ell\ll a$.
Because these perturbing particles reside on streamlines that are close by,
one can ignore the streamlines' curvature that only occurs at great
distances $\gg\ell$, and instead treat the streamlines as if they were straight
wires having a linear mass density $\lambda(a)$ that is essentially
constant\footnote{Actually, $\lambda$ varies around the orbital ellipse by a small
fractional amount that is of ${\cal O}(e)\ll1$, which may be neglected.}
about a streamline whose semimajor axis is $a$. Note, however, that
perturbations of the ring also changes its surface density $\sigma$, since these
streamlines can be compressed or rarefied in the radial direction. 
To see this, let $\delta m=\lambda\delta\ell$ be the total mass that
resides along a segment of a streamline whose tangential length is $\delta\ell$.
If the ring were unperturbed, 
$\delta m =  \lambda\delta\ell= \sigma_0\delta a\delta\ell$,
where $\delta a$ is the streamline's radial width, so $\lambda = \sigma_0\delta a$
is the streamline's linear density. If, however, the ring were perturbed,
then $\delta m=\lambda\delta\ell = \sigma\delta r\delta\ell$, {\it i.e.}
$\lambda = \sigma_0\delta a = \sigma\delta r$
where $\sigma(a,\theta)$ is the perturbed ring's surface density.
Consequently, mass conservation requires
$\sigma(a,\theta)=\sigma_0(a)/J$ where
$J(a,\theta)=\delta r/\delta a\rightarrow\partial r/\partial a$
in the limit that the streamline width
$\delta a\rightarrow0$. Thus
\begin{equation}
    \label{sigma}
    \sigma(a, \theta) = \frac{\sigma_0(a)}{J} 
\end{equation}
where
\begin{equation}
    \label{J}
    J(a, \theta) = \frac{\partial r}{\partial a}
        ={\Re e[1-{\cal R}'_me^{im(\theta-\theta_s)}]}
\end{equation}
is a measure of the streamlines' degree of compression \citep{BGT85}.
Note that this quantity is sensitive to gradients in the streamlines' orbit elements, since
\begin{equation}
    \label{R'}
    {\cal R}'_m = \frac{\partial{\cal R}_m}{\partial a} = 
        \left( R'_m - im\frac{\partial\tilde{\omega}}{\partial a}R_m \right)e^{-im\tilde{\omega}}
        = \left[\frac{\partial(ae)}{\partial a} + ikae\right]e^{-im\tilde{\omega}}
\end{equation}
where $R_m=ae$ and the wavenumber
$k\equiv-m(\partial\tilde{\omega}/\partial a)$ is the rate at which the streamline's
longitude's of periapse varies with semimajor axis $a$. The above can also be written
more compactly as
\begin{mathletters}
    \label{R'-q-eta}
    \begin{eqnarray}
        \label{R'q}
        {\cal R}'_m &=& qe^{i(\eta- m\tilde{\omega})}\\
        \label{q-delta}
         \mbox{where}\quad q &=& \sqrt{ (e+e')^2 + (kae)^2}
         \quad\mbox{and}\quad \tan\eta = \frac{kae}{e+e'}
    \end{eqnarray}
\end{mathletters}
since $\partial(ae)/\partial a = e +de/dx\equiv e + e'$, so
\begin{equation}
    \label{J2}
    J = 1 -q\cos(\phi + \eta)
        = 1 - \Re e\left[qe^{i(\phi + \eta)} \right]
        = \Re e\left\{1 - \frac{\partial}{\partial a}[ea\exp(i\phi)]\right\}.
\end{equation}
where $\phi=m(\theta-\theta_s - \tilde{\omega})$. The angle $\eta$
indicates when the ring's response to perturbations takes the
form of a spiral density wave. When the ring's responds is a tightly-wrapped wave,
$|kae|\gg|e + e'|$ and so $|\eta|\simeq\pi/2$, while $|\eta|\ll1$ when there ring's response
is not wavelike.

Note that $q=|{\cal R}'_m|$ is the nonlinearity parameter of
\cite{BGT85}, who point out that adjacent streamlines cross if
their separation $|\delta r|$ exceeds $|\delta a|$ at any longitude $\theta$,
which would occur if $q\ge1$. However previous studies have shown that
a ring's internal forces tend to adjust the streamlines so as to avoid
crossing (c.f., \citealt{SYL85}), and we expect $q<1$ here, too.

The ring's response to perturbations is said to be linear when
$q\ll1$. In that case, the fractional variations in the ring's surface
density, $(\sigma-\sigma_0)/\sigma_0\simeq{\cal R}'_me^{im(\theta-\theta_s)}$,
are sinusoidal and of low amplitude. However, when the ring is nonlinear, 
$q=|{\cal R}'_m|$ is not small, and large variations in $\sigma(a, \theta)$ can occur.
Although those variations in $\sigma(a, \theta)$ would be periodic in longitude $\theta$,
Eqn.\ (\ref{sigma}) shows that they would not be 
sinusoidal. Lastly, note that Eqn.\ (\ref{sigma}) also implies that a nonlinear ring's
surface density can drop by no more than $50\%$ since 
$|{\cal R}'_m|<1$, which also occurs in the troughs of
nonlinear spiral density waves \citep{SYL85}.

If this study were instead interested in spiral density waves, we would then note
that tightly wrapped spiral waves have wavenumbers $|ka|\gg1$ and
amplitudes $e(a)\sim$ constant. In this case, 
the second term in Eqn.\ (\ref{R'}) would dominate over the first,
so ${\cal R}'_m\simeq ik{\cal R}_m$. This is known as the
tight-winding approximation, and it is valid when $|kae| >> |e+e'|$ such
that $\eta\simeq\pm\pi/2$ in this case. If we were to insert
this into Eqn.\ (\ref{sigma}) and proceed further, we would
then arrive at the theory for nonlinear spiral density waves
(c.f.\ \citealt{BGT86}), which will in fact be considered in a followup study.

However this effort is interested in the motion of particles orbiting near
an inner Lindblad resonance, most of whom are on the non-wave side of the resonance.
Also recall the results of Section \ref{spm}, which suggests that the ring particles
there will have similar longitudes of perihelia $\tilde{\omega}$.
Since $\tilde{\omega}$ varies slowly with radial distance $x$,
the $|kae|$ term in the above is usually small in comparison to
$|e+e'|\simeq|e'|$ term, since the
streamline's eccentricities $e(a)$ grow rapidly with proximity to the resonance.
Thus ${\cal R}'_m \simeq (de/dx) e^{-im\tilde{\omega}}$,  $qe^{i\eta}\simeq e'$,
$\eta\simeq0$ or $\pi$, and $\sigma(a, \theta) = \sigma_0(a)/J \simeq
\sigma_0/[1-e'\cos m(\theta-\theta_s-\tilde{\omega})]$ in most
(but not all) of the scenarios considered here.

Finally, recall that the forced eccentricity of a single isolated ring particle is
$e(x)=\psi_s/x$ (Eqn.\ \ref{e_forced_single}), so
the threshold for streamline crossing would be
$|{\cal R}'_m|=|de/dx|=\psi_s/x^2=1$. Solving for $|x|\equiv x_{NL}$
then provides a rough estimate for the fractional distance from resonance
where the ring's surface density variations will be nonlinear, and where
the ring's internal forces should be significant (e.g., \citealt{BGT82}):
\begin{equation}
    \label{x_NL}
    x_{NL} = \sqrt{|\psi_s|} = \sqrt{ \frac{f_\epsilon^m\mu_s}{3(m - \epsilon)} }.
\end{equation}
This study is interested in Mimas' $m=2$ ILR that lies near the B ring's outer edge,
which has $\epsilon=1$ and $f_\epsilon^m=1.50$.
The perturbing satellite's  mass is 
$\mu_s=6.60\times10^{-8}$ Saturn masses, and its $m=2$
resonance lies at $a_r=117,553.7$km from Saturn's center, so
$\psi_s=3.30\times10^{-8}$, $x_{NL}=1.8\times10^{-4}$, and
$\Delta a_{NL}=x_{NL}a_r\simeq20$km is the physical distance from resonance where
nonlinear effects play a role in the ring's dynamics.
We also note that the outer edge of the A
ring is maintained by an m=7 ILR with the Janus/Epimetheus coorbital pair.
That resonance has $f_\epsilon^m=9.49$, 
$\mu_s=3.32\times10^{-9}$ (which is the mass of Janus, the larger of the
two coorbitals, \cite{PTW07}),
and a mean semimajor axis $a_r=136,773$km \citep{PDG84},
so $\psi_s=1.75\times10^{-9}$,
$x_{NL}=4.2\times10^{-5}$, and $\Delta a_{NL}\simeq6$km. It is over these
spatial scales where the ring's internal forces, such as self gravity, are expected to
play an important role.

\subsubsection{ring gravity}
\label{gravity}

In the local approximation, a perturbing streamline is an infinitely long,
straight wire having a linear mass density $\lambda(a')=\sigma_0(a')\delta a'$, where
$\sigma_0(a')$ is the unperturbed ring's surface density,
$a'$ its semimajor axis, and  $\delta a'$
is its radial width in semimajor axis space. 
The gravitational acceleration that
streamline exerts on a nearby particle is
$\delta a_g=2G\lambda(a')/d=2G\sigma_0(a')\delta a'/d$,
where $d=r'-r'$ is the separation between the perturbing streamline,
whose planetocentric distance is $r'$, and the perturbed ring particle at $r$.
Since the streamlines' eccentricities $e(a)$ are small, the streamlines
are all nearly parallel, so the gravitational forces that they exert are radial.
If the particle's semimajor axis is $a$, then that separation is 
$d=r'-r=a'-a-\Re e\{[R_m(a')e^{-im\Delta\tilde{\omega}'} -R_m(a) ]
e^{im(\theta-\theta_s-\tilde{\omega})}\}$ by Eqn.\ (\ref{r}), where
$\Delta\tilde{\omega}'\equiv\tilde{\omega}(a')-\tilde{\omega}$.
This separation is then written more compactly as
$d=a(x'-x)\Re e[1 - \Delta e^{im(\theta-\theta_s-\tilde{\omega})}]
=a(x'-x)(1-\delta\cos\phi)$, where 
$\phi = m(\theta-\theta_s-\tilde{\omega}) + \beta$ and
\begin{equation}
    \label{Delta}
    \Delta \equiv \frac{R_m(a')e^{-im\Delta\tilde{\omega}'} - R_m(a)}{a'-a} =
        \frac{e(x')e^{-im\Delta\tilde{\omega}'} - e(x)}{x'-x} \equiv \delta e^{i\beta}
\end{equation}
where
\begin{equation}
     \tan\beta = \frac{-e(x')\sin (m\Delta\tilde{\omega}')}
        {e(x')\cos (m\Delta\tilde{\omega}') - e(x)}
\end{equation}
and $\delta=|\Delta|$ is real.
The total acceleration that the entire ring exerts on the particle is then
\begin{equation}
    \label{a_g_exact}
    a_g(a, \theta) = \int_{\mbox{\scriptsize ring}}\delta a_g 
        = \int_{a_{\mbox{\scriptsize in}}}^{a_{\mbox{\scriptsize out}}}
        \frac{2G\sigma_0(a')da'}
        { a(x'-x)(1-\delta\cos\phi) }
\end{equation}
where $a_{\mbox{\scriptsize in}}$ and $a_{\mbox{\scriptsize out}}$ are the 
semimajor axes of those streamlines at the ring's innermost and outermost edges.

Next, Fourier-expand the longitude-dependent factor
$(1-\delta \cos\phi)^{-1}$ in Eqn.\ (\ref{a_g_exact}),  which is
\begin{equation}
    \label{fourier1}
    \frac{1}{1-\delta\cos\phi} = \frac{1}{\sqrt{1-\delta^2}}
        + \frac{2}{\delta}[(1-\delta^2)^{-1/2}-1]\cos\phi + \mbox{ other terms},
\end{equation}
provided $\delta<1$ such that streamlines do not cross.
Those other terms in the above are proportional to
$\cos(n\phi)=\cos[nm(\theta-\theta_s- \tilde{\omega}) + n\beta]$,
where the integer $n$ takes all values of $n\ge2$. Those other terms represent
non-resonant forcings of the ring, and they are negligible
since $R_{nm}$ will be small.
Inserting the surviving terms in Eqn.\ (\ref{fourier1}) back into 
Eqn.\ (\ref{a_g_exact}) then shows that the ring's gravity $a_g$ has
the desired form of Eqn.\ (\ref{a_r}):
\begin{mathletters}
    \label{A_g}
    \begin{eqnarray}
        \label{A_g_0}
        A_{gr}^0(x) &=&  
            2G\int_{a_{\mbox{\scriptsize in}}}^{a_{\mbox{\scriptsize out}}}
            \frac{\sigma_0(a')dx'}{(x'-x)\sqrt{1-\delta^2}}\\
        \label{A_g_m}
        A_{gr}^m(x) &=& 
            4Ge^{-im\tilde{\omega}}
            \int_{a_{\mbox{\scriptsize in}}}^{a_{\mbox{\scriptsize out}}}
            \frac{\sigma_0(a')[(1-\delta^2)^{-1/2}-1]e^{i\beta}}
           {(x'-x)\delta}dx'\\
            &=&4Ge^{-im\tilde{\omega}}
            \int_{a_{\mbox{\scriptsize in}}}^{a_{\mbox{\scriptsize out}}}
            \frac{\sigma_0(a')[(1-\delta^2)^{-1/2}-1]}
           {e(x')e^{im\Delta\tilde{\omega}'}-e(x)}dx'
    \end{eqnarray}
\end{mathletters}
where the $r$ subscripts indicate that these are radial accelerations.

For numerical work it will be convenient to put Eqns.\ (\ref{A_g}) in a dimensionless
form, which is obtained by dividing by $|a{\cal D}|=3(m-\epsilon)a\Omega^2$.
Also let $s(x)=\sigma(x)/\sigma_\infty$ be the ring's fractional surface density,
where the constant $\sigma_\infty$ is the ring's undisturbed surface density
far away from the resonance. Also introduce the so-called normalized disk mass
$\mu_d=\pi\sigma_\infty a^2/M_p= \pi G\sigma_\infty/a\Omega^2$,
which is a dimensionless measure of the ring's unperturbed surface density
that is quite convenient.
With these definitions in hand, the dimensionless version of the above
accelerations become
\begin{mathletters}
    \label{alpha_g}
    \begin{eqnarray}
        \label{alpha_g_0}
        \alpha_{gr}^0(x) &=& \frac{A_{gr}^0}{|a{\cal D}|} =
            \frac{2\mu_d}{3\pi(m-\epsilon)}
            \int_{x_{\mbox{\scriptsize in}}}^{x_{\mbox{\scriptsize out}}}
            \frac{s(x')dx'}{(x'-x)\sqrt{1-\delta^2}}\\
        \label{alpha_g_m}
        \mbox{and}\qquad \alpha_{gr}^m(x) &=& \frac{A_{gr}^m}{|a{\cal D}|} = 
            \bar{\alpha}_{gr}^me^{-im\tilde{\omega}}\\
         \mbox{where}\qquad \bar{\alpha}_{gr}^m(x) &=& 
            \frac{4\mu_d}{3\pi(m-\epsilon)} 
            \int_{x_{\mbox{\scriptsize in}}}^{x_{\mbox{\scriptsize out}}}
            \frac{s[(1-\delta^2)^{-1/2}-1]dx'}{e(x')e^{im\Delta\tilde{\omega}'}-e(x)},
    \end{eqnarray}
\end{mathletters}
with the integration variable $x'$ being the fractional distance
from the nominal resonance
that ranges over the ring's inner and outer boundaries $x_{\mbox{\scriptsize in}}$ 
and $x_{\mbox{\scriptsize out}}$.

\subsubsection{the hydrodynamic approximation}
\label{hydro_section}

Currently, there is some uncertainty in the rings literature
on how to handle the microscopic interactions that occur among ring particles.
Those interactions are important, since they 
control the dynamical heating of a ring as well as
the transport of angular momentum through the ring.
For many years it had been thought that 
collisions were the most important particle-particle interaction,
and sophisticated models were developed to handle the effects of collisions.
For instance, \cite{BGT83} solve the Boltzmann equation 
for the stress tensor that accounts for those interparticle interactions, 
which yielded a formalism known as the {\em particle-gas} model. 
\cite{BGT85} also derive an alternate stress tensor for an incompressible
{\em fluid-ring}. We also note that the {\em particle-jam} model of \cite{ME02}
appears to be an intermediate case, since their ring is quasi-incompressible.

However, we now know that self-gravitating wakes also play an important role
in stirring dense planetary rings \citep{DTI01}, and that the viscous overstability
can be another mechanism for viscous transport
in the ring \citep{SSS01}.
In light of the varied and uncertain ways in which ring 
particles interact on microscopic scales,
we have elected to treat these interactions in the simplest way possible,
via the hydrodynamic approximation, which assumes that the ring is 
a Newtonian fluid whose internal stresses manifest themselves via pressure and viscosity. 
This hydrodynamic approach has been described as
``qualitatively acceptable'' \citep{BGT82},
but we consider it to be the most reasonable and practical approach, 
given the current uncertainties.

A Newtonian fluid is one where stress is proportional
to strain. However, \cite{LO06} use kinetic theory to show that the
influence of wakes and other instabilities in a `dilute' planetary ring can in fact be
non-Newtonian. But \cite{LO06} also show
that the results of the kinetic theory do agree with hydrodynamics when the ring is dense and
collisions are frequent, as is the outer B ring.
It should also be noted that the many spiral density waves that are seen in Saturn's rings provides
observational evidence that the hydrodynamic treatment is in fact appropriate,
since the damping of linear spiral density waves is well described by simple
hydrodynamic viscosity \citep{TBN07}.
\cite{SSA08} also show that subtle variations seen in these waves'
dispersion relation are probably due to hydrodynamic pressure.
Both of these observations support the notion
that the hydrodynamic approximation is in fact appropriate when studying a planetary ring's
large-scale motion, and justify its usage here.

\subsubsection{particle collisions and pressure}
\label{pressure_section}

The following will account for the effects of particle collisions
in a simple quantitative way, with collisional effects being treated as 
if they were due to a pressure $p(\mathbf{r}, t)$ that
tends to repel particles away from regions of higher density (where collisions
will be more frequent and vigorous) and towards regions of lower density. 
To effect this, adopt a barotropic equation of state that assumes that 
the pressure in the ring can be expressed as the function
$p(\mathbf{r}, t)=p(\rho)$ where $\rho(\mathbf{r}, t)$ is the ring's volume density
that obeys $c^2=dp/d\rho$ where $c$ is the ring particles' dispersion velocity.
Then the radial acceleration that an individual ring particle experiences
due to collisional pressure is
\begin{equation}
    \label{pressure0}
    a_p=-\frac{1}{\rho}\frac{dp}{dr} 
        = -\frac{1}{\rho}\frac{dp}{d\rho}\frac{d\rho}{dr}
        =-\frac{c^2}{\rho}\frac{d\rho}{dr}=-\frac{c^2}{\sigma}\frac{d\sigma}{dr}
\end{equation}
since $\rho=\sigma/2h$ where $h=c/\Omega$ is the ring's vertical half-thickness.
Note, though, that this acceleration depends on a radial gradient, so this formula
will be problematic at the edge of a sharp-edged ring. However that problem is avoided
by first considering the linear momentum flux $G_p$ that pressure effects also 
transmit through the ring.

The linear momentum flux $G_p$ is defined as the force-per-length
(e.g., the two-dimensional pressure)
that one streamline exerts on that exterior to it due to pressure effects. Thus
$\delta f=G_p\delta\ell$ is the radial force that one streamline exerts on
a neighboring streamline segment whose tangential length is $\delta\ell$.
If those streamlines reside in the ring's interior (rather than at its edge),
then the {\em net} force on a segment is
$\Delta f =\delta f(r-\delta r) - \delta f(r)\simeq -(\partial\delta f/\partial r)\delta r
=-(\partial G_p/\partial r)\delta r\delta\ell$ where $\delta r$ is the segment's
radial width. And since $\Delta f = a_p\delta m$ where
$\delta m=\lambda\delta\ell=\sigma\delta r\delta\ell$, then
the acceleration due to pressure is related to the linear momentum flux via
\begin{equation}
    \label{pressure1}
    a_p=-\frac{1}{\sigma}\frac{\partial G_p}{\partial r}
        =-\frac{1}{\sigma_0}\frac{\partial G_p}{\partial a},
\end{equation}
noting that $\sigma=\sigma_0/J$ and that $\partial r = J\partial a$.
A comparison of Eqns.\ (\ref{pressure0}) and (\ref{pressure1})
shows that the ring's linear momentum flux due to pressure is
\begin{equation}
    \label{G_p1}
    G_p = c^2\sigma = \frac{c^2\sigma_0}{J},
\end{equation}
assuming that $c$ is constant.

Next, Fourier expand the
longitude dependent $J^{-1}$ factor  in Eqn.\ (\ref{G_p1}), which is
Eqn.\ (\ref{fourier1}) with $\delta\rightarrow q$ and
$\phi\rightarrow m(\theta-\theta_s-\tilde{\omega}) +\eta$ where $\eta$ is 
from Eqn.\ (\ref{q-delta}), so
\begin{equation}
    \label{G_p2}
    G_p \simeq \frac{c^2\sigma_0}{\sqrt{1-q^2}}\left[ 1 + 
        \frac{2(1-\sqrt{1-q^2})}{q}e^{i[m(\theta-\theta_s-\tilde{\omega}) +\eta]}\right]
\end{equation}
upon reverting to complex notation. Inserting this into Eqn.\ (\ref{pressure1})
shows that the acceleration due to pressure now has the desired form of 
Eqn.\ (\ref{perturbations}) where
\begin{mathletters}
    \label{A_p}
    \begin{eqnarray}
        \label{A_p_0}
        A_{pr}^0 &=& -\frac{c^2}{\sigma_0}\frac{\partial}{\partial a}
            \left(\frac{\sigma_0}{\sqrt{1-q^2}}\right) \\
        \label{A_p_p}
        A_{pr}^m &=& -\frac{2c^2}{\sigma_0}\frac{\partial}{\partial a}
            \left[\frac{\sigma_0(1-\sqrt{1-q^2})}{q\sqrt{1-q^2}}
            e^{i(\eta-m\tilde{\omega})}\right],
    \end{eqnarray}
\end{mathletters}
with the $r$ subscript indicating that these are radial accelerations.
The dimensionless versions of these accelerations are
\begin{mathletters}
    \label{alpha_p}
    \begin{eqnarray}
        \label{alpha_p_0}
        \alpha_{pr}^0(x) &=& \frac{A_{pr}^0}{|a{\cal D}|} 
            = -\frac{c'^2}{3(m-\epsilon)s}
            \frac{\partial}{\partial x}\left(   \frac{s}{\sqrt{1-q^2}}  \right)\\
        \label{alpha_p_m}
        \mbox{and}\qquad \alpha_{pr}^m(x) &=& \frac{A_{pr}^m}{|a{\cal D}|} = 
            \bar{\alpha}_{pr}^m e^{-im\tilde{\omega}}\\
        \mbox{where}\qquad \bar{\alpha}_{pr}^m(x) &=& 
             -\frac{2\ c'^2}{3(m-\epsilon)s}\left(\frac{\partial g_p}{\partial x}
             + ikag_p \right)\\
        \label{g_p}
        \mbox{and}\qquad g_p(x) &=&\frac{s(1-\sqrt{1-q^2})}{q\sqrt{1-q^2}}
            e^{i\eta},
    \end{eqnarray}
\end{mathletters}
with $c'=c/a\Omega=h/a$ being the particles' dispersion velocity in units of their
orbital velocity. The $\bar{\alpha}_{pr}^m$
quantity is introduced here for convenience, since it is a
real quantity when the ring is nearly peri- or apo-aligned.
Since $|m\tilde{\omega}|\sim0$  or $180^\circ$ in that case,  $qe^{i\eta}\simeq e'$ 
in Eqn.\ (\ref{g_p}), while the terms proportional
to $kag_p$ terms are negligible. Most (but not all) of the ring scenarios
considered here are in this limit.

Keep in mind that the accelerations in Eqns.\
(\ref{A_p}--\ref{alpha_p}) are only valid for a ring particle that
orbits in the ring's interior. Those equations do not apply to
particles inhabiting the streamline at the ring's outer edge, since
the pressure there is exerted only by 
the streamline that is orbiting just interior to it.
In that case, $\delta f=G_p\delta\ell=\delta m a_p=\lambda a_p\delta\ell$
is the force on a streamline segment having a tangential length $\delta\ell$
due to pressure, so the acceleration
on the ring particles there is $a_p=G_p/\lambda=G_p/\sigma_0\Delta a$
where $\Delta a$ is the outermost streamline's semimajor axis width.
Inserting Eqn.\ (\ref{G_p2}) into $a_p$ then
provides the coefficients for the acceleration due to pressure at the ring edge, 
whose dimensionless forms are
\begin{mathletters}
    \label{alpha_pr_outer}
    \begin{eqnarray}
        \label{alpha_pr_0_outer}
        \alpha_{pr}^0(x) &=& \frac{A_{pr}^0}{|a{\cal D}|} =
            \frac{c'^2(a/\Delta a)}{3(m-\epsilon)\sqrt{1-q^2}}\\ 
        \label{alpha_pr_m_outer}
        \mbox{and}\qquad\alpha_{pr}^m(x) &=& 
           \frac{A_{pr}^m}{|a{\cal D}|} = 
           \bar{\alpha}_{pr}^m e^{-im\tilde{\omega}}\\
         \mbox{where}\qquad\bar{\alpha}_{pr}^m(x) &=&  
             \frac{2\ c'^2\left(1-\sqrt{1-q^2}\right)}{3(m-\epsilon)q\sqrt{1-q^2}}
             \left(\frac{a}{\Delta a}\right)e^{i\eta}.
    \end{eqnarray}
\end{mathletters}
This derivation also illustrates the utility of deriving the acceleration due to
pressure from the linear momentum flux $G_p$, since this approach easily
handles the discontinuous pressure drop that will occur are the ring's outer edge.

\subsubsection{ring viscosity}
\label{viscosity-section}

The viscous acceleration ${\mathbf a}_\nu$ that a parcel of ring material 
experiences along one of the Cartesian axes will be denoted as $a_{\nu i}$,
where the $i=1,2,3$ subscript indicates one of the $x,y,z$ axes. That acceleration is 
\begin{equation}
    \label{a_v_xyz}
    a_{\nu i} = \frac{1}{\rho}\left\{ \sum_{j=1}^3\frac{\partial}{\partial x_j}\left[\eta
        \left( \frac{\partial u_i}{\partial x_j} + \frac{\partial u_j}{\partial x_i} 
        - \frac{2}{3}\delta_{ij}\sum_{k=1}^3 \frac{\partial u_k}{\partial x_k}
        \right)  \right]
        +  \frac{\partial}{\partial x_i}\zeta\sum_{k=1}^3
        \left(\frac{\partial u_k}{\partial x_k}\right) \right\}
\end{equation}
where $\rho$ is the ring's volume density, $\eta$ its shear 
viscosity\footnote{Note that the shear viscosity $\eta$ that is used only
in Section \ref{viscosity-section} is distinct from the angle $\eta$ of Eqn.\ (\ref{q-delta})
that appears elsewhere in this paper.}, $\zeta$
is its bulk viscosity, and $u_i$ is the velocity along
the $i^{\scriptsize th}$ axis \citep{LL87}. To transform this into cylindrical coordinates,
replace $x_1\rightarrow r$ so that the radial velocity $u_1\rightarrow v_r=\dot{r}$
and the tangential velocity $u_2\rightarrow v_\theta=r\dot{\theta}$  in the above,
while the differentials $\partial x_1\rightarrow\partial r$
and $\partial x_2\rightarrow r\partial \theta$.
Next, note that the ring's various quantities, such as its density or noncircular velocity,
are all expected to vary rapidly in the radial $j=1=k$ direction, but vary
slowly in the tangential direction. Consequently, only the $j=1=k$ terms need to
be preserved in the above, so the components of the
viscous acceleration simplify to
\begin{mathletters}
    \label{a_v_wrong}
    \begin{eqnarray}
        \label{rho-a_v_r}
        \rho a_{\nu r} &\simeq& \frac{\partial}{\partial r}\left[
            \left(\frac{4}{3}\eta +\zeta\right)\frac{\partial v_r}{\partial r}\right]\\
        \label{rho-a_v_theta}
        \rho a_{\nu \theta} &\simeq& \frac{\partial}{\partial r}\left(
            \eta\frac{\partial v_\theta}{\partial r}\right).
    \end{eqnarray}
\end{mathletters}

Note that Eqn.\ (\ref{rho-a_v_theta}) implies that
$a_{\nu \theta}\ne0$ in a rigidly rotating disk having
$v_r=r\dot{\theta}$ where $\dot{\theta}$ is constant. But this is unphysical, because
it implies that a rigid rotator would also experience
a viscous transport of angular momentum. However, this problem is easily fixed by
replacing the $\partial v_\theta/\partial r$ in Eqn.\ (\ref{rho-a_v_theta}) with
$r\partial \dot{\theta}/\partial r$, which provides a more physical
expression that is also in agreement
with other treatments of viscous astrophysical disks (e.g., \citealt{LP74}).
Also replace the viscosities in the above 
with $\eta=\nu_s\rho$ and $\zeta=\nu_b\rho$,
where $\nu_s$ is the kinematic shear viscosity and $\nu_b$ the kinematic
bulk viscosity, and integrate the repaired version of
Eqns.\ (\ref{a_v_wrong}) along the vertical direction so that
$\int\rho dz\rightarrow\sigma$, which results in
\begin{mathletters}
    \label{a_v}
    \begin{eqnarray}
        \label{a_v_r}
        a_{\nu r} &\simeq& \frac{1}{\sigma}\frac{\partial}{\partial r}\left[
            \left(\frac{4}{3}\nu_s +\nu_b\right)\sigma\frac{\partial v_r}{\partial r}\right]\\
        \label{a_v_theta}
        a_{\nu \theta} &\simeq&  \frac{1}{\sigma}\frac{\partial}{\partial r}\left(
            \nu_s\sigma r\frac{\partial \dot{\theta}}{\partial r}\right).
    \end{eqnarray}
\end{mathletters}
Note that the viscous acceleration also depends on radial gradients,
which would be problematic at a ring edge. But that difficulty is again
avoided by considering the ring's radial flux of angular and
linear momentum.

\subsubsubsection{angular momentum flux}

The viscous angular momentum flux $F_\nu(r,\theta)$ is the rate per-unit-length
that one streamline transmits angular momentum to that orbiting
just exterior to it via the ring's viscous friction. 
Thus $\delta t = F_\nu\delta\ell$ is the torque
that a segment of length $\delta\ell$ exerts on its exterior neighbor.
If that segment orbits in the ring's interior,
then it is also torqued by ring material orbiting just interior to it,
so the net torque on that segment is 
$\Delta t = \delta t(r-\delta r) -\delta t(r) \simeq
-(\partial\delta t/\partial r)\delta r=-(\partial F_\nu/\partial r)\delta\ell\delta r$
where $\delta r$ is the radial spacings between the adjacent streamlines.
And since the net torque on this streamline segment is also
$\Delta t = ra_{\nu\theta}\delta m$ where
$\delta m=\lambda\delta\ell=\sigma\delta r\delta\ell$ is the segment's mass,
this provides a relation between the viscous angular momentum flux $F_\nu$
and the tangential acceleration $a_{\nu\theta}$,
\begin{equation}
    \label{a_nu_theta2}
    a_{\nu\theta} =-\frac{1}{r\sigma}\frac{\partial F_\nu}{\partial r}
       =-\frac{1}{a\sigma_0}\frac{\partial F_\nu}{\partial a}.
\end{equation}
Comparing this to Eqn.\ (\ref{a_v_theta})
shows that the viscous angular momentum flux is
\begin{equation}
    \label{F_nu}
    F_\nu\simeq -\nu_s\sigma r^2\frac{\partial \dot{\theta}}{\partial r}
        \simeq-\frac{\nu_s\sigma_0 a^2}{J^2}\frac{\partial \dot{\theta}}{\partial a}.
\end{equation}

The ring's angular velocity is $\dot{\theta}=\Omega + \dot{\theta}_1
\simeq\Omega[1+2e\exp(i\phi)]$ where 
$\phi=m(\theta-\theta_s-\tilde{\omega})$ (see Eqn.\ \ref{dtheta2}), so
$\partial\dot{\theta}/\partial a\simeq\partial\Omega/\partial a + 2(\Omega/a)(1-J)$
since
\begin{equation}
    \label{1-J}
    1-J = \frac{\partial}{\partial a}\left(eae^{i\phi}\right)
        \simeq a\frac{\partial}{\partial a}\left[e\exp(i\phi)\right]
\end{equation}
by Eqn.\ (\ref{J2}) when small terms of ${\cal O}(e)$ are ignored.
The derivative of Eqn.\ (\ref{Omega_approx}) provides
$\partial\Omega/\partial a\simeq
-(3\Omega/2a)[1 + (m-\epsilon)\partial\alpha_r^0/\partial x]$
where $\alpha_r^0=A_r^0/|a{\cal D}|=\alpha_g^0+\alpha_p^0$
is the axisymmetric part of the ring's gravity + pressure, so
$\partial\dot{\theta}/\partial a\simeq-(3\Omega/2a)[4J/3 - 1/3
+ (m-\epsilon)\partial\alpha_r^0/\partial x]$. Inserting this into
Eqn.\ (\ref{F_nu}) then yields
\begin{equation}
    \label{F_nu2}
    F_\nu\simeq2\nu_s\sigma_0 a\Omega
      \left\{ \frac{1}{J} -\frac{1}{4J^2}
      \left[1-3(m-\epsilon)\frac{\partial\alpha_r^0}{\partial x}\right]\right\}
\end{equation}
to lowest order in the streamline's eccentricity $e$. This
expression is equivalent
to the viscous angular momentum flux that is given in 
\cite{BGT82} when the gradient in
the ring's radial acceleration, $\partial\alpha_r^0/\partial x$, is negligible.
Next, Fourier-expand
the longitude-dependent factors in $F_\nu$, which yields
$J^{-2}=[1+2q\cos(\phi+\eta)]/(1-q^2)^{3/2}+$ other terms,
with the $J^{-1}$ expansion obtained from Eqn.\ (\ref{fourier1}), so
\begin{equation}
    \label{F_nu3}
    F_\nu\simeq\frac{3}{2}\nu_s\sigma_0 a\Omega
      \left[ f_\nu + 2q\left(f_\nu-\frac{4/3}{1+\sqrt{1-q^2}} \right)
      e^{i(\phi+\eta)}\right]
\end{equation}
where $f_\nu$ is shorthand for
\begin{equation}
    \label{f_nu}
    f_\nu(x)=(1-q^2)^{-3/2}\left[ 1 + (m-\epsilon)\frac{\partial\alpha_r^0}{\partial x}
          -\frac{4}{3}q^2  \right],
\end{equation}
and remembering to preserve only the real part of Eqn.\ (\ref{F_nu3}).

The angular momentum luminosity through the ring is the
integral about a streamline,
\begin{equation}
    \label{L_nu}
    {\cal L}_\nu(x)=\oint F_\nu d\ell=3\pi f_\nu\nu_s\sigma_0 a^2\Omega.
\end{equation}
This is equivalent to that given in  \cite{BGT82} when
$\partial\alpha_r^0/\partial x=0$, which is probably
true for most planetary rings. Note that ${\cal L}_\nu\propto f_\nu(x)$,
and that $f_\nu<0$ when the nonlinearity parameter $q> q^\star$ where
$q^\star=\sqrt{3}/2\simeq0.866$ (again, provided that
$\partial\alpha_r^0/\partial x\simeq0$; see Eqn.\ \ref{f_nu}).
This is the threshold for the angular momentum flux reversal that was first
described in \cite{BGT82}. When streamlines become so disturbed that
$q>q^\star$, the viscous torque causes the streamline's
angular momentum to flow inwards
(${\cal L}_\nu<0$) rather than in the usual outwards direction. 
The ring particles orbiting
in this disturbed region then spiral inwards due to this angular momentum loss,
which also opens a gap in the ring.

\subsubsubsection{linear momentum flux}

The radial component of the ring's viscous acceleration $a_{\nu r}$ also transmits 
a flux of linear momentum $G_\nu$ where 
\begin{equation}
    \label{a_nur-Gnu}
    a_{\nu r} =-\frac{1}{\sigma}\frac{\partial G_\nu}{\partial r}
        =-\frac{1}{\sigma_0}\frac{\partial G_\nu}{\partial a}
\end{equation}
(see Section \ref{pressure_section}).
Comparison with Eqn.\ (\ref{a_v_r}) shows that the ring's viscous
linear momentum flux is
\begin{equation}
    \label{Gnu}
    G_\nu = -\left(\frac{4}{3}\nu_s + \nu_b\right)\sigma\frac{\partial v_r}{\partial r}
        =-\left(\frac{4}{3}\nu_s + \nu_b\right)\frac{\sigma_0}{J^2}
        \frac{\partial v_r}{\partial a}.
\end{equation}
The time derivative of Eqn.\ (\ref{r1}) provides the streamline's radial velocity,
which is $v_r=\dot{r}_1=-i\omega_mR_me^{i\phi}
\simeq-i\epsilon\Omega eae^{i\phi}$, so
$\partial v_r/\partial a\simeq -i\epsilon\Omega(1-J)=
-i\epsilon\Omega qe^{i(\phi+\eta)}$ by Eqns.\ (\ref{J2}) and (\ref{1-J}). Inserting
the real part into Eqn.\ (\ref{Gnu}) then yields
$G_\nu =  -\epsilon\left(\frac{4}{3}\nu_s + \nu_b\right)\Omega \sigma_0q
\sin(\phi+\eta)/J^2$.
The Fourier expansion of the longitude-dependent factor in $G_\nu$ is
$\sin(\phi+\eta)/[1-q\cos(\phi+\eta)]^2= 2\sin(\phi+\eta)/
[\sqrt{1-q^2}(1+\sqrt{1-q^2})]$ + other unimportant terms, so the viscous flux of
linear momentum is
\begin{equation}
    \label{Gnu3}
    G_\nu \simeq \frac{2i\epsilon\left(\frac{4}{3}\nu_s + \nu_b\right)
        \Omega \sigma_0q e^{i(\phi+\eta)}}{\sqrt{1-q^2}\left(1+\sqrt{1-q^2}\right)}
\end{equation}
upon switching back to the complex notation.

\subsubsubsection{acceleration due to viscosity}

Inserting the linear and angular momentum fluxes $G_\nu$ and $F_\nu$ into
Eqns.\ (\ref{a_nu_theta2}) and (\ref{a_nur-Gnu}) then provides the radial and
tangential acceleration that are
due to the ring's viscosity, which also have the same form
as Eqns.\ (\ref{perturbations}) with
\begin{mathletters}
    \label{A_nu}
    \begin{eqnarray}
        \label{A_nu_0}
        A_{\nu\theta}^0(a) &\simeq&  -\frac{3\Omega\nu_s}{2a\sigma_0}
            \frac{\partial}{\partial x}(\sigma_0f_\nu)\\
        \label{A_nu_theta_m}
        A_{\nu\theta}^m(a) &\simeq&-\frac{3\Omega\nu_s}{a\sigma_0}
            \frac{\partial}{\partial x}\left[  \sigma_0q
            \left( f_\nu-\frac{4/3}{1+\sqrt{1-q^2}}  \right) 
            e^{i(\eta-m\tilde{\omega})}\right]\\
        \label{A_nu_r_m}
        \mbox{and}\qquad A_{\nu r}^m(a) &\simeq&
           - \frac{2i \epsilon\Omega}{a\sigma_0}\left(\frac{4}{3}\nu_s + \nu_b\right)
            \frac{\partial}{\partial x}\left[
            \frac{\sigma_0q e^{i(\eta-m\tilde{\omega})}}
            {\sqrt{1-q^2}\left(1+\sqrt{1-q^2}\right)}\right],
    \end{eqnarray}
\end{mathletters}
where it is assumed that only  $\sigma_0$, $e$, and
$\tilde{\omega}$ might vary rapidly with distance $x$ while the $\nu$'s
are treated as constants. The dimensionless versions of the
tangential accelerations are
\begin{mathletters}
    \label{alpha_nu_theta}
    \begin{eqnarray}
        \label{alpha_nu_theta_0}
        \alpha_{\nu\theta}^0(x) &=& \frac{A_{\nu\theta}^0}{|a{\cal D}|}=
            -\frac{\nu_s'}{2(m-\epsilon)s}
             \frac{\partial(sf_\nu)}{\partial x}\\
        \label{alpha_nu_theta_m}
         \mbox{and}\qquad \alpha_{\nu\theta}^m(x) &=& 
           \frac{A_{\nu\theta}^m}{|a{\cal D}|} = 
           \bar{\alpha}_{\nu\theta}^m e^{-im\tilde{\omega}}\\
        \mbox{where}\qquad \bar{\alpha}_{\nu\theta}^m(x) &=& 
             -\frac{\nu_s'}{(m-\epsilon)s}
             \left( \frac{\partial g_\theta}{\partial x} + ikag_\theta\right)\\
        \mbox{with}\qquad g_\theta(x) &\equiv& 
            sq \left( f_\nu-\frac{4/3}{1+\sqrt{1-q^2}}  \right) e^{i\eta},
    \end{eqnarray}
\end{mathletters}
where $\nu_s'=\nu_s/a^2\Omega$ is a dimensionless version of the ring's
shear viscosity, and $s(x)=\sigma(x)/\sigma_\infty$ its fractional
surface density. Similarly, the dimensionless version of the radial acceleration is
\begin{mathletters}
    \label{alpha_nu_r}
    \begin{eqnarray}
        \label{alpha_nu_r_m}
         \mbox{and}\qquad \alpha_{\nu r}^m(x) &=& 
           \frac{A_{\nu r}^m}{|a{\cal D}|} = 
           i\bar{\alpha}_{\nu r}^m e^{-im\tilde{\omega}}\\
        \mbox{where}\qquad \bar{\alpha}_{\nu r}^m(x) &=& 
             -\frac{2 \epsilon\left(\frac{4}{3}\nu_s' + \nu_b'\right)}{3(m-\epsilon)s}
             \left( \frac{\partial g_r}{\partial x} + ikag_r\right)\\
        \mbox{with}\qquad g_r(x) &\equiv& 
           \frac{sqe^{i\eta}}{\sqrt{1-q^2}\left(1+\sqrt{1-q^2}\right)},
    \end{eqnarray}
\end{mathletters}
where $\nu_b'=\nu_b/a^2\Omega$. Also note that the $\bar{\alpha}_{\nu\theta}^m$
and $\bar{\alpha}_{\nu r}^m$ become
real quantities when the ring is peri- or apo-aligned.

Again, these accelerations are only valid for ring particles
orbiting in the ring's interior, and do not apply to
particles that inhabit the streamline at the ring's outer edge.
In that case, $\delta t=F_\nu\delta\ell=\delta m ra_\nu=\lambda ra_\nu\delta\ell$
is the viscous torque that is exerted
on a streamline segment on the edge that has a tangential length $\delta\ell$, 
so the viscous acceleration
on the ring particles there is $a_\nu=F_\nu/\lambda r\simeq F_\nu/\sigma_0a\Delta a$
where $\Delta a$ is the outermost streamline's semimajor axis width.
Inserting Eqn.\ (\ref{F_nu3}) into $a_\nu$ then
provides the coefficients for the tangential viscous acceleration, 
\begin{mathletters}
    \label{alpha_nu_theta_outer}
    \begin{eqnarray}
        \label{alpha_nu_0_outer}
        \alpha_{\nu\theta}^0(x) &=&\frac{A_{\nu\theta}^0}{|a{\cal D}|} = 
            \frac{\nu_s' f_\nu}{2(m-\epsilon)}\left(\frac{a}{\Delta a}\right)\\
        \label{alpha_nu_theta_m_outer}
         \mbox{and}\qquad \alpha_{\nu\theta}^m(x) &=& 
           \frac{A_{\nu\theta}^m}{|a{\cal D}|} = 
           \bar{\alpha}_{\nu\theta}^m e^{-im\tilde{\omega}}\\
         \mbox{where}\qquad\bar{\alpha}_{\nu\theta}^m(x) &=& 
            \frac{\nu_s'qe^{i\eta}}{m-\epsilon}
            \left(f_\nu - \frac{4/3}{1+\sqrt{1-q^2}} \right)\left(\frac{a}{\Delta a}\right),
    \end{eqnarray}
\end{mathletters}
for particles orbiting in the ring's outermost streamline. Similarly,
the radial component of the viscous acceleration is
$a_{\nu r} = G_\nu/\sigma_0\Delta a$ for particles in the outermost
streamline, so the dimensionless version of this acceleration is
\begin{mathletters}
    \label{alpha_nu_r_outer}
    \begin{eqnarray}
        \label{alpha_nu_r_m_outer}
        \alpha_{\nu r}^m(x) &=& 
           \frac{A_{\nu r}^m}{|a{\cal D}|} = 
           i\bar{\alpha}_{\nu r}^m e^{-im\tilde{\omega}}\\
         \mbox{where}\qquad\bar{\alpha}_{\nu r}^m(x) &=& 
            \frac{2\epsilon\left(\frac{4}{3}\nu_s' + \nu_b'\right)qe^{i\eta}}
            {3(m-\epsilon)\sqrt{1-q^2}\left(1+\sqrt{1-q^2}\right)}
            \left(\frac{a}{\Delta a}\right).
    \end{eqnarray}
\end{mathletters}

Finally, note that there
are several known sources of viscosity in planetary rings: inter-particle collisions
which have $\nu\propto\sigma$ in an optically thin ring \citep{GT82},
and self-gravitating wakes that result in $\nu\propto\sigma^2$ \citep{DTI01}. 
Nbody simulations also indicate that the viscous overstability
varies as $\nu\sim\sigma^2$ (see Fig.\ 13 of \citealt{SSS01}). 
Alternatively, one could also have
employed a more sophisticated viscous stress
tensor to account for the effects of collisions \citep{BGT83, ME02}.
However, previous studies have shown that the surface density
near a perturbed ring-edge  tends to be nearly constant \citep{BGT89}, 
so the results obtained here are not expected to be particularly sensitive to
any viscosity law. In light of this, a simple
constant-viscosity law is employed here. Nonetheless, if a
power-law $\nu\propto\sigma^c$ viscosity were instead preferred,
then the formulas derived here may be adapted.

\subsubsection{drag}
\label{drag-section}

Small ring particles are also susceptible to drag forces, such as
Poynting-Robertson (PR) and plasma drag \citep{GT82}, atmospheric
drag \citep{GP87}, and possibly the Yarkovsky effect \citep{R06}. 
Drag forces can cause particles to migrate radially, and are particularly
effective at damping orbital eccentricities. Although large ring particles are relatively
immune to drag forces, they can still experience the effects of drag
indirectly by colliding with smaller drag-sensitive ring particles.

Most drag forces vary with the particle's velocity relative to the local
circular speed $\mathbf{\Delta v}$, so this work 
will assume that the acceleration on a
ring particle due to drag has the generic form 
$\mathbf{a}_d = -C_d\Omega\mathbf{\Delta v}$
where $C_d$ is a dimensionless drag coefficient whose value will depend
on particulars of the unspecified drag force and the particle size.
Since $\mathbf{\Delta v} = \dot{r}\mathbf{\hat{r}} + 
(r\dot{\theta} - a\Omega)\mathbf{\hat{\theta}}
= (\mathbf{\hat{\theta}} -i\epsilon\mathbf{\hat{r}})ea\Omega e^{i\phi}$
where $\phi = m\theta_0 + \omega_mt - m\tilde{\omega}\simeq
m(\theta-\theta_s - \tilde{\omega})$
(from Eqns.\ \ref{r1} and \ref{dtheta2}), this drag acceleration
has the same form as Eqns.\ (\ref{perturbations}) whose radial
and tangential components are
$A_{d\theta}^m = -C_d ea\Omega^2 e^{-im\tilde{\omega}}$
and $A_{dr}^m = -i\epsilon A_{d\theta}^m$.
The dimensionless versions of these drag accelerations are
\begin{mathletters}
    \label{alpha_drag}
    \begin{eqnarray}
        \label{alpha_drag_r}
        \alpha_{dr}^m(x) &=& 
           \frac{A_{dr}^m}{|a{\cal D}|} = 
           i\bar{\alpha}_{dr}^m e^{-im\tilde{\omega}}\\
         \mbox{where}\qquad\bar{\alpha}_{dr}^m(x) &=& 
            \frac{\epsilon C_d e}{3(m-\epsilon)}\\
        \mbox{and }\qquad\alpha_{d\theta}^m(x) &=& 
           \frac{A_{d\theta}^m}{|a{\cal D}|} = 
           \bar{\alpha}_{d\theta}^m e^{-im\tilde{\omega}}\\
         \mbox{where}\qquad\bar{\alpha}_{d\theta}^m(x) &=& 
            -\frac{C_d e}{3(m-\epsilon)}.
    \end{eqnarray}
\end{mathletters}

This drag force will damp the ring particle's eccentricity at a rate that
may be obtained by inserting the drag accelerations $A_{dr}^m$
and $A_{d\theta}^m$ into the equation of motion (\ref{r_eom_final}),
which yields
\begin{equation}
    \label{de/dt}
    \frac{de}{dt} = -\frac{3}{2} C_d e\Omega.
\end{equation}
This corresponds to an $e$-damping timescale of
\begin{equation}
     \label{tau_e}
    \tau_e = \left|\frac{e}{de/dt}\right| = \frac{2}{3C_d\Omega} =
        \frac{P_{\mbox{\scriptsize orb}}}{3\pi C_d}
\end{equation}
where $P_{\mbox{\scriptsize orb}}=2\pi/\Omega$ is the particle's
orbit period.

\subsection{dimensionless equation of motion}
\label{eom-again}

The equation for the particle's motion is Eqn.\ (\ref{N2}) with
$\mathbf{a}=\mathbf{a}_g+\mathbf{a}_p+\mathbf{a}_\nu+\mathbf{a}_d$
being the acceleration due to the ring's
gravity, pressure, viscosity, and a possible drag.
Now that  the ring's internal forces are
suitably Fourier-decomposed (Eqns.\ \ref{A_g}, \ref{A_p}, and
\ref{A_nu}--\ref{alpha_drag}),
insert those as well as the anticipated solution for the particle's motion
(Eqn.\ \ref{r1}) into its equation of motion (\ref{r_eom_final}) to obtain
\begin{equation}
    \label{eom1}
    D{\cal R}_m = -\Psi_c^m = 
        -\Psi_s^m + \frac{2i\Omega}{\omega_m}A_\theta^m - A_r^m,
\end{equation}
where $D(r)=\kappa^2 - \omega_m^2$ is again the particle's
frequency-distance from resonance. The dimensionless version of this equation
is obtained by dividing by $|a{\cal D}|$ and noting that
${\cal R}_m=ea\exp(-im\tilde{\omega})$
and $\omega_m=\epsilon\kappa\simeq\epsilon\Omega$ near a resonance, so
\begin{equation}
    \label{eom2}
    \epsilon ede^{-im\tilde{\omega}}  - 2i\epsilon\alpha_\theta^m+ \alpha_r^m
        + \psi_s = 0
\end{equation}
where $d(x) = D/{\cal D}$ is the dimensionless version of $D$, and the constant
$\psi_s$ is the satellite's dimensionless forcing function, Eqn.\ (\ref{psi}).
The radial part of the dimensionless acceleration is
the sum of the contributions due to ring gravity, pressure, viscosity, and 
drag, so  $\alpha_r^m = A_r^m/|a{\cal D}|=[\bar{\alpha}_{gr}^m + \bar{\alpha}_{pr}^m
+ i(\bar{\alpha}_{\nu r}^m + \bar{\alpha}_{dr}^m)]e^{-im\tilde{\omega}}$,
while the tangential acceleration is  
$\alpha_\theta^m = A_\theta^m/|a{\cal D}|=
(\bar{\alpha}_{\nu \theta}^m + \bar{\alpha}_{d\theta}^m)
e^{-im\tilde{\omega}}$ by Eqns. (\ref{alpha_g}, \ref{alpha_p}--\ref{alpha_pr_outer},
\ref{alpha_nu_theta}--\ref{alpha_nu_r_outer}, and \ref{alpha_drag}). Then let
$\bar{\alpha}_{cr}^m = \bar{\alpha}_{gr}^m + \bar{\alpha}_{pr}^m$,
which is the sum of all the accelerations that are due to conservative forces
(gravity plus pressure) while
$\psi_d = -(\bar{\alpha}_{\nu r}^m + \bar{\alpha}_{dr}^m) +
2\epsilon(\bar{\alpha}_{\nu \theta}^m + \bar{\alpha}_{d\theta}^m)$
becomes the sum of all the dissipative accelerations (viscosity plus drag), so that the
equation of motion becomes
\begin{equation}
    \label{eom3}
    (\epsilon de  + \bar{\alpha}_{cr}^m -i\psi_d) e^{-im\tilde{\omega}} + \psi_s = 0,
\end{equation}
where $\psi_d$ is called the dissipative forcing function, 
in analogy with Eqn.\ (\ref{ff}). The factor $d(x)=D/{\cal D}$ in the above 
is obtained by inserting Eqns.\ (\ref{Omega_kappa_approx})
into $D$ to show that
\begin{equation}
    \label{d}
   d(x) = x - (3\epsilon-m)(\alpha_{gr}^0 + \alpha_{pr}^0),
\end{equation}
where $x$ is understood
to the fractional distance from the {\em nominal resonance},
which is where the resonance would be if the ring's internal accelerations
where zero. Equation (\ref{d}) thus shows how the 
axisymmetric part of the ring's radial acceleration will displace the resonance. 

The equation of motion
(\ref{eom3}) is complex, so its real and imaginary parts provide
two coupled equations for the streamline's forced orbital elements $e(x)$ and
$\tilde{\omega}(x)$. However those equations decouple when the
dissipation is weak, $|\psi_d|\ll|\psi_s|$, which also results in a ring
that is nearly peri- or apo-aligned such that $|m\tilde{\omega}|\ll1$.
In that case, the accelerations $\bar{\alpha}_{cr}^m$ and $\psi_d$
are all real and depend only on $e(x)$, so the real and imaginary 
parts of Eqn.\ (\ref{eom3}) yield
\begin{mathletters}
    \label{eom_approx}
    \begin{eqnarray}
        \label{eom_e}
        \epsilon ed &+& \bar{\alpha}_{cr}^m + \psi_s \simeq 0\\
        \label{eom_w}
        \mbox{and}\quad \tilde{\omega} &\simeq& \frac{\psi_d}{m\psi_s}.
    \end{eqnarray}
\end{mathletters}
As one might expect, a numerical algorithm that uses 
the approximate decoupled equations of motion  (\ref{eom_approx})
is will converge to a solution
much faster than one that attempts to solve the exact coupled Eqn.\ (\ref{eom3}).
But keep in mind that these approximate solutions to the equations of motion are only 
valid when $|m\tilde{\omega}|\ll1$ and $|kae|\ll|e'|$.

Lastly, we note that this problem also has a third
unknown, the ring's surface density $\sigma_0(a)$ or, equivalently,
its fractional surface density $s(x)$. To address this, a third equation for this quantity
is derived below, which is obtained by requiring that all of the torques
exerted on each streamline balance to zero.

\subsection{angular momentum transport}
\label{ang-mom-lum}

Two mechanisms transport angular momentum through the ring:
the ring's viscosity, and the satellite's gravitational torque. The rate
at which this transport occurs is the ring's angular momentum luminosity,
and it has two parts, ${\cal L}(a) = {\cal L}_\nu + {\cal L}_s$, 
where ${\cal L}_\nu$ is the viscous angular momentum luminosity
(Eqn.\ \ref{L_nu}), and ${\cal L}_s$ is the angular momentum luminosity 
that is due to the satellite's torque on the ring.
Note that ${\cal L}$ must also be conserved, {\it i.e.},
$\partial{\cal L}/\partial a =0$,
for if this were not the case, then streamlines would gain or
lose angular momentum over time, which would also cause them to
evolve radially since $\partial{\cal L}/\partial a \propto A^0_\theta$.
Consequently, static equilibrium requires  ${\cal L}(a)$ to be a
constant everywhere, which also provides that third equation for the ring's unknown
surface density $\sigma_0(a)$.

To calculate the ring's angular momentum luminosity ${\cal L}_s$
that is due to the satellite torque, first consider
the specific torque that the satellite exerts on a single ring particle, which is
$T_1(r, \theta)=\mathbf{r\times(-\nabla}\Phi_s)\cdot\mathbf{\hat{z}}
=-\partial\Phi_s/\partial\theta
=m\phi_s^m(r)\sin m(\theta-\theta_s)$. The time-averaged torque on
the particle is obtained by inserting $r(t)=a+r_1$ and
$\theta(t)=\theta_0 +\Omega t +\theta_1$ into $T_1$, Taylor-expanding to
first order in the small quantities $r_1$ and $\theta_1$
(which are Eqns.\ \ref{r1} and \ref{theta1_approx}), and then time-averaging,
which yields 
\begin{equation}
    \label{T1}
        T_1 \simeq \left<m\left(\phi_s^m + r_1 \frac{\partial\phi_s^m}{\partial a}\right)
            \left[ \sin m(\theta-\theta_s) + m\theta_1\cos m(\theta-\theta_s)\right]\right>
            =\frac{1}{2}mR_m\Psi_s^m\sin(m\tilde{\omega}),
\end{equation}
where the brackets indicate
a time-average over the forcing period $2\pi/|\omega_m|$.
An equivalent expression is also derived in \cite{HWR95}. The time-averaged
torque that the satellite exerts on an entire streamline is then $\delta T_s =T_1\delta m$,
where $\delta m= 2\pi\sigma_0 a\delta a$ is that streamline's mass
and $\delta a$ its radial width. Another useful quantity is the satellite's radial
torque density $\partial T_s/\partial a = \delta T/\delta a$, so 
\begin{equation}
    \label{dT/da}
    \frac{\partial T_s}{\partial a} = m\pi\sigma_0 a R_m\Psi_s^m\sin(m\tilde{\omega}).
\end{equation}
Also note that
the torque that a streamline exerts on the satellite is simply
$-\delta T_s$, so the total gravitational
torque that is exerted by all streamlines
having semimajor axis (sma) interior to $a$ is
\begin{equation}
    \label{T_r(a)}
    T_r(a)=-\oint_{\mbox{\scriptsize sma}<a}\delta T_s
        =-\int_0^a m\pi\sigma_0 a R_m(a')\Psi_s^m\sin\left(m\tilde{\omega}(a')\right)da'.
\end{equation}
This torque is a second-order effect since it
is the product of two small quantities---the satellite's forcing $\Psi_s^m$
and the streamline's epicyclic response
$R_m$. This explains why the ring-satellite torque was formally absent from the
linearized equations of motion (\ref{dh0/dt}). And since Eqn.\ (\ref{T_r(a)})
is the torque that the ring interior to $a$
exerts on the satellite, it is also the luminosity of
angular momentum ${\cal L}_s$ that the ring material interior to $a$ 
transmits gravitationally to the satellite, which becomes
\begin{mathletters}
    \label{L_s and gamma}
    \begin{eqnarray}
        \label{L_s(x)}
        {\cal L}_s(x)&=&T_r = -\epsilon\hspace*{0.2ex} 
            m f^m_\epsilon \mu_s\mu_d \gamma(x)M_p(a\Omega)^2\\
         \mbox{where}\quad \gamma(x) &=& \int^x s(x')e(x')\sin(m\tilde{\omega}(x'))dx'
    \end{eqnarray}
\end{mathletters}
when $\Psi_s^m$ is replaced by Eqn.\ (\ref{ff2-repeat}), 
$\sigma_0=s(x)\sigma_\infty$, and $R_m = e(x)a$.

The ring's total angular momentum luminosity is 
\begin{equation}
    \label{L}
    {\cal L}(x)={\cal L}_\nu+{\cal L}_g=3f_\nu s\nu_s'\mu_d M_p(a\Omega)^2
        -\epsilon\hspace*{0.2ex} m f^m_\epsilon \mu_s\mu_d \gamma M_p(a\Omega)^2,
\end{equation}
which is a constant everywhere when the ring is in static equilibrium.
At sites far from the resonance, $e$ and $\gamma\rightarrow 0$ while $s$ and
$f_\nu\rightarrow1$, so ${\cal L}=3\nu_s'\mu_d M_p(a\Omega)^2$ 
is the ring's angular momentum luminosity. This is also the rate at which the
unmodeled part of the ring, which lies interior to the simulated region, 
delivers angular momentum to the simulated part of the ring.
Inserting this into Eqn.\ (\ref{L}) and dividing by ${\cal L}$ then provides a third
equation of motion for the third unknown, the ring's fractional
surface density $s(x)$:
\begin{mathletters}
    \label{torque-balance}
    \begin{eqnarray}
        \label{torque-balance2}
        \ell_\nu(x) &+& \ell_s(x) = 1\\
        \label{ell_nu}
        \mbox{where} \quad \ell_\nu(x) &=& sf_\nu\\
        \label{ell_s}
        \mbox{and} \quad \ell_s(x) &=& 
            -\frac{\epsilon\hspace*{0.2ex} m f^m_\epsilon \mu_s}{3\nu_s'}\gamma(x)
    \end{eqnarray}
\end{mathletters}
are the dimensionless angular momentum luminosities due to viscosity and the
satellite's perturbations, in units of ${\cal L}$.  Equation (\ref{torque-balance2})
can  also be used to determine the location of the ring's edge, since the edge 
streamline at $x=x_{\mbox{\scriptsize edge}}$ is where the viscous torque is
counterbalanced by the satellite's torque. Or, equivalently, the edge is where
the ring's angular momentum luminosity ${\cal L}(x_{\mbox{\scriptsize edge}})$
is entirely due to the satellite's gravitational torque. Consequently,
the ring's edge is the streamline that satisfies
$\ell_s(x_{\mbox{\scriptsize edge}})=1$ and
$\ell_\nu(x_{\mbox{\scriptsize edge}})=0$.

Lastly, we note that if the dissipation in the ring is weak, {\it i.e.}, $|m\tilde{\omega}|\ll1$,
then $\tilde{\omega}\simeq\psi_d/m\psi_s$ (see Eqn.\ \ref{eom_w}), and
the fractional angular momentum luminosity due to satellite
perturbations  becomes
\begin{equation}
    \label{ell_s_approx}
    \ell_s(x) \simeq 
            -\frac{m(m-\epsilon)}{\nu_s'}\int^x s(x')e(x')\psi_d(x')dx',
\end{equation}
where $\psi_d$ is the dissipative forcing function. Most of the ring
scenarios considered below will be in this limit.

\section{Numerical simulations of a sharp-edged planetary ring}
\label{sims}

The streamline model described above provides three equations of motion,
the real and imaginary parts of Eqn.\ (\ref{eom3}) and the torque-balance
Eqn.\ (\ref{torque-balance}), which are to be solved 
for the system's three unknowns, the ring's eccentricity $e(x)$, 
longitude of periapse $\tilde{\omega}(x)$, and the ring's fractional surface density
$s(x)$, where $x$ is the radial coordinate in the ring. 
Note that this is a coupled set of nonlinear integro-differential equations,
since the ring's gravitational acceleration requires integrating over the
unknown $e(x)$, $\tilde{\omega}(x)$, and $s(x)$ (see Eqns.\ \ref{alpha_g}),
while the accelerations due to pressure and viscosity involve derivatives 
of these quantities (Eqns.\ \ref{alpha_p},  and \ref{alpha_nu_theta}---\ref{alpha_nu_r}). 
The next subsection describes how these equations are solved numerically, with
the streamline model then applied to the outer edge of Saturn's B ring.
The following subsections then show how the simulated outcomes vary with the model's
physical parameters, which are the ring's unperturbed surface density
$\sigma_\infty$, the ring particles' dispersion velocity $c$,
the ring kinematic shear and bulk viscosities $\nu_s$ and $\nu_b$,
and the drag coefficient $C_d$. 

\subsection{numerical method}
\label{methods}

To solve Eqns.\ (\ref{eom3}) and (\ref{torque-balance}),
treat the broad planetary ring as if it were composed of $N$ discrete
streamlines that are uniformly spaced with semimajor axes
$a_i=a_{\mbox{\scriptsize in}}+(i-1)\Delta a$, where the 
ring index $i$ ranges between 1 and $N$, 
and $\Delta a$ is the rings' width in semimajor axis.
The $i^{\mbox{\scriptsize th}}$ streamline's eccentricity is  $e_i=e(a_i)$, its
longitude of periapse $\tilde{\omega}_i=\tilde{\omega}(a_i)$,
and its fractional surface density $s_i=\sigma_0(a_i)/\sigma_\infty$. This 
discretization then allows integrals like Eqns.\ (\ref{alpha_g})
to be replaced with sums over the $e_i$, $\tilde{\omega}_i$, and $s_i$,
with finite differences used to calculate the derivatives that
appear in the accelerations due to pressure and viscosity.
This results in a coupled system of $3N$ nonlinear equations
for the $3N$ unknown $e_i$, $\tilde{\omega}_i$, and $s_i$,
which are then straightforward to solve for numerically.

For example, the $m=0$ component of the ring's dimensionless
gravitational acceleration, Eqn.\ (\ref{alpha_g_0}), is the sum
$\alpha_{gr}^0(x_i)=\sum_{j\ne i}^NG_{ij}$ where
$G_{ij}=2\mu_ds_j\Delta a/3\pi(m-\epsilon)(x_j-x_i)\sqrt{1-\delta_{ij}^2}$
is the gravitational acceleration that streamline
$j$ exerts on a particle in streamline $i$, where $\mu_d$ is the ring's
normalized mass (see Section \ref{gravity}), and with $\delta_{ij}$ calculated
via a similarly quantized version of Eqn.\ (\ref{Delta}). Note that $G_{ii}=0$ here,
because in the local approximation,
a straight wire-thin streamline exerts no net gravitational
force on the particles inhabiting that streamline. A similar strategy is also
used to calculate the dimensionless $m\ge1$ part
of the ring's gravity, $\bar{\alpha}_{gr}^m(x_i)$; see Eqn.\ (\ref{alpha_g_m}). Also
note that the accelerations due to pressure and viscosity,
Eqns.\ (\ref{alpha_p}) and (\ref{alpha_nu_theta}---\ref{alpha_nu_r}),
require derivatives of $e$, $\tilde{\omega}$, and $s$, which are calculated numerically
using a three-point Lagrangian interpolation scheme \citep{H56}.
Lastly, keep in mind that Eqns.\ (\ref{alpha_pr_outer}) and 
(\ref{alpha_nu_theta_outer}--\ref{alpha_nu_r_outer}) are the accelerations
on the $N^{\mbox{\scriptsize th}}$ streamline at the ring's edge
that is exerted by the adjacent $N-1$ streamline, so the right hand
sides of those equations are evaluated at $x_{N-1} = (a_{N-1} - a_r)/a_r$.

The number of streamlines $N$, as well as their radial widths $\Delta a$, are both
chosen so that the streamline model can readily resolve the disturbances that
Mimas excites at the outer edges of the B ring.
The following subsections will show that these disturbances are usually
confined to the rings' outermost 50 or so km, so most calculations
use $N=300$ streamlines that have radial widths of $\Delta a=0.5$ km
so that the total radial extent of the simulated region $w=N\Delta a$
is usually about $w=150$ km. For most ring models, this width 
is broad enough to show that the simulated ring's innermost part 
furthest from the resonance has in fact adopted the 
single particle solution, Eqn.\ (\ref{e_forced_single}). This indicates
that the ring's internal forces there are negligible, and
that a wider a ring need not be simulated.
However, the following simulations will
also show that the self-gravity in a more massive ring is able to transmit the
satellite's disturbance to greater distances inwards of the resonance.
To account for the greater reach of these ring's internal forces,
the streamlines' radial widths are increased to $\Delta a=1$ or 2 km
so that the total radial extent of the simulated region becomes
$w=300$ to 600 km in these more massive planetary rings.

The streamline model developed here, called {\tt NLsgvp}, is written in IDL, 
and it solves the coupled set of $3N$ nonlinear equations for the streamlines'
$3N$ orbit elements and surface densities. Although IDL does supply routines
for solving coupled systems of nonlinear equations, they are not particularly
robust, and sometimes fail to find a satisfactory solution to these equations. So
the MPFIT algorithm is instead used here, which is an IDL procedure written by 
Craig Markwardt that is available at 
http://cow.physics.wisc.edu/$\sim$craigm/idl/fitting.html.
MPFIT is a parameter search algorithm, and it was formally designed
to search parameter space for the purpose of fitting a parametrized model to a
dataset. However, the act of seeking solutions to a coupled set of nonlinear equations
is conceptually very similar to modeling data, since
we wish to find the set of $3N$ `parameters' $e_i$, $\tilde{\omega}_i$, and $s_i$,
that best satisfy the $3N$ equations (\ref{eom3}) and (\ref{torque-balance}),
whose right-hand sides are `data'.
MPFIT is quite efficient and very well-suited
for solving this type of problem.

MPFIT must be initialized by first providing it with a trial solution.
That initial guess is formed from the single particle solution,
Eqn.\ (\ref{e_forced_single}), which is also the solution to the
equation of motion  (\ref{eom_e})
at sites far from the resonance where the conservative part of
ring's internal acceleration, $\bar{\alpha}_{cr}^m$, is negligible.
Note, however, that Eqn.\ (\ref{e_forced_single}) diverges at the resonance where
$d(x)\simeq x\rightarrow0$, whereas $e(x)$ is expected
to stay finite inside the nonlinear region where $|x|<x_{NL}$ (see Eqn.\ \ref{x_NL})
due to the ring's internal forces. So to qualitatively mimic this effect, replace
$x$ in Eqn.\ (\ref{e_forced_single})  with
$x\rightarrow-\epsilon\sqrt{x^2+x_{NL}^2}$ such that
\begin{equation}
    \label{e_trial}
    e_{\mbox{\scriptsize trial}}(x)=\frac{\psi_s}{\sqrt{x^2+|\psi_s|}},
\end{equation}
where $\psi_s$ is the satellite's dimensionless forcing function, Eqn.\ (\ref{psi}).
This is the trial solution that is used to initialize MPFIT, which also
adopts $s(x)=1$ and $\tilde{\omega}(x)=0$. 

All of the calculations described below use double-precision arithmetic.
To assess the accuracy of the model results, we
insert all numerical solutions obtained by {\tt NLsgvp} back into the equation of motion
(\ref{eom3}) and divide by $\psi_s$, with those solutions also being inserted into
the torque-balance equation (\ref{torque-balance2})
subtracted by 1. The residuals on the right-hand side of those equations
are the solutions' fractional errors, which for the models described below
are all smaller than $4\times10^{-12}$.

\subsection{simulations of the outer B ring}
\label{Bring}

\subsubsection{variations with surface density}
\label{surface density variations}

Figure \ref{sigma_fig} illustrates how the ring's epicyclic amplitude $R_m(x)=ae(x)$
varies with $\sigma_\infty$, which is the ring's undisturbed surface density
far from the edge. These curves are solutions to Eqns.\ (\ref{eom3}) and
(\ref{torque-balance}), and were obtained using the numerical method
described in Section \ref{methods}. The main point of this figure is to show
that the epicyclic amplitude 
$R_m$ at the ring edge is larger when the ring has a lower mass.
Note that the ring's response
approaches the single particle result, Eqn.\ (\ref{R_epi_single2}), as
$\sigma_\infty\rightarrow0$. All of these simulations adopt viscosities of
$\nu_s=\nu_b=50$ cm$^2$/sec, which is comparable to the viscosity measured
in Saturn's A ring \citep{TBN07, PTW07}. There is no drag in these simulations
($C_d=0$), and the ring's outer edge is chosen to be at the nominal
resonance at $x=0=a-a_r$. The ring particles' dispersion
velocity $c$ is also chosen such that the ring is marginally gravitationally stable.
A gravitationally stable ring has a $Q>1$ \citep{T64}, where
$Q\simeq c\Omega/\pi G\sigma_\infty=c'/\mu_d$. 
However, Saturn's main A and B rings also exhibit brightness asymmetries
that are believed to be due to the presence of self-gravitating wakes,
but those wakes only occur when  $Q\lesssim2$ \citep{S92}. Consequently,
a higher-mass but marginally stable ring  also has
a larger dispersion velocity, and the simulations reported in
Fig.\ \ref{sigma_fig} have $c$ chosen so that $Q=2$.
The dotted horizontal line also
indicates the B ring's observed epicyclic amplitude,
$R_{\mbox{\scriptsize obs}}\simeq45$ km, which was measured in Cassini
spacecraft observations that were acquired in 2005 \citep{SP06}.
Figure \ref{sigma_fig} tentatively suggests that the outer edge of the B
ring might have a surface density as low as $\sigma_\infty\sim10$ gm/cm$^2$.

Figure \ref{q_fig} plots the nonlinearity parameter $q$ versus distance from
resonance for the simulations described in Fig.\ \ref{sigma_fig}. In all of these
simulations, $|kae|\ll|e'|$, so the nonlinearity parameter $q\simeq|e'|$
is also the ring's eccentricity gradient. Figure \ref{q_fig} shows that the disturbed
edges of lighter rings tend to be more nonlinear than heavier rings. 
The ring's epicyclic amplitude $R_m$ is also proportional to the satellite's mass,
so increasing the perturber's mass also makes the ring more nonlinear.

\subsubsection{variations in the ring-edge location}
\label{edge}

Section \ref{spm} notes that the outer edge of the B ring could lie as far as 24km
exterior to Mimas' nominal $m=2$ ILR. This displacement was first measured 
by \cite{PDG84} from Voyager observations, but with a large uncertainty.
However, a preliminary analysis of Cassini observations of the B ring,
which is described in \cite{SP06}, does indicate that the ring-edge
lies about 20 or so km beyond the resonance.

Figure \ref{edge_fig} illustrates the consequences of displacing the B ring's edge
radially outwards; the three curves show the ring's epicyclic amplitude 
$R_m(a)$ for simulated rings whose outermost semimajor axis lies a distance
$\Delta a=0, 10,$ and 25 km exterior to the nominal resonance location.
Figure \ref{edge_fig} also shows that the ring's epicyclic amplitude 
$R_m(a)$ grows linearly as the edge is approached,
even in the portion of the ring edge that lies exterior the resonance.
Consequently, letting the ring edge lie farther beyond the resonance will result in
a larger epicyclic amplitude. However, Section \ref{surface density variations}
shows that the increase in the epicyclic amplitude 
can be offset by increasing the ring's surface density $\sigma_\infty$.
So Fig.\ \ref{edge_fig} assigns a larger surface density to rings whose
outer edges are displaced farther outwards, but in a manner that keeps
the epicyclic amplitude at the edge comparable to the observed
value, $R_{\mbox{\scriptsize obs}}\simeq45$ km, which is
dotted line in Fig.\ \ref{edge_fig}. The rightmost curve in Fig.\ \ref{edge_fig},
which agrees with the B ring's observed epicyclic amplitude as well as the likely
upper limit on the edge's location, now shows that the B ring could
have a much higher surface
density of $\sigma_\infty\sim280$ gm/cm$^2$.

\subsubsection{variations with dispersion velocity}
\label{dispersion v}

Figure \ref{Q_fig}a shows that the ring's 
epicyclic amplitude $R_m$ is quite insensitive to the particles
dispersion velocity $c$. Those simulations are for a B ring that has a
surface density fixed at $\sigma_\infty=30$ gm/cm$^2$
and a stability parameter that ranges over $0\le Q\le 12$, which is a convenient
proxy for $c$ since $Q\simeq c\Omega/\pi G\sigma_\infty$. 
These $Q$ values correspond to dispersion velocities of
$c \le 4.9$ mm/sec and vertical scale heights $h=c/\Omega$ 
of $h \le 32$ m.  Note that the black $Q=0$ curve,
which represents a pressureless ring, 
is indistinct from the other $Q\le5$ models.
Figures \ref{Q_fig}b and \ref{Q_fig}c plot the simulated rings'
fractional surface densities $s(a)$ and
longitudes of periapse $\tilde{\omega}(a)$
versus semimajor axis $a$, which is able to
distinguish the dynamically hotter $Q\gtrsim10$ models
from the cooler $Q\le5$ models. Note the large variations
in  $s(a)$ and  $\tilde{\omega}$ that is seen
at the outer edges seen of the hotter rings that have $Q\ge10$; those
variations are due to the larger pressure drop that occurs across
the outermost streamline (e.g., Eqns.\ \ref{alpha_pr_outer}).
Those surface density excesses seen at the rings' edges
are also reminiscent of those occurring in models of narrow ringlets 
(e.g., \citealt{CG00, ME02}), but the variations seen in Figs.\ \ref{Q_fig}
are poorly resolved due to the models' radial sampling of $\Delta a=0.5$ km.
Nonetheless, Figs. \ref{Q_fig} do show that pressure effects,
if important at all, are confined to the ring's outermost
$\Delta a\lesssim0.5$ km. Since this study is interested in
the perturbed ring's state over a much broader radial scale, $\Delta a\sim40$ km
according to B ring models of Fig.\ \ref{Q_fig},
the following will for simplicity keep $Q=2$
for most of the  ring scenarios that are considered below.

Lastly, note that  $\tilde{\omega}(a)>0$ in the hotter $Q=12$ model
(Fig.\ \ref{Q_fig}c). Equation (\ref{dT/da}) indicates that the torque that the satellite
exerts on this model ring is positive. But that torque has
the wrong sign, since a satellite must exert a negative torque on the ring
if it is to be confined interior to an ILR.
This inability of the satellite to confine a 
viscous and dynamically hot ring is another
reason why hot ring models need not be considered further.
The confinement of the B ring by Mimas' gravitational
torque is also considered in greater detail below.

\subsubsection{ring viscosity, and the torque-balance problem}
\label{viscosity variations}

Figure \ref{v_fig} shows how the simulated outcomes depend on 
the ring's viscosity. These
models are similar to the B ring simulations shown in Fig.\ \ref{Q_fig},
except that the rings' viscosities take values of $\nu_s=\nu_b=5, 50$, and 
500 cm$^2$/sec while $Q=2$.
Note that the rings' epicyclic amplitudes $R_m=ea$ are all insensitive to the
viscosity $\nu_s=\nu_b$ (Fig.\ \ref{v_fig}a), as are the rings' fractional surface
densities $s$ (Fig.\ \ref{v_fig}b), and their angular momentum luminosities
$\ell_\nu$ and $\ell_s$ (Fig.\ \ref{v_fig}d). This is due to the ring being
nearly peri- or apo-aligned, {\it i.e.},
$|\tilde{\omega}|\simeq|\psi_d/m\psi_s|\ll1$ (see Eqn.\ \ref{eom_w}) in all of these
simulations. Since the dissipative forcing function $\psi_d$ is small
compared to the other terms in the equation of motion, it has little
effect on $e$ (see Eqns\ \ref{eom3} and \ref{eom_approx}).
However, Fig.\ \ref{v_fig}c demonstrates that the streamlines' longitudes of
periapse $\tilde{\omega}\propto\psi_d$ are proportional to
the ring viscosity, since
\begin{equation}
    \label{psi_v}
    \psi_d=-\bar{\alpha}_{\nu r}^m+2\epsilon\bar{\alpha}_{\nu\theta}^m
    \equiv\psi_\nu\propto(\nu_s+f\nu_b)
\end{equation}
where the factor $f$ is of order 1
(see Eqns.\ \ref{alpha_nu_theta}--\ref{alpha_nu_r}).

Figure \ref{v_fig}d also shows that the ring's angular momentum luminosity
$\ell_s$, which is due to the satellite's gravitational torque on the ring, is small, {\it i.e.},
$|\ell_s|\ll1$. This is very problematic, because Section \ref{ang-mom-lum}
showed that the ring's outer edge should be the site where the viscous torque
is counterbalanced by the satellite's torque, which is also
where $\ell_\nu\rightarrow0$ and $\ell_s\rightarrow1$. However, this
torque-balance requirement is not satisfied by {\em any} of the simulations
described in Figs.\ \ref{sigma_fig}--\ref{v_fig}. Although the simulations
described above provide useful illustrations of how a ring's simulated
outcome depend on its physical properties $\sigma_\infty$, $\nu$, and $c$,
they are all unphysical since they do not achieve a torque-balance at the ring's
outer edge. However, Section \ref{v_b scenario} does
explore an alternate ring scenario that does in fact achieve
the desired torque balance.

But first, a final comment on Fig.\ \ref{v_fig}b, which shows that the simulated
ring's surface density has a $\sim20\%$ excess in its outermost $\sim20$km.
This is a due the conservation of the ring's angular momentum
luminosity ${\cal L}\simeq{\cal L}_\nu\propto sf_\nu$ (see Eqn.\ \ref{L} and
note that ${\cal L}_s$ is negligible in these simulations) where the function
$f_\nu$ varies approximately as $1-4q^2/3$ (see Eqn.\ \ref{f_nu}) where
the nonlinearity parameter is
$q\simeq|de/dx|$, which is also plotted in  Fig.\ \ref{v_fig}d. Because the ring'
eccentricity gradient gets large near the resonance,
the quantity $f_\nu$ is diminished there,
but that must be offset by an increase in the ring's fractional surface density
$s=\sigma_0(a)/\sigma_\infty$ in order to conserve ${\cal L}\propto sf_\nu$.

\subsubsection{possible B ring solution}
\label{v_b scenario}

Section \ref{viscosity variations} notes that all of the ring simulations
considered thus far fail to balance the ring's viscous torque against the satellite's
gravitational torque, so the proximity of the B ring's outer edge near
Mimas' $m=2$ ILR is not yet accounted for.
Equation (\ref{L_s and gamma}) shows that the satellite's torque on the ring is
proportional to $\gamma(x)=\int^x se\sin(m\tilde{\omega}) dx'$,
which is the product of the streamline's forced eccentricity $e(x)$ and the angle
by which that streamline lags behind the satellite's longitude, $\tilde{\omega}(x)$.
That lag angle depends on the ring's shear and bulk viscosity via
$\sin m\tilde{\omega}\propto\psi_d\propto \nu_s+f\nu_b$ where 
$f\sim{\cal O}(1)$ (Section \ref{viscosity variations}), so the satellite's total torque
on the ring is controlled by the sum of these viscosities.
Although an increase in the ring's shear viscosity $\nu_s$ does increase the satellite's
torque, this is of no help here because the viscous torque (Eqn.\ \ref{L_nu}) would
increase by the same factor, and the torque mismatch would still persist.
However, Eqn.\ (\ref{L_s and gamma}) indicates that
an increase in the ring's bulk viscosity $\nu_b$
can increase the satellite's torque without altering the viscous torque.

But first, a comment on measurements of the viscosity of planetary rings.
The shear viscosity $\nu_s$ is, as the name suggests, a measure of the friction that results
from the ring's shearing motions. On the other hand, the ring's bulk viscosity $\nu_b$
is the friction that occurs due to the ring's compressive or decompressive motions,
which explains why $\nu_b$ enters into the equations of motion when
there is a gradient in the ring's radial velocity (Eqn.\ \ref{rho-a_v_r}).
One way to infer $\nu_s$ and $\nu_b$
is to examine spacecraft observations of spiral density waves;
these waves are damped by viscous effects over the radial scale
$\Delta r_\nu\propto(\nu_s+ \frac{3}{7}\nu_b)^{-1/3}$ \citep{GT78, S84},
so a measurement of $\Delta r_\nu$ provides an estimate of
the combined viscosities $\nu_s+ \frac{3}{7}\nu_b$.
\cite{TBN07} use Cassini observations of spiral density waves in Saturn's A ring
to infer viscosities of $30\lesssim\nu\lesssim300$ cm$^2$/sec there.
However, \cite{TBN07} dropped the bulk viscosity $\nu_b$  from their Eqn.\ (7),
so the viscosities quoted there should instead be interpreted as the combination
$\nu\rightarrow\nu_s+\frac{3}{7}\nu_b$. Similarly, Eqns.\ (\ref{alpha_nu_r})
and (\ref{alpha_nu_r_outer}) indicates that our
simulations will be sensitive to the combination $\nu_s+\frac{3}{4}\nu_b$.
\cite{PTW07} also use the widths of the Keeler and Encke gaps
in Saturn's A ring to infer viscosities of
$20\lesssim\nu\lesssim87$ cm$^2$/sec, but again these viscosities are likely
measurements of some combination of $\nu_s$ and $\nu_b$.
The upshot is that current ring observations only constrain the sum of
$\nu_s$ and $\nu_b$, and that $\nu_s$ and $\nu_b$ are not known
individually.

With this in mind, Fig.\ \ref{vb_fig}
shows the results of three B ring simulations
whose viscosities sum to $\nu_s+\nu_b=100$ cm$^2$/sec (which is roughly the
A ring's total viscosity), while their
ratios obey $\nu_b/\nu_s=0, 1, 1000,$  and $7873$. That last
ratio was chosen so that the viscous and satellite torques do indeed
balance at the ring's outer edge, which is where $s\rightarrow0$ while
$\ell_\nu\rightarrow0$ and $\ell_s\rightarrow1$
(black curves in Fig.\ \ref{vb_fig}b and d).
These simulations' other parameters are similar to that adopted previously
in Fig.\ \ref{v_fig}, with $\sigma_\infty=30$ gm/cm$^2$, $Q=2$, $C_d=0$,
and with the ring's outer edge stationed at the nominal resonance position at
$a=a_r$.

But note that the B ring's mean outer edge actually lies about 24km exterior to
the nominal resonance, and that an outwards shift in the resonance position 
would also result in a larger epicyclic amplitude $R_m$ (Section \ref{edge}).
But that can compensated for
with a larger surface density $\sigma_\infty$, which reduces $R_m$
(Section \ref{surface density variations}). Figure \ref{vb_edge_fig}
illustrates one possible B ring solution that does 
achieve a balance of the viscous and satellite torques at the ring's outer edge
at $a-a_r=24$ km, which is where
$s\rightarrow0$ while  $\ell_\nu\rightarrow0$ and $\ell_s\rightarrow1$,
and also has the observed epicyclic amplitude of $R_m\simeq45$ km at the
ring's edge. This simulation has $\sigma_\infty=226$ gm/cm$^2$, $Q=2$,
$C_d=0$, a total kinematic viscosity of $\nu_s+\nu_b=100$ cm$^2$/sec,
a shear viscosity of $\nu_s=0.00603$ cm$^2$/sec, and a viscosity
ratio of $\nu_b/\nu_s\simeq1.6\times10^4$. Note that
these simulations depend only on the viscosity ratio $\nu_b/\nu_s$.
For instance, when the simulation of Fig.\ \ref{vb_edge_fig}
is executed again using shear and bulk viscosities that are
$\times10$ larger, the same Figs.\ \ref{vb_edge_fig}a, b, and d are obtained,
while the longitude of periapse $\tilde{\omega}$ (Fig.\ \ref{vb_edge_fig}c)
is larger by $\times10$, as expected (e.g., Eqn.\ \ref{eom_w}).

Figure \ref{vb_edge_fig}b shows that the simulated ring is also very nonlinear,
with $q=0.868$ at its outer edge. This is actually just slightly larger
the expected maximum possible value of $q^\star\simeq0.866$, but keep in mind
that this $q^\star$ threshold is approximate; see just below Eqn.\ (\ref{L_nu}).
This large nonlinear parameter
also results in significant longitudinal variations in the ring's surface density $\sigma$.
These variations are demonstrated in Fig.\ \ref{s_r_fig}, which plots radial profiles of the
ring's surface density $\sigma(r,\theta)$ along the satellite's longitude $\theta=\theta_s$,
as well as along longitude $\theta=\theta_s\pm45^\circ$ and $\theta=\theta_s\pm90^\circ$.
Those curves are calculated via 
\begin{mathletters}
    \label{sigma(r,theta)}
    \begin{eqnarray}
        \label{sigma(a,theta)}
        \frac{\sigma(a, \theta)}{\sigma_\infty} &=& \frac{s(a)}{1-q\cos(\phi+\eta)}\\
        \label{Delta r}
        \Delta r(a, \theta)&=& r - a_r\simeq a_r(x-e\cos\phi)\\
        \mbox{and}\quad \phi(a, \theta) & = & m(\theta-\theta_s - \tilde{\omega})
            = \phi^\star + m[\tilde{\omega}(a_{\mbox{\scriptsize edge}}) - \tilde{\omega}(a)]
    \end{eqnarray}
\end{mathletters}
which may be obtained from Eqns.\ (\ref{r}), (\ref{sigma}), and  (\ref{J2}), where
$\Delta r(a, \theta)=r-a_r$ is the radial distance of streamline 
$a$ at longitude $\theta$ from the nominal resonance $a_r$, and
$\tilde{\omega}(a_{\mbox{\scriptsize edge}})$ is the longitude of periapse
at the ring's outer edge.
Setting the angle $ \phi^\star=
m[\theta-\theta_s - \tilde{\omega}(a_{\mbox{\scriptsize edge}})]=0$ 
generates a radial profile along the ring-edge's
longitude of periapse, while setting
$\phi^\star=\pm180^\circ$ results in a surface-density
profile towards the ring-edge's apoapse. Figure \ref{s_r_fig} shows that the
periapse profile has a large surface density excess at the ring's outer edge,
which is due to the satellite's perturbation having shoved ring material inwards
and compressing the streamlines there. Conversely,
the outer edge of the apoapse profile
shows a broad low-density shoulder, which is due to the
streamlines being more distended there. Since the $|\tilde{\omega}|$ and $|\eta|$
are all small in this model, the intermediate $\phi^\star=\pm90^\circ$
profile, which is along longitude $\theta=\theta_s\pm45^\circ$,
is also where $J\simeq1$ and $r\simeq a$, so
the surface density along this longitude equals its so-called
undisturbed surface density $\sigma_0(a)$ that the ring has adopted in order
in order to achieve a torque balance.

So to summarize the results of this model of a viscous gravitating B ring:
if the B ring is indeed a viscous Newtonian fluid,
then its edge near Mimas' $m=2$ ILR can only be accounted for
when the ratio of the ring's bulk/shear viscosities at the ring's edge
takes extreme values of $\nu_b/\nu_s\gtrsim10^4$.
The ring's bulk viscosity $\nu_b$ is a measure of the friction that results
when the ring is compressed radially by the satellite's perturbation.
When that is large enough,
the shepherding torque exerted by the satellite is then strong
enough to counterbalance the ring's viscous torque, thereby confining the ring
in the vicinity of the resonance. However, it was a surprise to find 
that a more conventional model, one having a $\nu_b$ that was comparable or less than
$\nu_s$, failed by a wide margin to balance the viscous and satellite torques.
Also keep in mind that the ring viscosity $\nu$ 
that is inferred from studies of planetary rings (e.g., \citealt{TBN07, PTW07})
is actually a linear combination of $\nu_s$ and $\nu_b$.
So if the preceding scenario is correct, then the ring-edge's viscosity
is dominated by its bulk viscosity $\nu_b$, while the ring's shear viscosity $\nu_s$
is negligible in comparison. 
If this finding is correct, then the B ring can shear freely with little dissipation, 
while its compressed radial motions are very dissipative.

This is an unexpected finding. Nonetheless, the viscosity
requirement $\nu_b/\nu_s\gtrsim10^4$ might be satisfied if the B ring's compressed regions
are so densely packed that the ring particles there are `shoulder to shoulder',
with little voidspace between them.
In this case, the ring's volumetric density would be 
nearly incompressible, as envisioned by \cite{BGT85}. This volumetric 
incompressibility means that as the satellite attempts to
compress the ring's streamlines further in the radial direction, ring particles must
rise vertically and roll or tumble over each other as they are shoved inwards.
This could be a very lossy process, one that might satisfy $\nu_b/\nu_s\gtrsim10^4$,
since the particles at the ring-edge would effectively
experience $2m$ `avalanches' during each orbit.

However it is unknown whether the compressed regions in a confined ring-edge
are in fact close-packed. And the suggestion that tumbling close-packed ring particles
might satisfy $\nu_b/\nu_s\gtrsim10^4$ is at this stage mere speculation. Indeed, the reviewer of
this paper suggests that the non-Newtonian properties of self-gravitating wakes might
instead play an important role here.
However, the relevant physics, such as the kinetic theory of \cite{LO06},
or the results of Nbody simulations \citep{DTI01, SSS01},
are not easily adapted to a semi-anlytic treatment like ours,
and so the consequences of non-Newtonian behavior are not explored here. In light of these
uncertainties, we also consider an alternate ring confinement mechanism below,
to demonstrate that there may be more than one way for a satellite to resonantly
confine a sharp-edged planetary ring.

\subsubsection{drag in a planetary ring}
\label{B-ring-drag-section}

Section \ref{drag-section} notes that small ring particles are also susceptible to drag forces, 
such as PR, plasma, and atmospheric drag, and possibly the
Yarkovsky effect. Although a drag force tends to have the greatest influence
among smaller ring particles, the
smaller particles can still communicate the effects of this
drag to the larger particles via collisions.
This is due to the enhanced eccentricity damping that a smaller particle experiences
(Eqn.\ \ref{de/dt}); its lower $e$ will then favor collisions 
with the larger, more eccentric particles.
The cumulative effect of these collisions
would then resemble a drag force that also acts on the larger particles, too. 
But this scenario only works if the smaller ring particles are also
sufficiently abundant, {\it i.e.}, if the ring particles' size distribution is sufficiently steep.
Obviously, the outcome of this scenario will depend on the
particular drag force that is operative in the ring, as well as the 
particles' size distribution. And this initial study of drag is not 
prepared to deal with these uncertain details in a rigorous, quantitative way.
Nonetheless, the effects of drag can still be assessed
qualitatively by using the simple generic drag acceleration that is described
Section \ref{drag-section}, and by assuming that there is a single effective
drag parameter $C_d$ that adequately describes how this drag effects the
entire ensemble of ring particles.

When the drag is the dominant source of dissipation, a particle's
dissipation function becomes 
\begin{equation}
    \label{psi_d_drag}
    \psi_d=\psi_{\mbox{\scriptsize drag}}
        =-\bar{\alpha}^m_{dr} + 2\epsilon\bar{\alpha}^m_{d\theta}
        =-\frac{\epsilon C_d e}{m-\epsilon}
\end{equation}
(see Sections \ref{drag-section}---\ref{eom-again}). 
This drag force also causes the streamlines' longitudes of perihelia
to  trail behind the satellite's longitude $\theta_s$ by an angle
$\tilde{\omega}\simeq\psi_{\mbox{\scriptsize drag}}/m\psi_s=-3C_d e/mf_\epsilon^m\mu_s$
(from Eqns.\ \ref{psi} and \ref{eom_w}). The torque $T_s$ that the
satellite exerts on the ring is $-1\times$ Eqn.\ (\ref{L_s(x)}), so
$T_s\propto\int_{\mbox{\scriptsize ring}} \epsilon se\sin m\tilde{\omega}dx'$
and $\mbox{sgn}(T_s)=-\epsilon$.
Consequently, this drag force also enables a shepherding torque
that tends to drive ring particles radially away from the satellite's orbit,
since $T_s<0$ at an  $\epsilon=+1$ ILR and
$T_s>0$ at an $\epsilon=-1$ OLR.  And if the shepherding torque is
strong enough, it can counterbalance the viscous torque
and maintain the ring's edge near the resonance\footnote{Of course, drag forces such as 
plasma drag, Yarkovsky, {\it etc.},
can also exert an axisymmetric $m=0$ torque on a ring particle
that would cause it to migrate radially. In that case, a sharp ring-edge would instead
indicate a balance between
the satellite's shepherding torque and the $m=0$ part of the drag torque.}.
This  torque-balance is also known as resonance trapping, and it 
can occur in a wide variety of disk-perturber systems that also have some
dissipation, such as dust trapped at a planet's OLR due to
PR drag \citep{RSS94}, and the capture of planetesimals at a protoplanet's
OLR due to the solar nebula's aerodynamic drag \citep{WD85, M93a, HWR95}.

Ring confinement due to drag is illustrated in Fig.\ \ref{drag_fig}, which is for
a B ring that has a surface density $\sigma_\infty=265$ gm/cm$^2$ far from the
resonance,  a kinematic shear viscosity of $\nu_s=46$ cm$^2$/sec, 
and a drag parameter $C_d=1.0\times10^{-4}$. These
parameters were chosen so that the epicyclic amplitude
at the ring's outer edge has the observed value of $R_m=45$ km, and that the
ring's viscous torque balances the satellite's torque
at a distance of $24$ km beyond the nominal resonance.
Keep in mind that Fig.\  \ref{drag_fig} represents just one possible solution for the
B ring, since other judiciously chosen combinations of ring parameters will also lead  
to a torque balance at $24$ km beyond the nominal resonance.

Figure  \ref{drag_fig}c also 
shows that when drag is the dominant source of dissipation,
then a large longitudinal offset becomes possible, with $\tilde{\omega}\simeq-23^\circ$
at the ring's outer edge in this example.
But keep in mind that $\tilde{\omega}\propto-C_d$, so stronger drag would result 
in a larger $\tilde{\omega}$. Interestingly, \cite{SP06} measured an offset of
$\tilde{\omega}\simeq-28^\circ$ in Cassini observations of
the B ring's orientation, yet Voyager
observations revealed no significant offset \citep{PDG84}.

Figure \ref{drag_s_r_fig} uses Eqns.\ (\ref{sigma(r,theta)}) to
calculate the ring's relative surface density
$\sigma(r, \theta)/\sigma_\infty$ as a function of radial distance $r$ along selected 
longitudes $\theta=\theta_s$ (which is along the ring's longitude
of periapse), $\theta_s\pm90^\circ$ (towards the ring's apoapse), and
along the intermediate longitude $\theta_s\pm45^\circ$.
Again, the model predicts a large surface density
enhancement at the ring--edge's periapse, as well as a
low surface density shoulder at the ring--edge's apoapse

Evidently, one of this model's main predictions is that
the ring-edge's surface density should increase by $\sim50\%$ at periapse; 
see Figs.\ \ref{s_r_fig} and \ref{drag_s_r_fig}. 
Presumably, this periapse enhancement would have been seen by the Cassini spacecraft, 
if not by Voyager. One possible explanation for
this non-detection can be found in \cite{PWR08}, which describes a sophisticated 
photometric model of a swarm of ring particles that are illuminated by the Sun
and Saturn, and also imaged by a nearby spacecraft. That model
predicts that the optical surface brightness $I/F$ of Saturn's main rings should
saturate when the ring's optical depth exceeds about 0.3. Since the
optical depth of the outer B ring exceeds that threshold \citep{CES07},
this $I/F$ saturation could account for the absence of any detection of the 
expected periapse enhancement of the ring-edge's surface density.
However, the ring should be less optically thick when observed at longer
wavelengths, so the anticipated periapse enhancement
might be detectable when observing the ring-edge during a radio occultation.

The main purpose of this Section is to illustrate how
other forms of ring dissipation might enable a satellite to confine a ring at 
a Lindblad resonance. For instance, drag forces might play a role in the confinement
of the inner Uranian rings, which do orbit close enough to that planet to
experience a drag with that planet's extended upper atmosphere \citep{GP87}.
However, it should be noted that the differential particle size distribution
in Saturn's B ring varies as $dN(s)\propto s^{-q}$, where $s$ is the particle radius
and $q\simeq 2.75$ \citep{FN00}. Consequently,
the B ring's differential mass distribution varies as
$dM(s)\propto s^{3-q\simeq0.25}$, which 
is weighted towards the larger particles, so the B
ring is probably immune to the effects of drag forces. 

\subsubsection{the satellite's torque}

As was noted above in Section \ref{viscosity variations}, 
all of the B ring models that are 
described in Sections \ref{surface density variations}--\ref{viscosity variations}
and Figs.\ \ref{sigma_fig}--\ref{v_fig} failed to find a solution that
balances the satellite's torque on the ring
against the ring's own viscous torque.
This was rather surprising since most of those models
were constructed (or so we thought) so that the satellite's torque would counter-balance
the viscous torque. The maximum torque that a satellite can exert on material orbiting
at its $m^{\mbox{\scriptsize th}}$ Lindblad resonance is
\begin{equation}
    \label{T_gt}
    T_{\mbox{\scriptsize GT}} = -\frac{\epsilon m\pi^2\sigma_\infty a^2(\Psi_m^s)^2}
        {|{\cal D}|}
        =-\frac{\epsilon\hspace*{0.1ex}\pi (f_\epsilon^m\mu_s)^2}
           {3(1-\epsilon/m)}\mu_dM_p(a\Omega)^2
\end{equation}
\citep{GT78, GT82}. The viscous torque is the ring's viscous
angular momentum luminosity,
${\cal L} =3(\nu_s/a^2\Omega)\mu_d M_p(a\Omega)^2$
(see just below Eqn.\ \ref{L}), so one might expect a torque-balance to be possible
when $|T_{\mbox{\scriptsize GT}}|>T_\nu$, or when
\begin{equation}
    \nu_s<\frac{\pi(f_\epsilon^m\mu_s)^2}{9(1-\epsilon/m)} a^2\Omega,
\end{equation}
which evaluates to $\nu_s<145$ cm$^2$/sec for Mimas' $m=2$ ILR. Thus a 
torque-balance at the B ring's outer edge should seemingly be possible if the B
ring's shear  viscosity were comparable or less than that inferred
for the A ring. However, the following review of the derivation of the
Goldreich-Tremaine torque formula,
Eqn.\ (\ref{T_gt}), will show why it can overestimate by a large margin the
torque that Mimas exerts at a ring's outer edge.

The easiest way to derive Eqn.\ (\ref{T_gt}) is to consider a drag-dominated
ring whose dissipation function is Eqn.\ (\ref{psi_d_drag}); that scenario
was also considered in \cite{MVS87}.  The real and imaginary parts of a
ring particle's equation of motion (\ref{eom3}) are
\begin{equation}
    \label{sincos}
    \sin m\tilde{\omega} =\frac{\psi_d}{\psi_s}
        \qquad\mbox{and}\qquad
        \cos m\tilde{\omega} =-\frac{\epsilon de  + \bar{\alpha}^m_{cr}}{\psi_s},
\end{equation}
where it is assumed that  $\bar{\alpha}^m_{cr}$ is real.
Squaring and summing Eqns.\ (\ref{sincos}) also shows that 
$(\epsilon e d + \bar{\alpha}^m_{cr})^2 + \psi_d^2 = \psi_s^2$,
where $\psi_d$ is Eqn.\ (\ref{psi_d_drag}) and
the $\bar{\alpha}^m_{cr}$ is the acceleration on a particle
that is due to gravity plus pressure. If that $\bar{\alpha}^m_{cr}$
term can be neglected, then the above expressions  provide the particle's
orbit elements in a drag-dominated ring, which are
\begin{equation}
    \label{e-drag}
    e(x)\simeq\frac{|\psi_s|}{\sqrt{x^2 + c_d^2}}
        \qquad\mbox{and}\qquad
        \sin m\tilde{\omega}(x)\simeq-\frac{c_d}{\sqrt{x^2 + c_d^2}}
\end{equation}
where $d(x)\simeq x$ (see Eqn.\ \ref{d}) and with $c_d=C_d/(m-\epsilon)$.
The satellite's radial torque density is Eqn.\ (\ref{dT/da}),
so the total torque that the satellite exerts on the ring is the integral
\begin{mathletters}
    \label{T_s}
    \begin{eqnarray}
        \label{T_s1}
        T_s &=& \int_{\mbox{\scriptsize ring}} \frac{\partial T_s}{\partial a}da=
            m\pi\sigma_\infty a^3\Psi_s^m\int_{\mbox{\scriptsize ring}} 
            es\sin(m\tilde{\omega})dx'\\
        \label{T_s2}
        &=&-\epsilon m \pi\sigma_\infty a^3\Psi_s^m\psi_s c_d
            \int_{-\infty}^{\infty}\frac{dx}{x^2+c_d^2}
            =-\frac{\epsilon m\pi^2\sigma_\infty a^2(\Psi_m^s)^2}{|{\cal D}|}
            =T_{\mbox{\scriptsize GT}}
    \end{eqnarray}
\end{mathletters}
when it is assumed that the ring has a constant surface density ($s=1$)
that extends everywhere. But if the ring instead had a sharp edge at the resonance,
then the satellite's torque would be half that.

The integrand in Eqn.\ (\ref{T_s2}) indicates that the satellite's torque on the ring is
exerted over a fractional radial scale $|x|_{\mbox{\scriptsize torque}}\sim c_d$,
which makes sense since this is
where the streamlines' $e$ and $|\tilde{\omega}|$ are maximal (Eqns.\ \ref{e-drag}).
But keep in mind that Eqns.\ (\ref{e-drag}) also ignored the ring's internal forces
$\bar{\alpha}^m_{cr}$,
which Section \ref{surface density} showed to be important in the nonlinear
zone whose radial extent is $x_{NL}=\sqrt{|\psi_s|}$ from the resonance
(Eqn.\ \ref{x_NL}). Those internal ring forces also
tend to inhibit streamline-crossing,
which they achieve by reducing the streamline's eccentricities below
that given in Eqn.\ (\ref{e-drag}). This in turn reduces the satellite torque
to something less than Eqn.\ (\ref{T_s2}). But this torque reduction should still
be insignificant whenever $x_{NL}\ll |x|_{\mbox{\scriptsize torque}}$,
{\it i.e.}, when the revised drag parameter satisfies
$c_d\gg\sqrt{|\psi_s|}$. If, however,
this criterion is not satisfied, then the Goldreich-Tremaine torque formula,
Eqn.\ (\ref{T_gt}), will overestimate the satellite's torque on the ring.

Of course, the dissipation in the ring models of Sections 
\ref{surface density variations}--\ref{viscosity variations} and 
Figs.\ \ref{sigma_fig}--\ref{v_fig} was due to viscosity rather than drag.
However, the preceding discussion suggests that Eqn.\ (\ref{T_gt})
will provide a reliable estimate of the satellite's torque whenever
$\Delta a_{\tilde{\omega}}$, which is defined here as the radial distance
over which the satellite substantially excites the streamlines' ${\tilde{\omega}}$,
satisfies $\Delta a_{\tilde{\omega}}\gg\Delta a_{NL}$ where 
$\Delta a_{NL}=\sqrt{|\psi_s|}a_r$ is the width of the nonlinear zone in
physical units. Note that the width of Mimas' 
$m=2$ ILR is $\Delta a_{NL}\simeq20$ km (Section \ref{surface density}),
while Figs.\ \ref{Q_fig}--\ref{v_fig} show that these viscous ring models
have $\Delta a_{\tilde{\omega}}\lesssim10$ km. Consequently, the satellite's torque
is reduced below Eqn.\ (\ref{T_gt}), which explains those models' difficulty
in achieving a torque balance at the ring's outer edge.

The torque $T_s$ that the satellite exerts on each model ring is also
reported in all of the figure captions.
Those captions show that the ring's internal
forces---gravity and pressure---reduce
the satellite's torque  below Eqn.\ (\ref{T_gt}) by three to
seven orders of magnitude. This explains why
the viscous ring model of Fig. \ref{vb_edge_fig} requires a viscosity ratio
$\nu_b/\nu_s\sim10^4$, which also boosts the satellite's torque by that factor,
in order to achieve a torque-balance at the
B ring's outer edge. The exception of course is the drag-dominated
ring of Fig.\ \ref{drag_fig}. In that model, the satellite's torque is only about
three times smaller than $T_{\mbox{\scriptsize GT}}$,
with the greater torque efficiency being
due to the drag's ability to excite larger ${\tilde{\omega}}$
across a wider radial span in the ring
such that $\Delta a_{\tilde{\omega}}>\Delta a_{NL}$.

\subsubsection{assessing the ring's internal forces}
\label{forces_section}

The relative importance of the various accelerations that a ring particle
experiences due to ring gravity, pressure, viscosity, etc., is assessed by
dividing the real part of the complex equation of motion (\ref{eom3})
by the satellite's forcing function $\psi_s$, which yields
\begin{equation}
    \label{real_eom}
    ({\mathcal A}_{cp} + {\mathcal A}_g + {\mathcal A}_p)\cos(m\tilde{\omega})
        -  ({\mathcal A}_\nu + {\mathcal A}_d)\sin(m\tilde{\omega}) + 1 \simeq 0
\end{equation}
where ${\mathcal A}_{cp}=\epsilon de/\psi_s$, 
${\mathcal A}_{g}=\Re e(\bar{\alpha}^m_{gr})/\psi_s$, 
${\mathcal A}_{p}=\Re e(\bar{\alpha}^m_{pr})/\psi_s$, 
${\mathcal A}_{\nu}=\Re e(\psi_\nu)/\psi_s$, and
${\mathcal A}_{d}=\Re e(\psi_{\mbox{\scriptsize drag}})/\psi_s$,
where $\psi_\nu$ and $\psi_{\mbox{\scriptsize drag}}$ are Eqn.\ (\ref{psi_v})
and (\ref{psi_d_drag}). Note that Eqn.\ (\ref{real_eom}) is only approximately
true because the accelerations
appearing in the equation of motion (\ref{eom3}) are complex, yet Eqn.\
(\ref{real_eom}) neglects their smaller imaginary parts.
The quantities ${\mathcal A}_{g}$, ${\mathcal A}_{p}$, ${\mathcal A}_{\nu}$,
and ${\mathcal A}_{d}$ represent the strength of the 
$m^{\mbox{\scriptsize th}}$ component of a particle's acceleration
due to ring gravity, pressure, viscosity, and drag, all in units of the satellite's forcing
$\psi_s$, while ${\mathcal A}_{cp}=\epsilon e d/\psi_s$ is the relative strength of the
centrifugal and Coriolis accelerations that the central planet exerts on the particle
due to its noncircular motions. 

The relative accelerations ${\mathcal A}$ are shown in 
Fig.\ \ref{forces_fig} for the two models that achieved a torque-balance
at the ring's outer edge: the model described in Section \ref{v_b scenario}
and Figs.\ \ref{vb_edge_fig}--\ref{s_r_fig} that
invokes an  extreme viscosity ratio $\nu_b/\nu_s\gg1$ to achieve its torque-balance
(see Fig.\ \ref{forces_fig}a),
and the model of Section \ref{B-ring-drag-section} and 
Figs.\ \ref{drag_fig}--\ref{drag_s_r_fig} that
relies on drag to balance the ring-satellite torques (Fig.\ \ref{forces_fig}b).
Both models show that the B ring's internal accelerations are small
interior to the ring's outermost $\sim40$ km. There, the particle's motion
balances the satellite's forcing against the central planet's
centrifugal/Coriolis forces, {\it i.e.}, ${\mathcal A}_{cp}\simeq-1$
since $|\tilde{\omega}|\ll1$, 
which is equivalent to the single particle solution, Eqn.\ (\ref{e_forced_single}).
Figure \ref{forces_fig} also shows that, in the model B ring's outermost $\sim40$ km,
the ring's gravity is the dominant internal ring-force whose strength is comparable to 
the satellite's forcing and the central planet's
centrifugal/Coriolis forces. Those curves show
that the acceleration due to ring pressure is small everywhere except at
the ring's outer edge, whose effects there are only marginally resolved in these models
(see Section \ref{dispersion v}).
Figure \ref{forces_fig} also shows that viscosity has no direct effect on the
ring's epicyclic amplitude; instead, its influence enters indirectly via
the torque-balance Eqn.\ (\ref{torque-balance}). Figure \ref{forces_fig}b
also illustrates how strong the 
drag force must be if it  is indeed responsible for a torque-balance
at the B ring's outer edge. 

Lastly, recall Eqn.\ (\ref{d}), which shows that
the $m=0$ component of the ring's conservative accelerations,
$\alpha_{gr}^0 + \alpha_{pr}^0$, can displace
the location of the resonance, which is the site where
$d(x)=0$.
It turns out that this displacement is quite small---too small to be resolved in
the models considered here. However, that displacement is easily inferred
from a linear interpolation of the $d(x_i)$
data that the streamline model generates.
For instance, an interpolation of the data generated by the two B ring models in
Figure \ref{forces_fig} shows that the ring's internal forces
displaces Mimas' $m=2$ ILR inwards about 15m. That small displacement is 
due to ring gravity, since the ring pressure is negligible there.

\section{Discussion}
\label{discussion}

\subsection{Heating the ring's outer edge}
\label{heating}

Another issue that merits consideration is the viscous heating of the ring's outer
edge (e.g, \citealt{BGT82}), which might be important when collisions are the
dominant source of viscosity, since $\nu\propto c^2$ in this case. 
To assess this heating, one has to calculate the rate at which the ring's
dissipative forces do work on each ring particle.
That quantity is probably positive, because
the viscous delivery of orbital energy to the ring's outer edge likely exceeds the
rate at which the satellite withdraws orbital energy from the ring edge,
resulting in a dynamical heating of the ring-edge \citep{BGT82}.
Of course, other processes also tend to cool the ring particles' random velocities,
such as dissipative collisions that convert impact energies into thermal heat,
mechanical grinding and fracturing of ring particles, and thermal radiation.
Those other cooling mechanisms have not yet been considered,
but do need to be included in the ring's energy-balance equation, which
could be used to relate the ring's viscosity $\nu(a)$ to
the ring particles' dispersion velocity $c(a)$.
But many of the terms in that equation will be difficult to quantify,
and will be deferred to a followup study.
One might have to resort to an order-of-magnitude type analysis,
or perhaps settle for an upper limit on the heating that occurs at the ring's edge. 
However, upcoming Cassini observations of the rings during the August 2009
equinox will be helpful here, since that is when the Sun will pass through
the ring plane. If the ring-edge is dynamically hot and thick, then the
shadows it will cast onto the ring-plane should be observable,
which would allow one to infer the ring-edge's 
vertical scale height $h$ and the ring particles'
dispersion velocity $c=h\Omega$ there.

\subsection{Future applications}
\label{applications}

The outer edge of Saturn's A ring is maintained by an $m=7$ ILR with
the coorbital satellites Janus and Epimetheus \citep{PDG84},
whose semimajor axes differ only by about $0.03\%$ \citep{JSP08}.
These satellite's mutual attractions cause their orbits to swap about
every four years \citep{YCS83}, such that
only one coorbital appears to have its $m=7$ ILR in the A ring
at any instant of time \citep{SPH08}. One might be tempted to use this streamline
model to calculate the ring's response to each individual satellite, and then to
superimpose the model outputs to obtain the ring's total response
to both satellites. However this might not lead to reliable results,
due to the satellites' time-varying orbits. Because the torque that they exert
on the A ring changes periodically with time, the outer A ring never experiences
a balance of the viscous and satellite torques at any given instant, and
so the model's torque balance Eqn.\ (\ref{torque-balance}) does not apply here.
Instead, a {\em time-average} of the outer A ring's viscous torque should be balanced
against the time-averaged torque that the coorbitals exert on the ring. However,
the calculation of those time-averaged torques is subtle, and will be saved
for a followup study of the A ring \citep{SPH08}.

The streamline model can also be used to study narrow
eccentric ringlets, but with some revision. For instance, if ringlets 
are maintained by small unseen shepherd satellites, then the
impulse approximation should be used to calculate the torque that they
exert on the nearby ringlet. Also, the differential precession that is due to 
planetary oblateness, which is important in a ringlet system, will also
need to be accounted for in a revised version of the streamline model.

Lastly, the streamline model can also be modified so that it can simulate
linear as well as nonlinear spiral density waves.
Spiral waves transport angular momentum through a ring via the tangential
accelerations that particles experience due to ring gravity and pressure.
Those terms are negligible in this study of a nearly peri-aligned B ring,
but their inclusion will be needed in order to handle the spiral waves'
angular momentum transport.
Those and other related problems will be considered in followup studies of planetary rings.

\section{Summary of Results}
\label{results}

The preceding describes a model of a broad, sharp-edged planetary 
ring that is confined by a satellite's $m^{\mbox{\scriptsize th}}$
Lindblad resonance. This model utilizes the
streamline formalism of \cite{BGT82, BGT85}, which makes
the calculation of the ring's internal forces---ring gravity, pressure, and 
viscosity---quite tractable.
The model also includes a simple prescription for handling the drag
forces, such as  such as plasma, Yarkovsky, atmospheric,
and/or PR drag,  that small ring particles might experience.
The model's main inputs are the ring's surface density $\sigma_\infty$,
the ring particle's dispersion
velocity $c$, the ring's kinematic shear and bulk viscosities
$\nu_s$ and $\nu_b$, and a dimensionless drag coefficient $C_d$.
This streamline model solves a nonlinear form of Newton's second
law of motion to obtain the streamlines' orbital eccentricities $e(a)$ and
longitudes of perihelia $\tilde{\omega}(a)$ as functions of the streamlines'
semimajor axis $a$. The model also balances the ring's viscous torque
against the satellite's gravitational torque in order to calculate the ring's surface density
$\sigma(r,\theta)$ as a function of radius $r$ and longitude $\theta$. 
That analysis also shows how to use linear and angular momentum fluxes
to calculate the effects of viscosity and pressure, both of which are discontinuous
at a ring's sharp edge.

The streamline model is then applied to the outer edge of Saturn's B ring,
which is maintained by an $m=2$ ILR with the satellite Mimas. A suite
of B ring scenarios are examined in order to illustrate how the model outcomes
depend upon the the ring's physical properties $\sigma_\infty$, $c$,
$\nu_s$ and $\nu_b$, and possibly $C_d$, with the main findings listed below.

1. As one might expect, increasing the ring's surface density tends to decrease
the ring-edge's epicyclic amplitude $R_m$. 

2. Pressure in the ring is controlled
by the ring particle's dispersion velocity $c$, but increases in $c$
are manifest only at the ring's outer edge, which is where the ring's
pressure drop is greatest. However the consequences of that pressure drop
are not fully resolved in the models considered here, which have
radial samplings of 0.5 to 2 km.

3. The magnitude of the B ring's internal forces are compared, and it is shown
that the ring's gravity dominates over its other internal forces that are due to pressure
and viscosity.  In the B ring's outer $\sim40$ km, the gravitational force that the
ring exerts on a particle is comparable to the satellite's forcing. However, at sites
well interior to the ring's outer $\sim40$ km, ring gravity is relatively small, so the ring
particles adopt the familiar single-particle solution $e=|\psi_s/x|$ there.

4. The ring's viscous torque is controlled by its shear viscosity $\nu_s$,
which also governs the rate of the ring's radial spreading. 
However the satellite's torque on a viscous ring, which opposes that spreading,
is proportional to the angle by which the streamline's periapse lags behind the satellite's
longitude. That lag angle is a linear
combination of the ring's shear $\nu_s$ and bulk $\nu_b$ viscosities, so
the satellite's torque on the ring is sensitive to the sum of those  viscosities.
Interestingly, a conventional ring model that has a bulk viscosity that
is comparable or less than the shear viscosity
fails, by a very wide margin, to balance the satellite's torque against the
ring's viscous torque, so that scenario fails to account
for the B ring's sharp edge near a resonance. Nonetheless, increasing $\nu_b$ 
does strengthen the satellite's torque on the ring, 
and models show that a torque balance
becomes possible if $\nu_b/\nu_s\sim10^4$. 
In other words, the ring particles' shearing tangential motions must 
be relatively free of frictional dissipation, while the ring particles' 
compressed radial motions must result in heavy frictional dissipation.
Such a scenario might be possible if particles in the B ring's compressed
regions are confined shoulder-to-shoulder so that there is little voidspace between
the ring particles there, and ring's volumetric density becomes incompressible there.
In this case, ring particles must rise and then fall vertically,
sliding or perhaps tumbling past one another as the satellite's perturbations
drives the ring-edge radially inwards and then outwards.
This periodic avalanche of ring particles could be quite lossy,
possibly resulting in a very large bulk viscosity $\nu_b$.
However it is unclear whether the B ring edge actually behaves in this way,
and whether its viscosity can satisfy this remarkable requirement
of $\nu_b/\nu_s\sim10^4$, so this finding is speculative. Note also
that current ring observations only provide measurements of the combined viscosity
$\nu_s +f\nu_b$ where $f\sim{\cal O}(1)$, so the ratio $\nu_b/\nu_s$
in a perturbed planetary ring-edge is actually unknown.

5. Drag forces provide an alternate means of boosting the satellite's torque on the ring.
A generic drag force is considered, one that is proportional to a ring particle's
noncircular velocity, and its main effect is to damp the particle's eccentricity
while causing its longitude of perihelia $\tilde{\omega}$ to trail
behind the satellite's longitude. Since the satellite's torque on the particle
is proportional to $\sin(m\tilde{\omega})$, this allows the satellite to torque
a wide annulus in the ring, which also makes a torque-balance at the ring-edge
favorable. However this torque-balance due to drag can only be effective if the ring particles'
mass distribution is dominated by small particles, which is probably not the
case for Saturn's B ring.

6. The torque that the satellite exerts on the entire ring is calculated for a
variety of viscous B ring models, and it is shown to be
3-7 orders of magnitude smaller than that anticipated by the Goldreich-Tremaine
torque formula $T_{\mbox{\scriptsize GT}}$ \citep{GT78, GT82}. 
This is due to the ring's self-gravity, which
suppresses the streamline's eccentricities in the nonlinear zone
that is near the resonance, which also reduces the
satellite's torque there. However, a drag-dominated ring can experience a much
larger torque. This is due
to the streamlines' trailing longitudes of perihelia $\tilde{\omega}$,
which communicates the satellite's torque beyond the nonlinear zone, and 
can result in a torque that is comparable to $T_{\mbox{\scriptsize GT}}$.

7. The outer edge of the B ring could lie as far as $\Delta a\simeq24$ km 
beyond Mimas' $m=2$ inner Lindblad
resonance, which is the site where the ring's viscous torque
precisely balances the satellite's torque on the ring. 
Models of the B ring are adjusted so that
the simulated ring-edge's epicyclic amplitude agrees with the observed
$R_m=45$ km amplitude, and that a torque
balance is achieved at distances of $0\le\Delta a\le24$ km beyond Mimas
$m=2$ ILR. This requires the outer B ring to have a surface density of
$10\lesssim\sigma_\infty\lesssim280$ gm/cm$^2$
in the ring's outermost $\sim40$ km, with the larger
surface densities required for rings whose edges lie farther beyond the resonance.
A more detailed comparison of models to Cassini observations, which is ongoing,
will lead to a more precise measurement of that ring's surface density.
Note, though, that the displacement
of the ring-edge from the resonance is {\em not} due to
the ring's internal forces altering the resonance location.
For instance, the displacement of the resonance due to ring self-gravity and pressure
is tiny, only about 15m for the models considered here. But if Mimas
had instead been more or less massive, then the B ring's outer edge would then
lie interior or exterior to its present location.

8. Models of the B ring predict that the ring-edge's surface density should be
enhanced by $\sim50\%$ at the ring's longitude of peripase. The ring's outer
edge should also exhibit a low surface density shoulder at the ring's longitude of
apoapse. These surface density variations are due to the satellite's perturbations,
which compresses the streamlines at periapse and rarefies them at apoapse.
It is curious that these periapse enhancements have not been reported in
spacecraft observations of the B ring edge, but this
non-detection may be due to a saturation
of the ring's surface brightness $I/F$ that is expected to occur when
the ring's optical depth exceeds about 0.3 \citep{PWR08}.

9. Modifications to the streamline model will also allow its application to
other dense planetary rings, such as the outer edge of Saturn's A ring, and the many
narrow and sometimes eccentric ringlets that
orbit both Saturn and Uranus. And with additional physics, this model
will also provide a useful tool that can be used to simulate nonlinear spiral density waves.
Detailed comparisons of models to spacecraft observations of Saturn's rings
are ongoing ({\it c.f.} \citealt{SPH08}), and that activity should yield better estimates of,
or else place limits on, the ring's physical parameters
$\sigma_\infty$, $c$, $\nu_s$ and $\nu_b$, and $C_d$. Such studies will also
lead to a better understanding of the mutual interactions
that are exerted in these very interesting ring-satellite systems.

\acknowledgments

\begin{center}
  {\bf Acknowledgments}
\end{center}

J.M.H.'s contribution to this work was supported by grant NNX07-AL44G
issued by NASA's Science Mission Directorate via its Outer Planets
Research Program. We also thank Glen Stewart for his review of this work.

\bibliography{biblio}

\newpage
\begin{deluxetable}{rl}
\renewcommand{\arraystretch}{1.0}
\tablewidth{0pt}
\tablecaption{\label{Saturn-table} Physical properties of the Saturn--Mimas
    system\tablenotemark{a}}
\tablehead{}
\startdata
Saturn's $GM_p$\hspace*{2ex} 	& 	$3.79312077\times10^7$ km$^3$/sec$^2$   \\
Saturn's zonal harmonic $J_2$\hspace*{2ex} 	& 	$1.629071\times10^{-2}$		\\
$J_4$\hspace*{2ex} 	& 	$-9.3583\times10^{-4}$	 	\\
$J_6$\hspace*{2ex} 	& 	$ 8.614\times10^{-5}$	 	\\
$J_8$\hspace*{2ex} 	& 	$-1.0\times10^{-5}$		 \\
Mimas' angular velocity\tablenotemark{b} $\Omega_s$\hspace*{2ex}	&	381.9944522 degree/day   \\
Mimas' fractional mass $\mu_s$\hspace*{2ex}	&	6.5969$\times10^{-8}$    \\
\enddata
\tablenotetext{a}{from \cite{JAB06}, except where noted otherwise}
\tablenotetext{b}{as reported by the JPL Solar System Dynamics website
http://ssd.jpl.nasa.gov/?sat\_elem on June 26, 2008}
\end{deluxetable}


\newpage
\begin{figure}
\epsscale{1.0}
\plotone{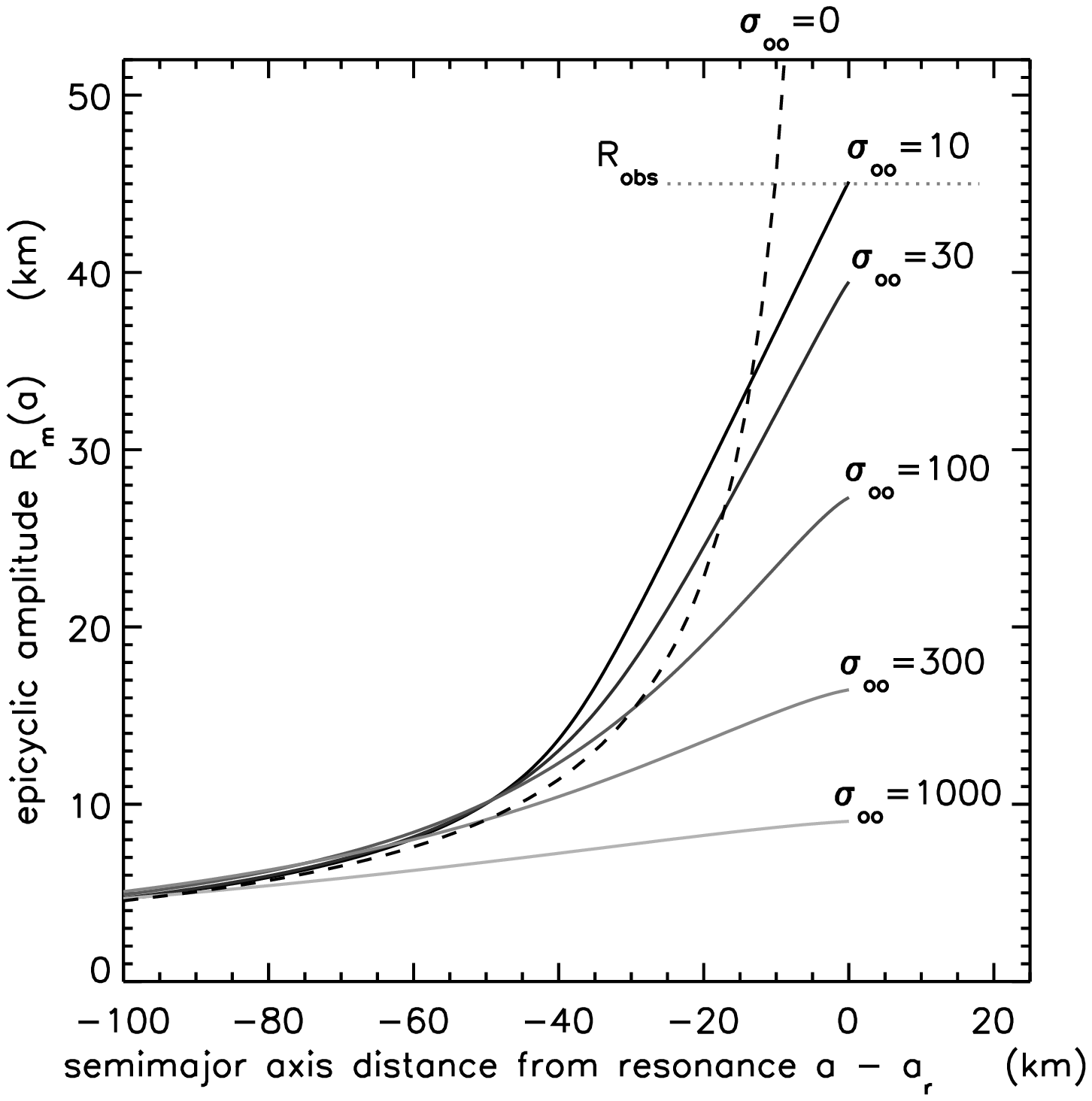}
\end{figure}

\newpage
\begin{figure}
\figcaption{
    \label{sigma_fig}
    Simulations of the B ring's epicyclic amplitude $R_m$ are plotted versus $a-a_r$,
    which is the radial distance from Mimas' nominal $m=2$ ILR.
    Solid curves are for simulated rings that  have a variety
    of surface densities $\sigma_\infty$ that are indicated above in units of gm/cm$^2$.
    The dashed $\sigma_\infty=0$ curve is the single particle solution, 
    Eqn.\ (\ref{R_epi_single2}), and the horizontal dotted line indicates the B ring's
    observed epicyclic amplitude of $R_{\mbox{\scriptsize obs}}\simeq45$ km
    \citep{SP06}. The model rings all have viscosities
    of $\nu_s=\nu_b=50$ cm$^2$/sec, a stability parameter of $Q=2$, 
    and no drag ($C_d=0$.) The rings' outer edges 
    are also forced to reside at the nominal resonance where $x=0=a-a_r$; see 
    Section \ref{surface density variations} for details.
    All of these simulations use $N=300$ streamlines that are distributed uniformly
    at the B ring's outer edge. The streamlines' radial widths differ in each model, with
    $\Delta a=0.50, 0.50, 0.67, 1.0$, and $2.0$ km, with wider streamlines
    used in the heavier rings for reasons explained in Section \ref{methods}.
    The total radial extent of the modeled region
    is $w=N\Delta a=150$ km in the lightest ring and $w=600$ km
    in the heaviest ring.
    The torque $T_s$ that the satellite exerts on the ring is Eqn.\ (\ref{T_r(a)}) 
    evaluated at the ring's outer edge, and it is quoted in units of $T_{GT}$ (Eqn. \ref{T_gt}).
    Those torque ratios range over $|T_s/T_{GT}| = 7.6\times10^{-4}$
    for the lighter $\sigma_\infty=10$ gm/cm$^2$ ring to 
    $|T_s/T_{GT}| = 1.7\times10^{-7}$ in the heavier  $\sigma_\infty=1000$ gm/cm$^2$
    ring.
}
\end{figure}

\newpage
\begin{figure}
\epsscale{1.0}
\plotone{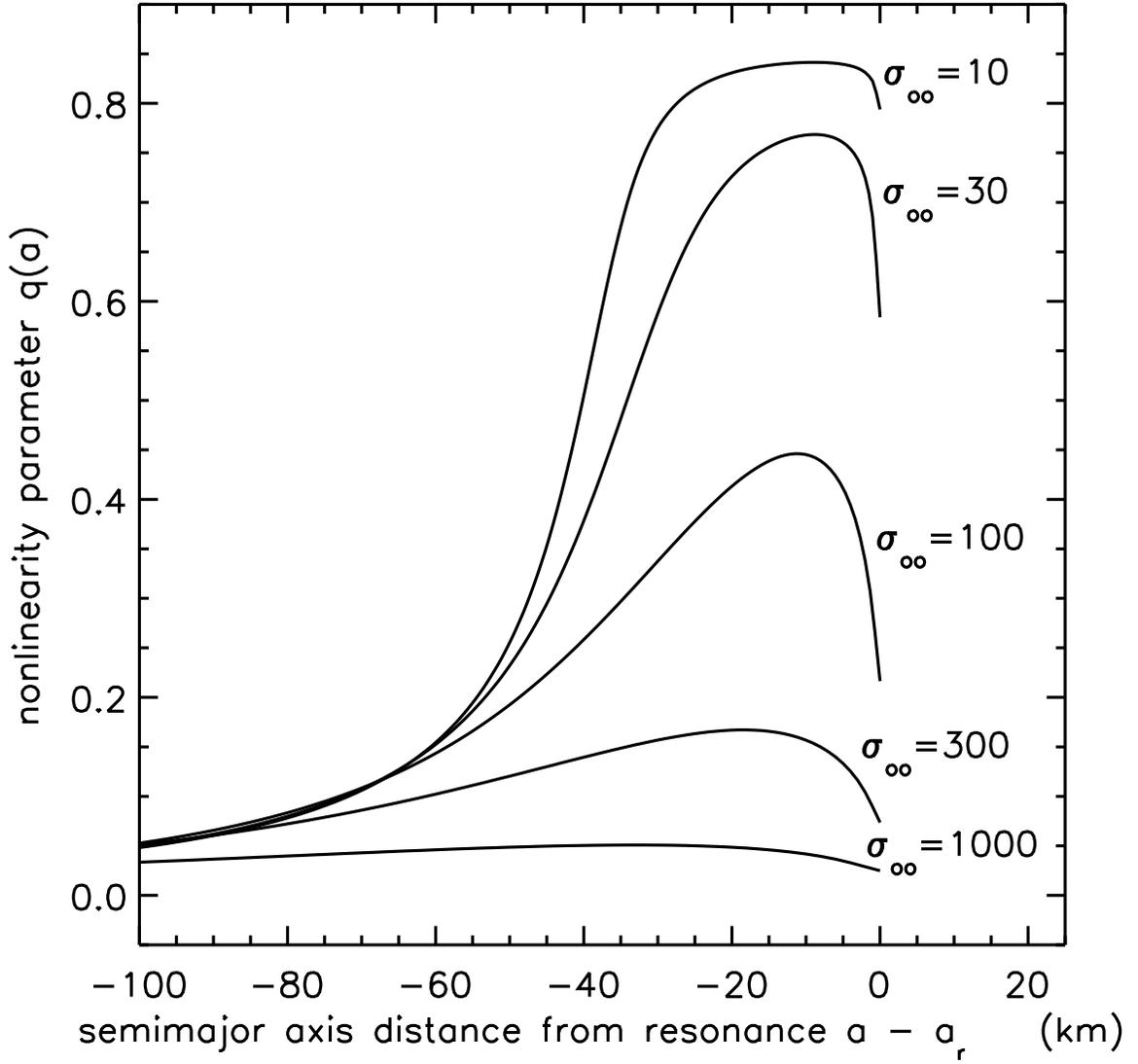}
\figcaption{
    \label{q_fig}
    The nonlinear parameter $q$ is plotted versus distance from resonance
    for the B ring models of Fig.\ \ref{sigma_fig}.
}
\end{figure}

\newpage
\begin{figure}
\epsscale{1.0}
\plotone{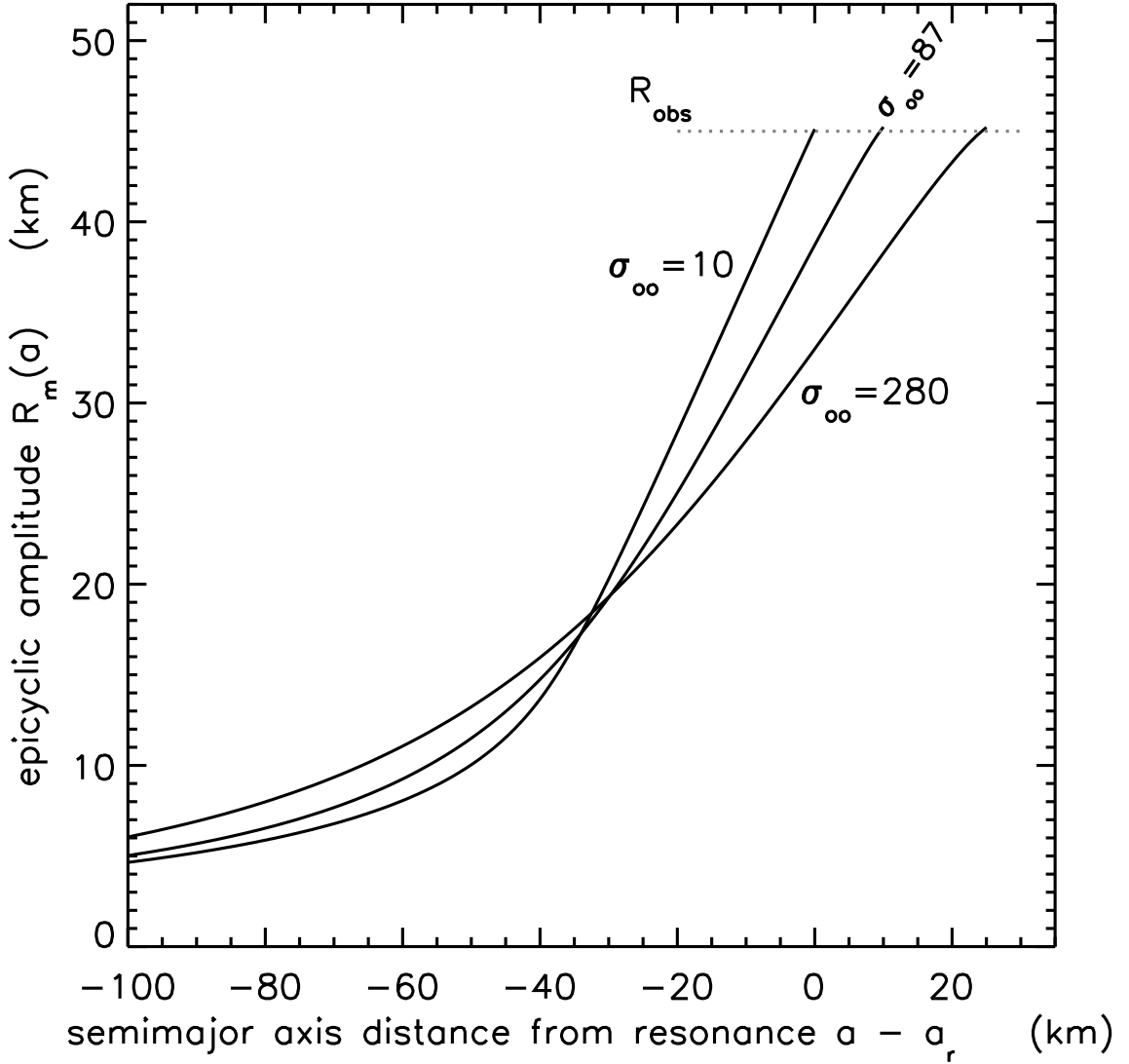}
\figcaption{
    \label{edge_fig}
    Results of three simulations of the outer B ring as it is perturbed by
    Mimas' $m=2$ ILR. The three simulated rings have outer edges that lie
    at radial distances of $a-a_r=0, 10,$ and 25 km beyond the nominal $m=2$ 
    resonance position. Labels indicate each ring's surface density $\sigma_\infty$
    in units of gm/cm$^2$, which is adjusted so that the epicyclic amplitude
    $R_m$ at the ring's outer edge is comparable to the observed value
    $R_{\mbox{\scriptsize obs}}\simeq45$ km, which is indicated by the dotted line.
    These models use $N=300$ streamlines having widths of $\Delta a = 0.50, 0.87$,
    and 1.42 km, with wider streamlines being used in the heavier rings, so the
    total radial extent of the modeled regions are 150, 260, and 425 km.
    The remaining model parameters are identical to that used in 
    Figs.\ \ref{sigma_fig}--\ref{q_fig}, with $\nu_s=\nu_b=50$ cm$^2$/sec,
    $Q=2$, and $C_d=0$. The torque that the satellite exerts on the ring
    ranges over $|T_s/T_{GT}| = 7.6\times10^{-4}$ in the lighter 
    $\sigma_\infty=10$ gm/cm$^2$ ring to 
    $|T_s/T_{GT}| = 2.9\times10^{-5}$ in the heavier  $\sigma_\infty=280$ gm/cm$^2$
    ring.
}
\end{figure}

\newpage
\begin{figure}
\epsscale{1.0}
\plotone{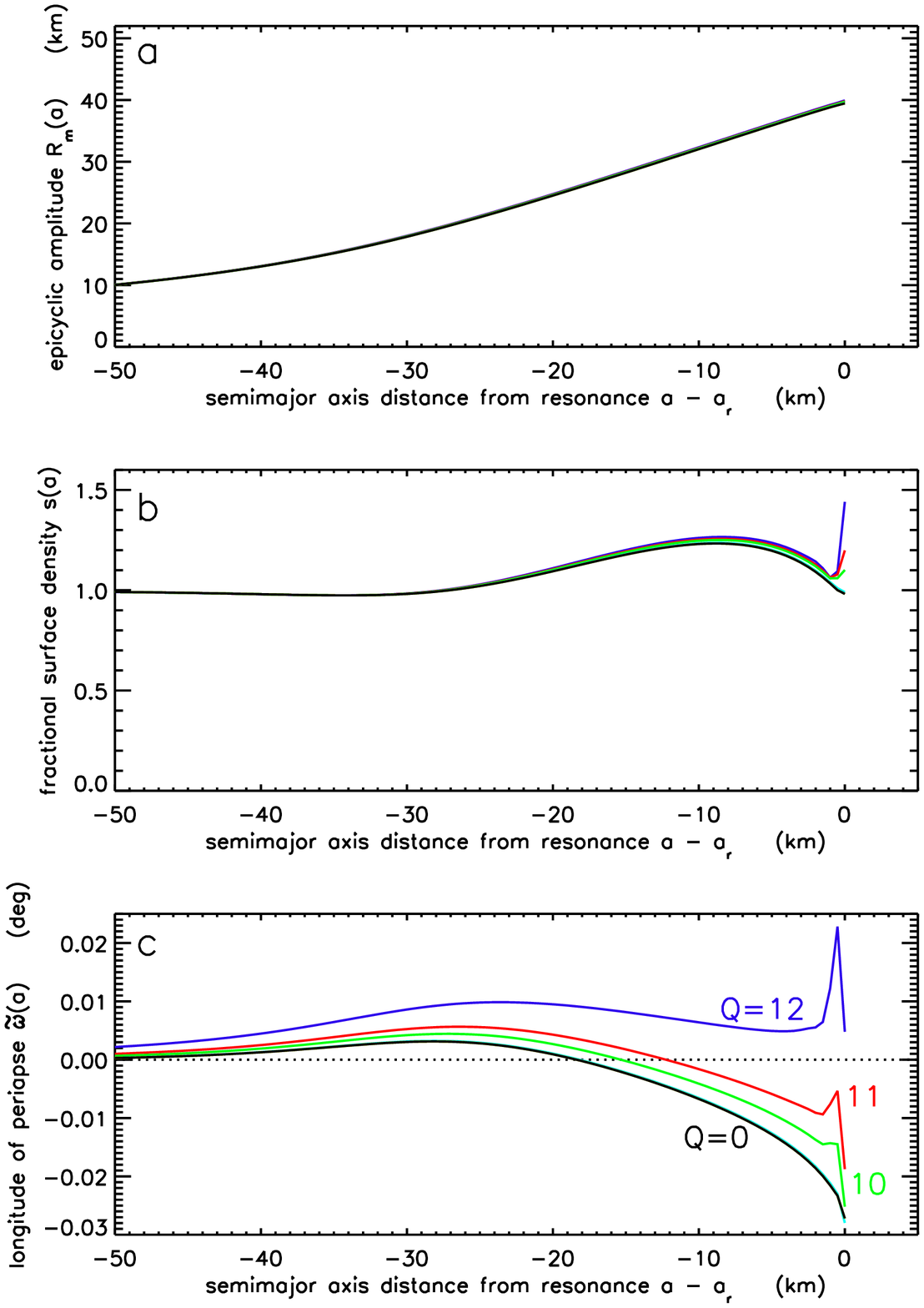}
\end{figure}

\newpage
\begin{figure}
\epsscale{1.0}
\figcaption{
    \label{Q_fig}
    The B ring's epicyclic amplitude $R_m(a)$, its fractional surface density $s(a)$, 
    and its longitude of periapse
    $\tilde{\omega}(a)$ are plotted versus distance from resonance
    $a-a_r$ for models that all have the same surface density
    $\sigma_\infty=30$ gm/cm$^2$, viscosities $\nu_s=\nu_b=50$ cm$^2$/sec,
    and stability parameters of 
    $Q=0$ (which corresponds to a pressureless ring, black curve),
    $Q=1$ (yellow), $Q=5$ (cyan), $Q=10$ (green), 
    $Q=11$ (red), and $Q=12$ (blue). The $Q=0$ to 5
    simulations are nearly indistinguishable and lie under the black curves. 
    There is no drag in these simulations
    ($C_d=0$), and the rings' outer edges 
    are at the nominal resonance where $a=a_r$. These simulations use $N=300$
    streamlines that are distributed uniformly over the ring's outermost 150 km, so the
    radial resolution here is $0.5$ km. The torque that the satellite exerts on the ring
    ranges is $|T_s/T_{GT}| = 1.5\times10^{-4}$ in the $0<Q\le 5$ models,
    $7.7\times10^{-5}$ in the $Q=10$ model, $6.1\times10^{-6}$ for $Q=11$, 
    and $2.3\times10^{-4}$ for $Q=12$.
}
\end{figure}

\newpage
\begin{figure}
\epsscale{1.25}
\hspace*{-15ex}\plotone{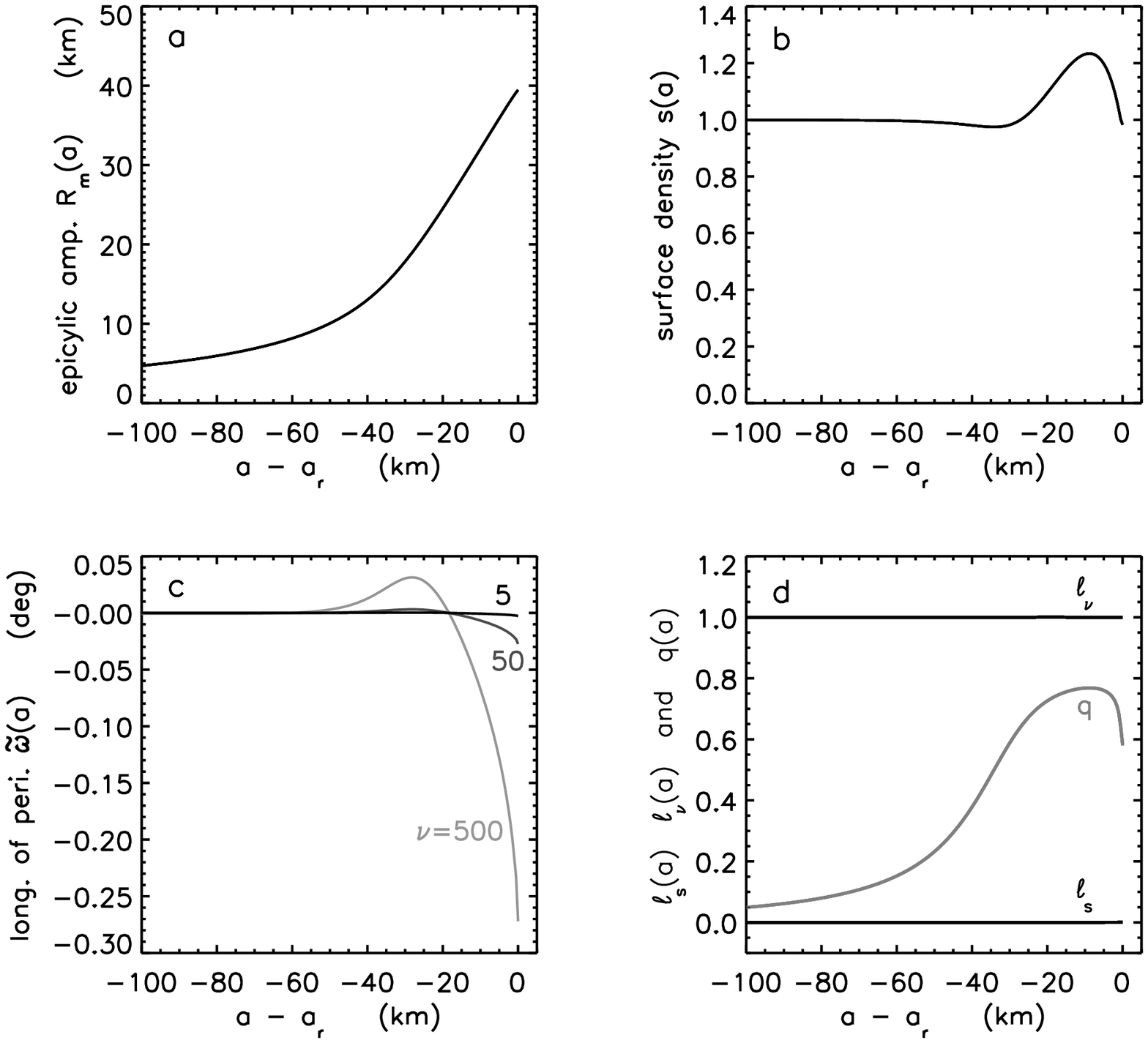}
\end{figure}

\newpage
\begin{figure}
\figcaption{
    \label{v_fig}
    The simulations of the B ring's epicyclic amplitude $R_m$ (Fig.\ a),
    fractional surface density $s$ (Fig.\ b), and
    longitude of periapse $\tilde{\omega}$ (Fig.\ c) are all plotted
   against the radial distance from resonance $a-a_r$.
   Figure d plots the rings' dimensionless angular momentum luminosity
   due to viscosity $\ell_\nu$ and the satellite's gravitational torque $\ell_s$,
   as well as the rings nonlinearity parameter $q$. Three simulations are shown,
   and they all have an undisturbed surface density $\sigma_\infty=30$ gm/cm$^2$,
   stability parameter $Q=2$ (which corresponds to particle dispersion
   velocity of $c=0.82$ mm/sec and a ring vertical half-thickness of $h=5.4$ meters).
   The ring's outer edge is also placed at  resonance at $a=a_r$, and 
   the drag coefficient $C_d=0$. $N=300$ streamlines are used to model
   the ring's outermost $w=N\Delta a=150$ km with a spatial sampling of $\Delta a=0.5$ km.
   These three simulations do have distinct
   viscosities $\nu_s = \nu_b = 5, 50$, and 500 cm$^2$/sec,
   which are indicated by the labels in Fig.\ c.
   Note also that the curves in  Figs.\ a, b, and d
   all lie on top of each other for each of the three simulations, due to 
   $R_m$, $s$, $q$, and $\ell$ being insensitive to the choice of $\nu_s = \nu_b$.
   The torque that the satellite exerts on the ring
   is $|T_s/T_{GT}|=1.5\times10^{-5}, 1.5\times10^{-4},$ and $1.5\times10^{-3}$
   for  the $\nu_s = \nu_b = 5, 50$, and 500 cm$^2$/sec ring models.
}
\end{figure}

\newpage
\begin{figure}
\epsscale{1.25}
\hspace*{-15ex}\plotone{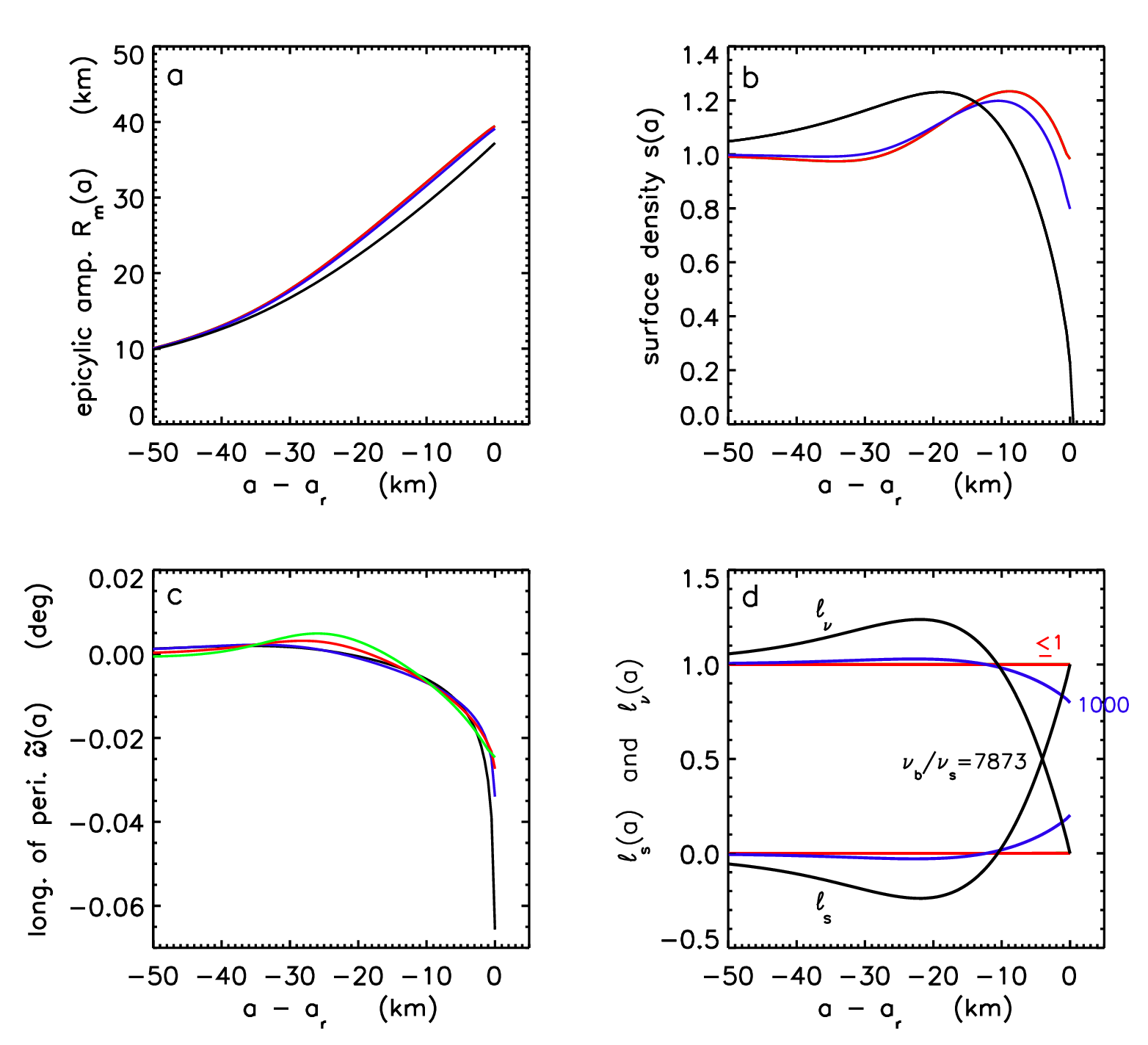}
\end{figure}

\newpage
\begin{figure}
\figcaption{
    \label{vb_fig}
    The epicyclic amplitude $R_m$ (Fig.\ a) is plotted versus
     radial distance $a-a_r$ from the simulated B ring's outer edge, as well as its
    fractional surface density $s$ (Fig.\ b),  longitude of periapse $\tilde{\omega}$
    (Fig.\ c), and the angular momentum luminosities $\ell_\nu$ and $\ell_s$ (Fig.\ d).
    Shown are the results of three simulations that all have the same
   undisturbed surface density $\sigma_\infty=30$ gm/cm$^2$,
   stability parameter $Q=2$, no drag ($C_d=0$), and an edge at the
   nominal resonance where $a=a_r$. $N=300$ streamlines are used to model
   the ring's outermost $w=150$ km with a spatial sampling of $\Delta a=0.5$ km.
   The simulated rings' shear and bulk
   viscosities all satisfy $\nu_s + \nu_b = 100$ cm$^2$/sec
   while having distinct ratios $\nu_b/\nu_s=0$ (green curve), 1 (red), 1000 (blue), and 
   7873 (black curve, which is the only simulation that satisfies the torque-balance
   requirement). Note that simulations having $\nu_b/\nu_s=0$ and $\nu_b/\nu_s=1$
   are indistinguishable in Figs.\ a,b, and d, so the green curves are hidden
   under the red. The torque that the satellite exerts on these model rings
   ranges over $8.8\times10^{-5}<|T_s/T_{GT}|=1.5\times10^{-4}$.
}
\end{figure}

\newpage
\begin{figure}
\epsscale{1.25}
\hspace*{-12ex}\plotone{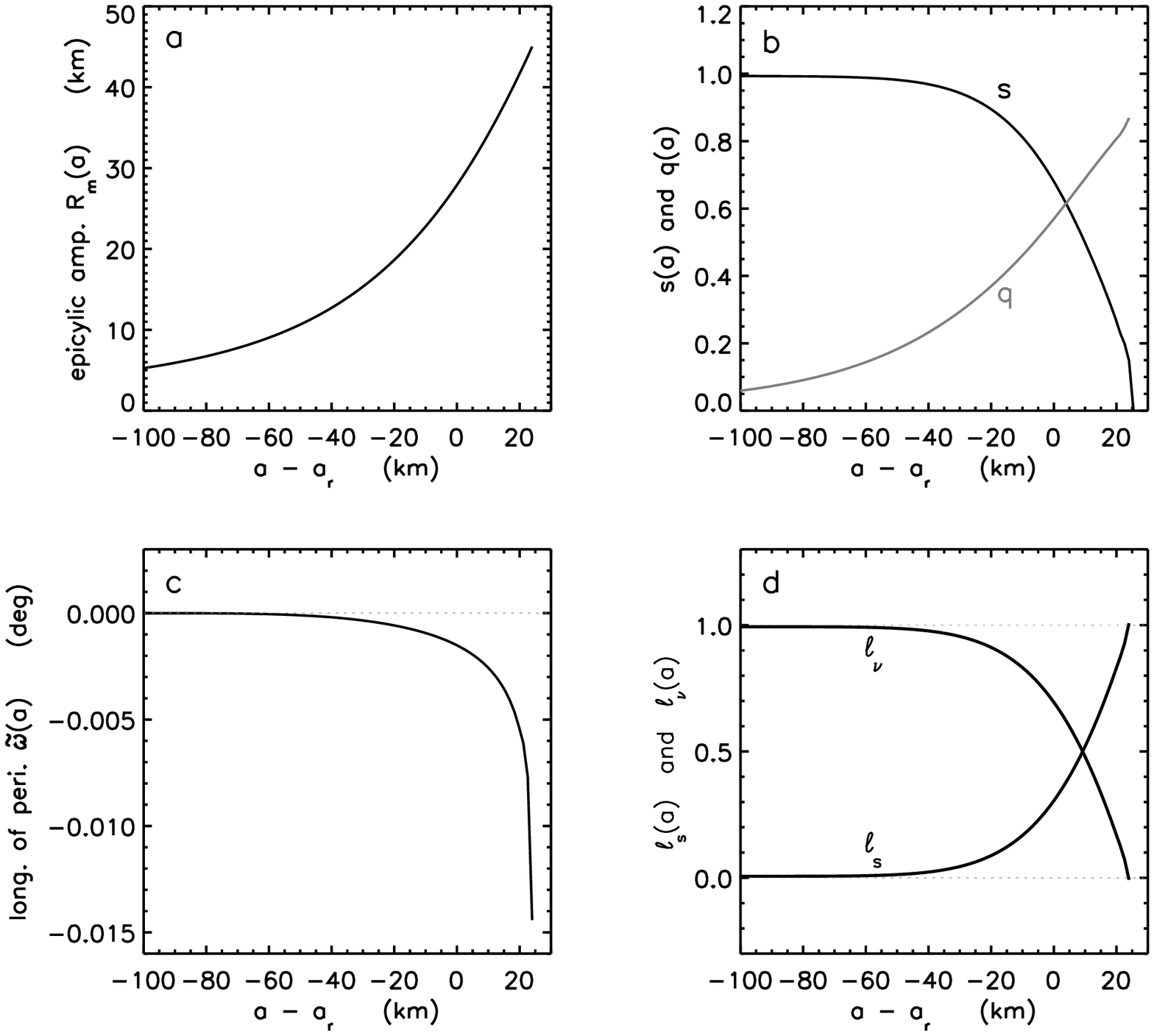}
\end{figure}

\newpage
\begin{figure}
\figcaption{
    \label{vb_edge_fig}
    Epicyclic amplitude $R_m$ (Fig.\ a), fractional surface density $s$
    and nonlinearity parameter $q$ (Fig.\ b),  
    longitude of periapse $\tilde{\omega}$ (Fig.\ c), and the 
    angular momentum luminosities $\ell_\nu$ and $\ell_s$ (Fig.\ d)
    are all plotted versus radial distance $a-a_r$ for a simulated B ring
    that has an undisturbed surface density $\sigma_\infty=226$ gm/cm$^2$,
    stability parameter $Q=2$, a kinematic shear viscosity of
    $\nu_s=0.00603$ cm$^2$/sec, kinematic bulk viscosity 
    $\nu_b = 100$ cm$^2$/sec,  no drag ($C_d=0$),
     and with an outer edge that lies 24 km beyond the nominal resonance
    position. $N=300$ streamlines
    are used over the ring's outer 424 km, so this calculation
    has a spatial sampling of $\Delta a=1.41$ km. The
    torque that the satellite exerts on the model ring is 
    $|T_s/T_{GT}|=4.2\times10^{-5}$.
}
\end{figure}

\newpage
\begin{figure}
\epsscale{1.0}
\plotone{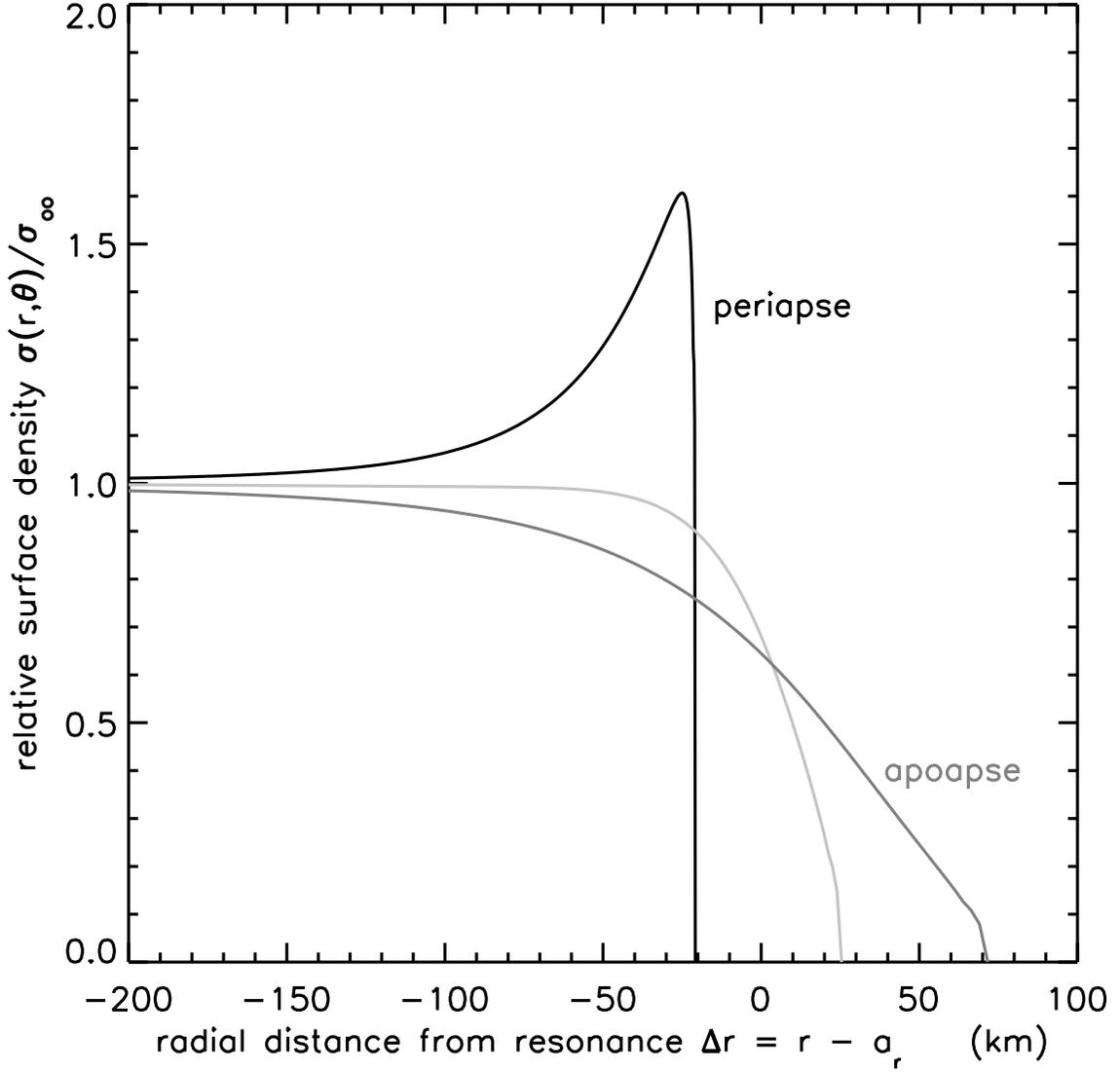}
\figcaption{
    \label{s_r_fig}
    Equation (\ref{sigma(a,theta)}) is used to
    calculate the ring's relative surface density
    $\sigma/\sigma_\infty$ as a function of semimajor axis $a$
    for the viscous B ring model that is described in Fig.\ \ref{vb_edge_fig}.
    Equation (\ref{Delta r}) is then used to convert that surface density profile into 
    a function of planetocentric distance $r$, which is plotted here
    versus distance $\Delta r = r-a_r$ from the nominal resonance.
    These curves give the simulated B ring's relative surface density
    along longitude $\theta=\theta_s$ (which is along the ring's longitude of periapse),
    longitude $\theta=\theta_s\pm90^\circ$ (along the ring's longitude of apoapse),
    and along the intermediate longitude $\theta=\theta_s\pm45^\circ$.
}
\end{figure}

\newpage
\begin{figure}
\epsscale{1.25}
\hspace*{-15ex}\plotone{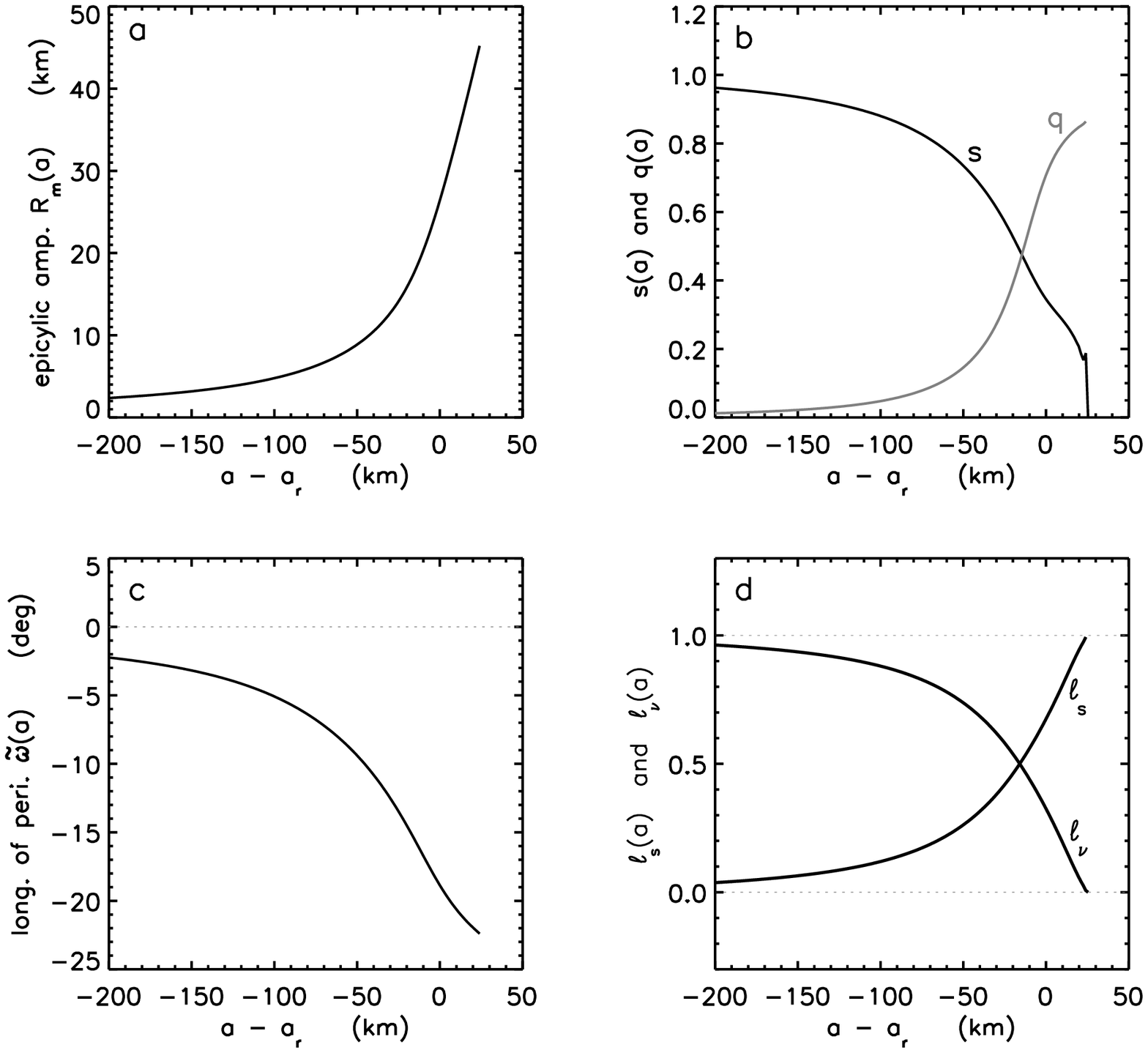}
\end{figure}

\newpage
\begin{figure}
\figcaption{
    \label{drag_fig}
    This simulated B ring has a surface density $\sigma_\infty=265$ gm/cm$^2$,
    stability parameter $Q=2$, a kinematic shear viscosity of
    $\nu_s=46$ cm$^2$/sec, kinematic bulk viscosity 
    $\nu_b=0$,  and a drag parameter $C_d=1.0\times10^{-4}$,
    with these parameters chosen so that epicyclic amplitude
   at the ring's outer edge is $R_m=45$ km, and that the
   ring's viscous torque balances the satellite's torque
   at a distance of $a-a_r=24$ km beyond the ring's nominal
    resonance.  $N=300$ streamlines were used to model the ring's
   outermost 424 km, so the spatial sampling here is $\Delta a = 1.41$km.
   The torque that the satellite exerts on the ring is $|T_s/T_{GT}|=0.32$.
}
\end{figure}

\newpage
\begin{figure}
\epsscale{1.0}
\plotone{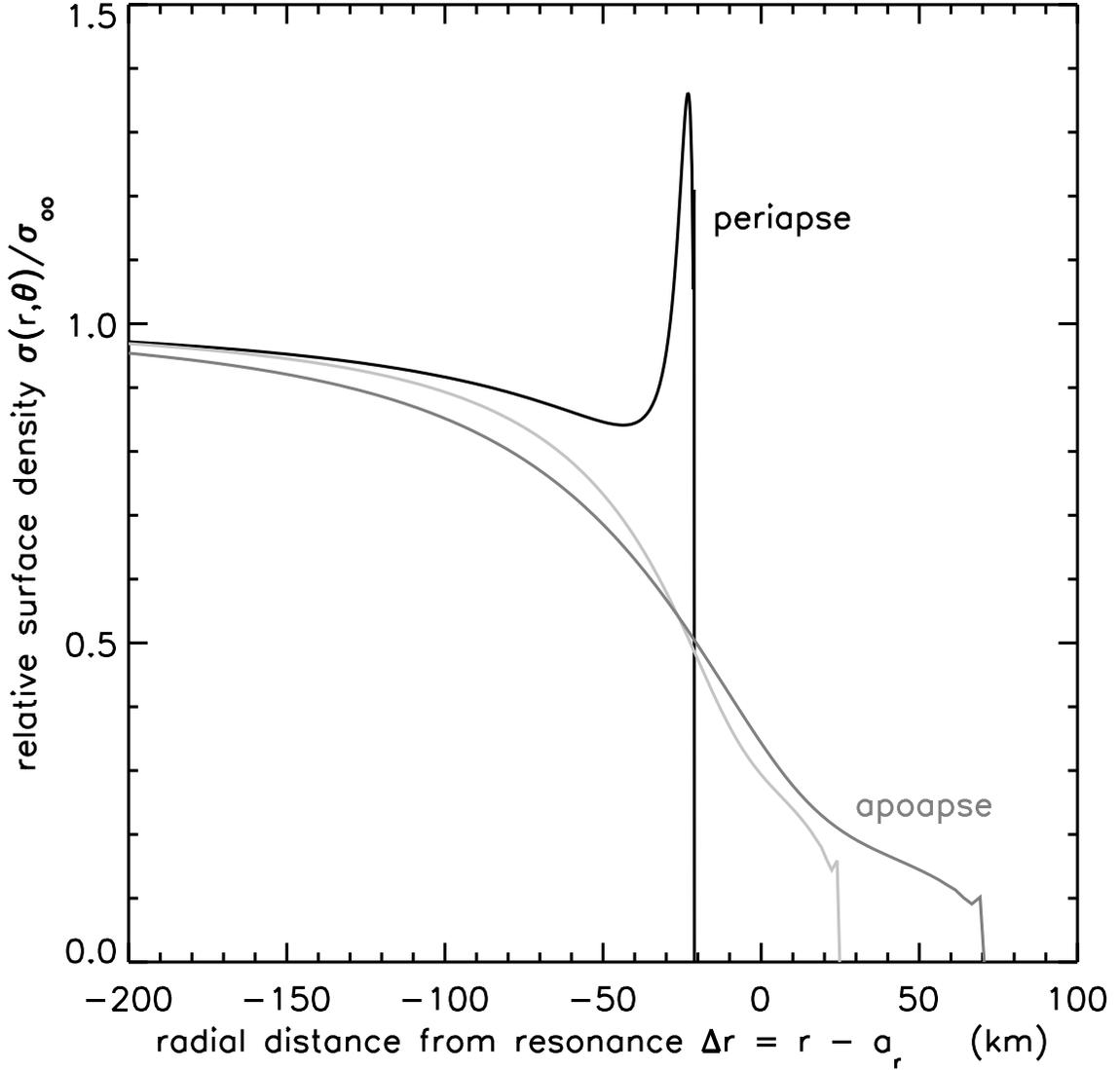}
\figcaption{
    \label{drag_s_r_fig}
    Equations (\ref{sigma(r,theta)}) are used to
    calculate the ring's relative surface density
    $\sigma(r,\theta)/\sigma_\infty$ as a function of radial distance $\Delta r = r-a_r$
    from the nominal resonance for the model B ring that is described in 
    Fig.\ \ref{drag_fig}. These curves give the simulated ring's relative surface density
    along the ring-edge's longitude of periapse, its longitude of apoapse, and
    at an intermediate longitude. Note also
    the small bumps seen at the rightmost part of these curves. They are due to
    the small, marginally resolved surface density excess that is barely seen
    to the right in Fig.\ \ref{drag_fig}b. That bump is due to the pressure drop that 
    the outermost streamline experiences (e.g, Eqn.\ \ref{alpha_pr_outer}), 
    and it disappears when the ring is pressureless with $Q=0$.
 }
\end{figure}

\newpage
\clearpage
\begin{figure}
\epsscale{1.0}
\hspace*{-15ex}\plotone{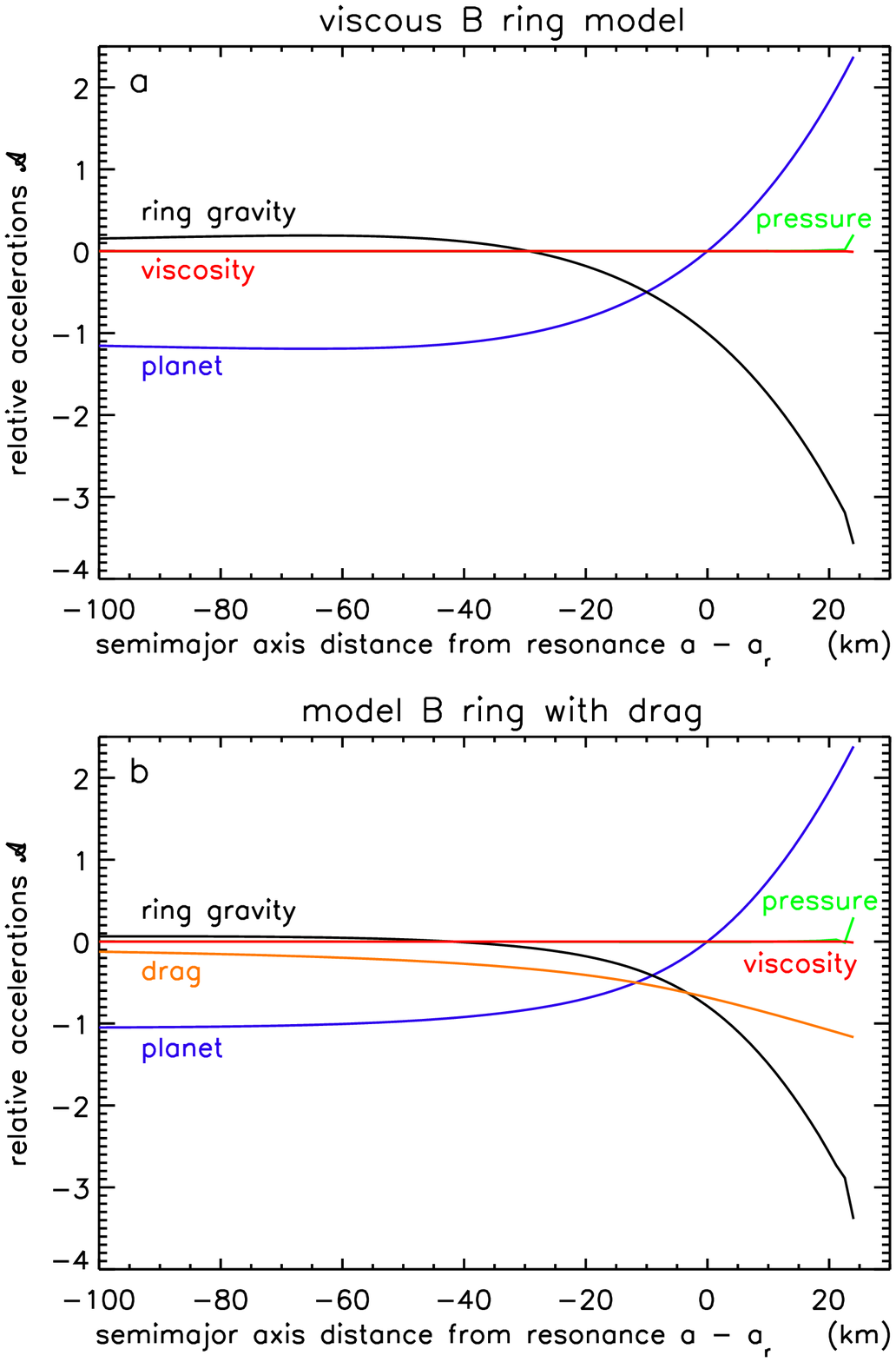}
\end{figure}

\newpage
\begin{figure}
\figcaption{
    \label{forces_fig}
    These figures plot the relative accelerations
    that a ring particle experiences due to ring gravity ${\mathcal A}_{g}$ (black curve),
    pressure ${\mathcal A}_{p}$ (green curve), 
    viscosity ${\mathcal A}_{\nu}$ (red curve),
    drag ${\mathcal A}_{d}$ (orange curve), and ${\mathcal A}_{cp}$,
    which is the acceleration that the central planet exerts on the particle; see
    Section \ref{forces_section} for details. The upper figure is for the model
    reported in Figs.\ \ref{vb_edge_fig}--\ref{s_r_fig} that
    achieved its torque-balance at the
    ring's outer edge via an extreme viscosity ratio $\nu_b/\nu_s=8473$.
    The lower Figure is for the model described in Figs.\ \ref{drag_fig}--\ref{drag_s_r_fig},
    which relies on drag to enable a torque-balance.
}
\end{figure}

\end{document}